\documentclass[aps,prd,twocolumn,floatfix,altaffilletter,superscriptaddress,preprintnumbers, tightenlines,showpacs,showkeys,nofootinbib,notoccite]{revtex4-2}
\usepackage{amssymb,amsmath,amsfonts}
\usepackage{natbib}
\usepackage{graphicx}
\usepackage{dcolumn}
\usepackage{tensor}
\usepackage{bm}
\usepackage{float}
\usepackage{color}
\usepackage[dvipsnames]{xcolor}
\usepackage{listings}
\usepackage{fancyvrb}
\usepackage{slashed}
\usepackage{ulem}
\usepackage{threeparttable}
\usepackage{enumitem}
\usepackage{listings}
\usepackage{mathtools}
\usepackage{slashed}
\usepackage{verbatim}
\usepackage{hyperref}
\usepackage{braket}
\usepackage{comment}

\def\ee{\end{eqnarray}}

\newcommand{\MeV}{{\,\rm MeV}}

\newcommand{\be}{\begin{eqnarray}}
\newcommand{\en}{\end{eqnarray}}
\newcommand{\bea}[1]{\left(\begin{array}{#1}}
\newcommand{\ena}{\end{array}\right)}
\newcommand{\ba}{\begin{eqnarray}}
\newcommand{\ea}{\end{eqnarray}}

\newcommand{\gs}{{\rm gs}}  
\newcommand{\ex}{{\rm ex}}  
\newcommand\myshade{80}
\colorlet{mylinkcolor}{Blue}
\colorlet{mycitecolor}{Red}
\colorlet{myurlcolor}{violet}
\hypersetup{
	linkcolor  = mylinkcolor!\myshade!black,
	citecolor  = mycitecolor!\myshade!black,
	urlcolor   = myurlcolor!\myshade!black,
	colorlinks = true
}

\begin{document}

\title{Revisiting Turner Window Axions: The Untapped Potential \\ of NaI Dark Matter Detectors}
\author{W. C. Haxton}
\email{haxton@berkeley.edu}
\affiliation{Department of Physics, University of California, Berkeley, CA 94720, USA}
\affiliation{Institute for Nuclear Theory, University of Washington, Seattle, WA 98195, USA}
\affiliation{Lawrence Berkeley National Laboratory, Berkeley, CA 94720, USA}
\author{Xing Liu} 
\email{xingyzt@berkeley.edu}
\affiliation{Department of Physics, University of California, Berkeley, CA 94720, USA}
\author{Anupam Ray}
\email{anupam.ray@queensu.ca}
\affiliation{The Arthur B. McDonald Canadian Astroparticle Physics Research Institute, Department of Physics, Engineering Physics, and Astronomy, Queen’s University, Kingston, ON, K7L3N6, Canada}
\affiliation{Perimeter Institute for Theoretical Physics, Waterloo, ON, N2L2Y5, Canada}
\author{Evan Rule} 
\email{erule@lanl.gov}
\affiliation{Theoretical Division, Los Alamos National Laboratory, Los Alamos, NM 87545, USA}

\date{\today}
            
\begin{abstract}
The ``Turner window" corresponds to axions with masses $\gtrsim$ 1 eV that have sufficiently strong couplings to matter to evade limits from the cooling of SN1987A. This window, through which the trajectories for the KSVZ and DFSZ QCD axions run, has been thought to be largely closed because of (1) the floor 
established by SN1987A cooling, (2) the absence of SN1987A-associated photons in the Kamioka II 
detector, and (3) the limit on neutrons produced by solar axions in the Sudbury Neutrino Observatory. We show that a more complete treatment of the axion opacity in SN1987A,  significantly weakens (2).  Consequently, for axion or axion-like particles with hadronic couplings, $g_{ann}$ and $g_{app}$, significant regions within the Turner window now become viable. We describe a new opportunity to constrain such hadronically coupled axions via their resonant absorption in NaI detectors. The source is the Milky Way's carbon-burning stars --- the progenitors of ONeMg white dwarfs as well as electron-capture and core-collapse supernovae --- which synthesize significant quantities of $^{23}$Na, keeping it at temperatures $\sim 10^9$K for periods up to tens of thousands of years.  $^{23}$Na acts as a thermal pump to convert stellar energy into axions, which arrive at the Earth as a thermally broadened line at 440 keV.  These axions can be detected via resonant absorption in NaI,
with the needed detector arrays already in place, developed by DAMA/LIBRA and other collaborations to search for the elastic scattering of light WIMPs. In axion detection, NaI serves as both the target, producing $\gamma$'s following resonant absorption, and the detector for those $\gamma$'s. With current array
masses and backgrounds, we find that the coupling range $|g_{app}| \sim 10^{-6.5}$--$10^{-2}$ can be covered after two years of data, including  
QCD axions with $m_a \gtrsim 10$ eV.
\end{abstract}

\maketitle
\preprint{N3AS-26-007, INT-PUB-26-016, LA-UR-26-21773}

\section{Introduction}
\label{introduction}
This paper provides details of the calculations first described in the companion letter \cite{HLRR}, in which it is argued that existing NaI dark matter detectors, currently being
used in searches for a low-energy signal from light WIMP scattering, have a new application --- the resonant capture of 440 keV line axions produced by the carbon-burning stars in our galaxy \cite{HLMR}.

It was noted some time ago~\cite{PhysRevLett.66.2557}  that stars can produce axions~\cite{PhysRevLett.38.1440,
PhysRevLett.40.223,PhysRevLett.40.279,Caputo:2024oqc} as thermally broadened lines. This occurs via a mechanism in which certain nuclei effectively serve as ``pumps,” converting thermal energy into particle emission through repeated cycles of photo-excitation and axio-deexcitation of a nuclear excited state. For this process to be efficient, several conditions must be satisfied: the relevant isotope must be sufficiently abundant, the excitation energy should not greatly exceed the stellar core temperature $T_c$, and the transition probability for axion emission must be favorable. When all these criteria are fulfilled, significant axion emission continues throughout the period in which thermal conditions produce a non-negligible Boltzmann occupation of the excited state.

These requirements are rather restrictive, as they call for an odd-$A$ isotope with appreciable abundance, a relatively small excitation gap, and a sufficiently strong magnetic transition probability. Stellar compositions, however, are dominated by H and by $\alpha$-stable fusion products such as $^4$He, $^{12}$C, and $^{16}$O with large gaps and E2 transition probabilities, rendering these isotopes inefficient axion emitters. In contrast, odd-$A$ abundances are typically low, and may be zero for isotopes that burn at high temperatures. When both the excitation energy and transition strength are taken into account, $^{57}$Fe and $^{23}$Na were identified in~\cite{PhysRevLett.66.2557} as the most efficient axion emitters at low ($T_8 \lesssim 1$) and high ($T_8 \gtrsim 4$) stellar core temperatures, respectively where $T_8$
is the temperature in 10$^8$K. Their relevant properties are summarized in Table~\ref{tab:isotopes}.

Over the years, significant attention has been given to $^{57}$Fe in this context. Its 14.4 keV M1 transition provides one of the most extensively studied channels for axion production in stellar environments. The mass fraction of $^{57}$Fe in stars of solar metallicity is $\approx 3 \times 10^{-5}$.  This isotope is most effective in producing axions at temperatures of $T_8 \approx 1$, a value typical of the cores of red giants.  At this temperature the Boltzmann occupation of the 14.4 keV $^{57}$Fe excited state is $\approx 0.27$. In \cite {PhysRevLett.66.2557}, red giant cooling was used to constrain the axion-nucleon effective coupling $g^\mathrm{eff~^{57}Fe}_{aNN}$ of Table \ref{tab:isotopes}.
\begin{table}[h!]
	\centering
    \resizebox{\columnwidth}{!}{%
	\begin{tabular}{ccccc}
		\hline
		& & & & \\[-4pt]
		Source &$\Delta$E (keV) & T Range & B(M1) & ~$g_{aNN}^\mathrm{eff}$~ \\[4pt]
		\hline
		& & & & \\[-4pt]
		$^{57}$Fe & 14.4 & $ T_8 \lesssim 1$ & 0.0078 & $g_{ann}+0.088 g_{app}$ \\
		$^{23}$Na & 440.2 & $4 \lesssim T_8$ & 0.225 & $g_{app} -0.062g_{ann}$ \\[4pt]
		\hline
	\end{tabular}%
    }
	\caption{Comparative properties of the axion emitters $^{57}$Fe and $^{23}$Na. These 
		metals are expected to dominate line axion emission for stellar core temperatures in the indicated ranges \cite{PhysRevLett.66.2557}.}
	\label{tab:isotopes}
\end{table}

Solar $^{57}$Fe has also been used as a source of axions for detection on the Earth~\cite{PhysRevLett.75.3222,KRCMAR199838,NAMBA2007398,Derbin}.  The flux is limited by the Sun's 
low central temperature, $T_c \approx 0.15 \times 10^8$\,K, which leads to a Boltzmann occupation of the 14.4 keV excited state of
just $5 \times 10^{-6}$ (averaged over the core).  The core mass of $^{57}$Fe is also modest, $\approx 7.7 \times 10^{-6}$ $M_{\odot}$.  Consequently, the mass of radiating $^{57}$Fe$^*$ is
only $10^{-11} M_\odot$.  Nevertheless, the Sun's proximity
makes it an interesting source. CAST (CERN Axion Solar Telescope) has been used in a search for $^{57}$Fe axions
through their $a \rightarrow \gamma$ conversion in the 9T CAST magnet~\cite{CASTcollaboration_2009}. The signal depends on the product of the axion-nucleon
$g_{aNN}^\mathrm{eff~^{57}Fe}$ (production) and axion-photon $g_{a \gamma \gamma}$
(detection) couplings. 
The CUORE \cite{CUORE:2012ymr} and Xenon1T \cite{XENON:2020rca} collaborations have conducted similar searches,
and there are upcoming plans to extend CAST limits with the International Axion Observatory (IAXO) \cite{DiLuzio:2021qct}. (See \cite{Fleury:2022plh,Ning:2025kyu} for other recent proposals for $^{57}$Fe axion searches.) 

The 14.4 keV $^{57}$Fe transition is widely used in
recoilless M\"{o}ssbauer spectroscopy.  As discussed by Moriyama
\cite{PhysRevLett.75.3222}, the analogous process of resonant absorption $a+{^{57}\mathrm{Fe}} \rightarrow {^{57}\mathrm{Fe}^*}$ followed by $\gamma$ emission  ---
the inverse of the stellar production process of \cite{PhysRevLett.66.2557} --- can be used to directly detect astrophysical $^{57}$Fe axions, because of thermal
broadening of the axion line. The cross section is enhanced by the ratio of the axion energy
to the thermal width of the line.
This signal has been exploited in several experiments \cite{KRCMAR199838,NAMBA2007398,Derbin}. It
has the important advantage of
isolating a single axion-nucleon coupling, as $g_{aNN}^\mathrm{eff~^{57}Fe}$ governs both production and detection.

We argue below that when all factors are considered in both production and detection, galactic $^{23}$Na axions are significantly
more attractive as a resonant-absorption target-of-opportunity than $^{57}$Fe axions --- despite the enormous advantage in having the Sun as a nearby source
of the latter. First, the nuclear physics of $^{57}$Fe is not optimal for resonant absorption: the transition is relatively weak, with a B(M1) value of just 0.0078 W.u.  In addition, the axion coupling is dominantly to the unpaired valence neutron in $^{57}$Fe, while for KSVZ \cite{PhysRevLett.43.103,SHIFMAN1980493} and DFSZ \cite{DINE1981199,osti_7063072} axions, the coupling
to protons is typically much stronger, e.g., $(g_{app}/g_{ann})^2 \approx 550$  for KSVZ axions.  This factor enters quadratically, affecting both the production and the resonant absorption on Earth. Second, the Sun is not an optimal source of $^{57}$Fe, due to its low core temperature and
modest radiating mass of $^{57}$Fe, as noted above.  Third, and perhaps most important, are the experimental challenges that arise for $^{57}$Fe. The isotope comprises only 2.2\% of natural Fe, so substantial target enrichment is required. The experiments performed have utilized thin foils, with the produced photons viewed in external Si(Li) detectors or silicon PIN photodiodes.  The use of foils --- necessary to avoid photon attenuation ---  places practical limits on target masses.

These experimental conditions contrast with those for $^{23}$Na axions, where 1) the transition strength is significantly larger, B$(\mathrm{M1}) \approx 0.225$
 W.u.; 2) the nucleus contains an unpaired proton; 3) no enrichment is required as $^{23}$Na is the only naturally occurring sodium isotope; 4) a NaI detector can serve as both $\gamma$-source and $\gamma$-detector; and most important, 5) the needed detectors have already been developed by the dark matter community for WIMP searches, with masses ranging up to 250 kg.

The rest of the paper is organized as follows. In Sec.~\ref{sec:II}, we briefly describe the basics of carbon burning and outline the key factors that lead to a $^{23}$Na axion flux at Earth comparable to the solar $^{57}$Fe axion flux. We contrast two detection strategies: the conversion of ultralight $^{23}$Na ALPs with masses $\lesssim$ 1 neV  into 440 keV $\gamma$ rays via mixing in the Galactic magnetic field --- detectable by all-sky $\gamma$-ray instruments such as COSI \cite{Tomsick:2023aue} --- and the approach pursued here and in \cite{HLRR}, namely the direct detection of heavier $^{23}$Na axions ($m_a \gtrsim$ few eV) through their resonant absorption in NaI.

In Sec.~\ref{sec:III} we discuss the nuclear physics of axion absorption and emission, including the low-energy cross section for the 3/2$^+\mathrm{(g.s.)}  \rightarrow 5/2^+$(440 keV) resonant absorption transition.  Following \cite{PhysRevD.37.618,PhysRevLett.66.2557},
we express the absorption cross section and axion decay rate in terms of the known $\gamma$ decay rate of the $5/2^+$ level, which limits the nuclear
physics input to the evaluation of two matrix-element ratios.  The resonant enhancement of the absorption is captured in the ratio $q_a/\sigma_\mathrm{TH}$ where $q_a$ is the magnitude of the axion three-momentum and $\sigma_\mathrm{TH}$ the thermal width of the axion line.  In addition to the $J^\pi =1^+$ longitudinal-spin allowed amplitude that dominates axion emission
from $^{23}$Na, axions can be absorbed through other abnormal-parity axial charge and longitudinal spin multipoles, $0^-$, $2^-$, $3^+,\ldots$.  These become important at higher axion energies, affecting certain axion observables that we examine
in Sec. \ref{sec:VI}.

In Sec.~\ref{sec:IV} we summarize the galactic modeling that was carried out in \cite{HLMR} to determine the $^{23}$Na axion flux at
Earth.  As $\sim$ 100 stellar sources contribute at any given time, the flux is a statistical quantity, determined by the locations and numbers of the carbon-burning stars, their masses, and the specific phase of carbon burning in each star. The probability distribution, determined by repeating the simulation
many times, is Gaussian on the low flux side and Lorentzian on the high flux side.  We also discuss a recent paper \cite{Saio} arguing, on the basis
of its observed brightness variations, that
Betelgeuse is in the late stage of carbon burning, in which case it would dominate today's axion flux at Earth due to its proximity.  Finally, we consider mechanisms by which the $^{23}$Na axions produced in a star can be
reabsorbed in that star, and thus removed from the flux. While we evaluate standard mechanisms such as Compton scattering and the Primakoff process, 
resonant reabsorption on $^{23}$Na dominates the opacity. As a result, at larger axion–nucleon couplings, emission originates from a narrow axio-sphere that forms at the edge of the carbon-burning region, with the escaping axions radiated radially outward from that shell.

In Sec.~\ref{sec:V} we consider axion detection through resonant absorption in existing NaI arrays, assuming an exposure of 500 kg yrs.  Using backgrounds
typical of current detectors, we determine the $3\sigma$ threshold for axion detection.  We note that our simple analysis can likely be improved, e.g., by
experimentalists employing a maximum likelihood analysis in combination with background models they have developed.  We find that current
detectors are capable of ruling out a range of  effective hadron couplings $|g_{aNN}^\mathrm{eff~^{23}Na}|$ that spans over four decades, including
QCD axions with masses $m_a \gtrsim 10$ eV.  We note that some detectors are capable of operating at both high and low gains simultaneously \cite{reina},
thus enabling WIMP and axion searches to proceed in parallel.

In Sec \ref{sec:VI} we discuss existing astrophysical constraints on axions with hadronic couplings that reside in the ``Turner window", a term often used to describe axions with masses
$m_a \gtrsim 1$\,eV and nucleon couplings at or above weak interaction 
strength. These constraints come from the impact of axion emission on the cooling time of supernova SN1987A; the counting of $\gamma$'s in the Kamioka II detector produced by absorption of SN1987 axions on $^{16}$O followed by nuclear de-excitation; and the detection of neutrons produced from 
the breakup of deuterium in the SNO detector after absorption of 
energetic solar axions.  We show that the axion flux reaching the 
Kamioka II water detector is greatly diminished by axion 
reabsorption on nuclei within the SN1987A progenitor. The principal absorbers are the nuclei that are abundant near the axio-sphere: 
axions are absorbed in the breakup of $^4$He, $^{56}$Ni (and other iron-group elements), and $^{28}$Si, and in resonant absorption on $^{16}$O.
Consequently, the constraints imposed by the non-observation of Kamioka II $\gamma$'s are significantly weakened.  This change plus a careful
treatment of the isospin dependence of limits on $g_{aNN}$ show that
significant regions of the Turner-window axion parameter space remain open.  These parameter regions are the targets for NaI detectors. 

In Sec.~\ref{sec:results} we describe the potential impact of these NaI experiments, assuming an accumulation of data equivalent to 500 kg-yrs. Finally, in Sec.~\ref{sec:summary} we summarize our results and point to additional work that should be completed to ensure that every corner of the Turner window --- including the QCD axion
trajectories that pass through the window --- has been thoroughly probed.

\section{Galactic Line Axions from Carbon-Burning}
\label{sec:II}
It was recently shown that $^{23}$Na is a major source of galactic axions \cite{HLMR}.  In contrast to primordial $^{57}$Fe, the production
comes from {\it in situ} production of $^{23}$Na during the carbon-burning phase of stars
with masses $\gtrsim 7.5~M_\odot$. These are the progenitors of ONeMg white dwarfs (WDs) and electron-capture as well as core-collapse supernovae (SNe).  The production is quite substantial, with each star typically generating $\sim$ 0.1 $M_\odot$ of $^{23}$Na, prior to the onset of $^{20}$Ne burning. This is very unusual, as odd-$A$ nuclei often burn up immediately in stellar cores.  As there are several hundred Milky Way stars currently
burning carbon, the galactic $^{23}$Na axion flux is a statistical quantity,
with an expected value and a Gaussian distribution (on the low-flux side) that can be used to bound flux from below
to any desired confidence level \cite{HLMR}.

The relevant burning reactions are
\begin{equation}
{^{12}\mathrm{C}}+{^{12}\mathrm{C}} \rightarrow {^{24}\mathrm{Mg}}^* \rightarrow \left\{ \begin{array}{ll} {^{20}\mathrm{Ne}}+\alpha & +4.62 \MeV \\ {^{23}\mathrm{Na}}+p & + 2.24 \MeV \\
^{23}\mathrm{Mg} + n & - 2.60 \MeV \end{array} \right.~.
\end{equation}
The branch to $^{23}$Na is exothermic because the Coulomb force is sufficiently strong to favor $N>Z$.
At temperatures characteristic of carbon burning --- $(0.7-1.7) \times 10^9$K, depending on the progenitor and burning stage ---
the Boltzmann occupation of the 440.2 keV $^{23}$Na excited state is significant.
The mass of radiating $^{23}\mathrm{Na}^*$ at the end of carbon burning reaches $(0.001-0.01)~M_\odot$, 
depending on the progenitor, exceeding the solar mass of radiating $^{57}\mathrm{Fe}^*$ by typically eight orders of magnitude. Further, as discussed
in \cite{HLMR}, there are several hundred such sources contributing to the galactic $^{23}$Na axion inventory at any given time.
These factors combine with others we have already discussed to
favor $^{23}$Na axion production.

Consequently, while one might have na\"ively assumed that the galactic $^{23}$Na/solar $^{57}$Fe flux ratio at Earth might scale as the distance ratio $(d_\mathrm{sun}/d_\mathrm{gal})^2 \approx 10^{-19}$,
the actual ratio is $\approx 1/350$ (KSVZ couplings). Detection considerations then strongly swing the balance in favor of galactic axions.
If detection is via $a \rightarrow \gamma$, the conversion probability scales like $(B_T d)^2$ where $d$ is the distance to the source and $B_T$
is the transverse component of the Galactic magnetic field.  This factor favors galactic conversion over conversion in the 9T CAST magnetic
by a factor of $\approx 10^{17}$ \cite{HLMR}.  Alternatively, if the detection is by resonant absorption, $^{23}$Na is favored by its
B$(\mathrm{M1})$ value, proton coupling, isotopic abundance, and large-volume detector potential.

Given that direct detection of $^{23}$Na axions would not make use of a critical advantage --- their efficient conversion in the galactic magnetic field --- why would such detection be of interest?
First, such a measurement would greatly extend the range of axion masses probed.  The axion-photon conversion of galactic axions is most effective for $m_a \lesssim 0.1$ neV: above this mass the conversion
probability rapidly diminishes. In fact, the impact of this constraint is lost by $m_a \sim 10$ neV~\cite{HLMR}.  In contrast, the resonant absorption cross section 
contributes at full strength for all $m_a \ll 440$ keV.  Second, in resonant absorption both the stellar production
and detection depend solely on $g_{aNN}^\mathrm{eff~^{23}Na}$, thus cleanly isolating one coupling.  Third, the
strong B$(\mathrm{M1})$ and more favorable phase space that contributes to the strength of stellar production 
also enhances the resonant absorption cross section.   Fourth, we will later find, after re-analyzing existing constraints, that experiments exploiting the galactic $^{23}$Na flux can probe certain regions of axion parameter
space that are currently not adequately constrained. The regions 
of interest lie above $m_a \gtrsim$ few eV and apply to
axion or axion-like particles with hadronic couplings
$g_{ann}$ and $g_{app} \gtrsim 10^{-7}$.

Although resonant absorption can be exploited for both $^{23}$Na and
$^{57}$Fe axions, it is very difficult to achieve large exposures
with the latter.  As the abundance of $^{57}$Fe is just 2.12\%, expensive target
enrichment is required. Even with enrichment, the target must be 
sufficiently thin to
allow the $\gamma$-rays to escape into external detectors.  In contrast, $^{23}$Na is
the only stable sodium isotope, and can be used in the form of NaI(Tl):
there is no need for an external detector as NaI is both the $\gamma$-ray source and the detector. Further, the field is already heavily invested in this technology, with NaI(Tl) crystals employed as 
underground WIMP detectors by the ANAIS \cite{ntnl-zrn9}, COSINE \cite{yu2025limitswimpdarkmatter}, DAMA/LIBRA \cite{2021arXiv211004734B}, KIMS \cite{kim2015statuskimsnaiexperiment}, NAIAD \cite{Alner_2005}, NEON \cite{PhysRevLett.134.021802} and SABRE \cite{Milligan_2025}
collaborations.  The high-purity crystals and sophisticated vetoes
developed for light WIMP searches are also ideal for $^{23}$Na
axion experiments.

\section{The Resonant Absorption Cross Section} 
\label{sec:III}
Here we derive the cross section for resonant absorption: despite previous works, we have not found a general
result of the form of Eq. \eqref{eq:resonant2} in the literature. As the width of the detector's 440.2 keV
level is much smaller than the width of the thermally broadened axion profile, we can treat the
excited state energy as discrete.  Neglecting very small corrections due to nuclear target recoil, we find the cross section
\begin{equation}
    \sigma(\epsilon_a) = \frac{\pi}{q_a} |\mathcal{M}|^2 \, \delta(\epsilon_a-E_\ex),
    \label{eq:cross}
\end{equation}
where $\epsilon_a$ and $q_a$ are the incident axion energy and three-momentum, $E_\mathrm\ex$ is the nuclear excitation energy,
and $\mathcal{M}$ is the invariant amplitude connecting the initial and final nuclear states.  As we will discuss both emission and absorption processes, it is helpful to use state labels that distinguish the two directions. For resonant absorption we set $|J_i M_i \rangle =
|J_\gs M_\gs \rangle$ and $|J_f M_f \rangle = |J_\ex M_\ex \rangle$.

The ALP-nucleon interaction is
\begin{eqnarray}
	\mathcal{L} &=& \frac{1}{2M_N}  \bar{N}\gamma^\mu \gamma_5 (g^0_{aNN} +g^3_{aNN} \tau_3 ) N \, \partial_\mu a \nonumber \\
	&=& \frac{1}{2M_N} \bar{N}\gamma^\mu \gamma_5 \left[ g_{app}\textstyle{ \left(\frac{1+\tau_3}{2} \right) }+ g_{ann} \left(\frac{1-\tau_3}{2} \right) \right] N \, \partial_\mu a,~~~~
	\label{eq:ncoupling}
\end{eqnarray}
where $M_N$ is the nucleon mass and $\tau_3$ is the third component of isospin, with $g_{app}= g^0_{aNN}+g^3_{aNN}$ and $g_{ann}=g^0_{aNN}-g^3_{aNN}$. 
Identifying the nuclear current as the corresponding sum over single-nucleon currents, performing a nonrelativistic
reduction, while averaging over initial nuclear spins and summing over final spins, one obtains to leading order
in $1/M_N$, 
\begin{align}
&\frac{1}{2J_\gs +1} \sum_{M_\gs M_\ex}|\mathcal{M}|^2 =\frac{\pi}{2 J_\gs+1} \frac{q_a^2}{M_N^2} ~ \times ~~~\nonumber \\
&\,~\sum_J|\langle J_\ex \| \sum_{i=1}^A \left( \Sigma^{\prime \prime}_{J}(q_a x_i) -\frac{\epsilon_a}{M_N} \Omega^\prime_J(q_a x_i) \right) \times \nonumber \\
&\,~~~~~~~~~~~~~~~~~~~~~~~~~~~~\left( g^0_{aNN} +g^3_{aNN} \tau_3(i) \right) \|  J_\gs \rangle |^2 ~.
\label{eq:invariant}
\end{align}
The longitudinal spin $\Sigma_J^{\prime \prime}$ and axial charge $\Omega_J^\prime$ are abnormal parity operators \cite{DONNELLY1979103}, so that
axion absorption on a $0^+$ nuclear ground state generates transitions to $J^\pi=0^-$, $1^+$, $2^-$, $\dots$ states.
Here $||$ denotes a matrix element reduced in angular momentum (but not isospin). For the case of axion emission, one reverses the sign on the second term.

We will later evaluate these responses in the nuclear shell model, with Slater determinants constructed in a harmonic 
oscillator basis, where the oscillator parameter $b$ is related to the nuclear radius.  Defining
$y=(q_a b/2)^2$, one finds in the long-wavelength
limit (where only the leading term in $y$ is retained)
\begin{equation}
    \frac{\epsilon_a}{M_N} \Omega_J^\prime \sim \frac{\epsilon_a}{M_N} y^{(J-1)/2},~~~~~\Sigma_J^{\prime \prime} \sim \left\{ \begin{array}{ll} y^{(J-1)/2} & J \ge 1 \\ ~y^{1/2} & J=0 \end{array} \right.~~
    ~.~~
\end{equation}
It follows that for $J \ge 1$, the contribution of $\Omega_J^\prime$ relative to $\Sigma_J^{\prime \prime}$ is suppressed by $\epsilon_a/M_N \sim \frac{1}{50}$, and thus can be ignored.  In this limit, the $J=1$ longitudinal spin operator reduces to its Gamow-Teller (GT) form
\begin{equation}
    \Sigma_1^{\prime \prime}(q_a x_i)  \rightarrow \frac{1}{\sqrt{12 \pi}} \sigma(i),
    \label{eq:LWL}
\end{equation}
while the first-forbidden corrections to the rate come from $\Sigma_0^{\prime \prime} - 
\frac{\epsilon_a}{M_N} \Omega_0^{\prime}$ and $\Sigma_2^{\prime \prime}$.  Consequently, for a nucleus with a $0^+$ ground state, one typically expects the cross section to be dominated by $0^+ \rightarrow 1^+$ transitions for low-energy axions.  The contributions from $0^+ \rightarrow 0^-$ and $0^+ \rightarrow 2^-$ transitions can become
significant for higher axion energies, as well as for closed-shell nuclei where the
allowed Gamow-Teller strength is suppressed.  For $0^+ \rightarrow 0^-$ transitions,
both the longitudinal-spin and axial-charge contributions must be retained, as they are typically of similar strength.

The 3/2$^+$ (g.s.) $\rightarrow 5/2^+$ (440.2 keV) resonant absorption of axions in $^{23}$Na is a low-energy process that can be evaluated in the allowed approximation.
The incoming line axions are thermally broadened, as they originate from stellar cores.  We can write the flux
as the total flux $\bar{\phi}_a$ multiplied by a (normalized) thermal probability distribution,
\begin{align}
    \phi_a(\epsilon_a) &= \bar{\phi}_a \, P(\epsilon_a) \nonumber \\ ~~P(\epsilon_a) &= \frac{1}{\sigma_\mathrm{TH} \sqrt{2\pi}}\, \exp\left[-\frac{(\epsilon_a-\bar{\epsilon}_a)^2 }{ 2\sigma_\mathrm{TH}^2}\right], \nonumber \\
   \sigma_\mathrm{TH}(T)  &= \bar{\epsilon}_a \sqrt{ \frac{k T }{ M_T}}~\Rightarrow ~\sigma_\mathrm{TH}(T_8=10)  \approx 0.88 \,\mathrm{keV},
   \label{eq:thermal}
\end{align}
where $M_T$ is the target mass, $\bar{\epsilon}_a \approx 440.2$ keV is the nuclear transition energy
in $^{23}$Na, and $\sigma_\mathrm{TH}$ is the thermal width, evaluated above for a temperature characteristic of carbon burning.  This Gaussian profile will break down very far from the peak, where it will heal to a Lorentzian reflecting the natural line profile widened by collisional effects, but this will not be an issue for our applications.

There are two effects that, in principle, shift $\bar{\epsilon}_a$ to the low-energy side of the na\"ive value $E_\ex$, the energy gap between the excited and
ground levels of the nucleus. First, the energy lost due to nuclear recoil in the emission --- already implicitly neglected in deriving the absorption cross section of Eq. \eqref{eq:cross} --- is readily found to be $E_\ex^2/2M_T \approx 4.5 \,\mathrm{eV}$.
Second, the gravitational red shift of the escaping axions is somewhat larger and
depends both on the point at which the axion is produced as well as the time, as the
star is evolving.  We have calculated the gravitational red shift as a function of position and time for the 9 and 10 $M_\odot$ progenitors, finding that it evolves from 100$\pm$5 eV at the
start of carbon burning to 200$\pm$30 eV at the end.  Evaluating the red-shift correction at 150 eV 
yields a correction to the flux density of less than 1.5\%. Consequently, both the recoil and the red shift are sufficiently small that the cross section can be evaluated with $P(\epsilon_a) \approx P(\bar{\epsilon}_a)$.  The flux density is then $ \approx 1/\sqrt{2 \pi} \, \sigma_\mathrm{TH}$.

Thus, combining Eqs. (\ref{eq:cross}), (\ref{eq:invariant}), (\ref{eq:LWL}), and (\ref{eq:thermal}) and integrating over the line profile, one obtains the flux-averaged cross section in the long-wavelength approximation
\begin{align}
   \langle  \sigma \rangle &=  \frac{{\pi} }{ 12 M_N^2} \frac{1 }{ \sqrt{2 \pi}\sigma_\mathrm{TH}} ~ q_a~\frac{1 }{ 2 J_\gs+1}~
    \times \nonumber \\
 & ~~|\langle J_\ex || \sum_{i=1}^A \sigma(i)(g^0_{aNN} +g^3_{aNN} \tau_3(i) ) ||  J_\gs \rangle |^2 \nonumber \\
&=\frac{{\pi} }{ 12 M_N^2} \frac{1 }{ \sqrt{2 \pi}\sigma_\mathrm{TH}}~ {q_a} ~\frac{1 }{ 2 J_\gs+1}~\times
   \nonumber \\
& ~~  |\langle J_\ex || \sum_{i=1}^A \sigma(i) \tau_3(i)  ||  J_\gs \rangle |^2 \left[ g^0_{aNN} \beta + g^3_{aNN}\right]^2~.
   \label{eq:resonant1}
\end{align}
The resonant enhancement is captured in the dimensionless ratio $q_a/\sqrt{2 \pi}\sigma_\mathrm{TH}$.
In the final two lines of Eq. \eqref{eq:resonant1}, the cross section has been expressed in terms of the nuclear matrix element ratio
\[ \beta \equiv \frac{\langle J_\ex || \displaystyle{\sum_{1=1}^A} \sigma(i) ||J_\gs \rangle  }{ \langle J_\ex || \displaystyle{\sum_{1=1}^A} \sigma(i)\tau_3(i) ||J_\gs \rangle } 
= \frac{\langle J_\gs || \displaystyle{\sum_{1=1}^A }\sigma(i) ||J_\ex \rangle  }{ \langle J_\gs || \displaystyle{\sum_{1=1}^A }\sigma(i)\tau_3(i) ||J_\ex \rangle },\]
and the isovector spin response.  

The $\gamma$ decay rate for the transition from $|J_i M_i \rangle = |J_\ex M_\ex \rangle $ to $|J_f M_f \rangle =|J_\gs M_\gs \rangle$ can be calculated similarly
\begin{align} 
&\omega_\gamma =(1+\delta^2) \frac{\alpha k_\gamma^3 }{ 12M_N^2}~\frac{1 }{ 2 J_\gs+1}~ \times \nonumber \\
&|\langle J_\gs || \sum_{1=1}^A \sigma(i)[\mu_0 +\mu_1 \tau_3(i)]
+ \ell(i) [1 + \tau_3(i)] ||J_\ex \rangle |^2
\end{align}
where $\delta$ is the $E2/M1$ mixing ratio and $\mu_0$ ($\approx 0.88$) and $\mu_1$ ($\approx 4.706$) are the isoscalar
and isovector nucleon magnetic moments, respectively.
As the total angular momentum operator $\vec{j}= \vec{\ell} +\vec{s} = \vec{\ell}+\vec{\sigma}/2$ cannot
generate transitions, the above can be rewritten
\begin{align} 
&\omega_\gamma =(1+\delta^2) \frac{\alpha k_\gamma^3 }{ 12M_N^2}\frac{1 }{ 2J_\ex +1}\times \nonumber \\
&|\langle J_\gs || \sum_{1=1}^A \sigma(i) \tau_3(i) ||J_\ex \rangle |^2
\left[(\mu_0-\textstyle\frac{1 }{ 2}) \beta + \mu_1 -\eta \right]^2,
\label{eq:gamma} 
\end{align}
where $k_\gamma$ is the photon momentum and
\[ \eta \equiv- \frac{\langle J_\gs || \displaystyle{\sum_{i=1}^A }\ell(i) \tau_3(i) ||J_\ex \rangle }{ \langle J_\gs || \displaystyle{\sum_{i=1}^A} \sigma(i) \tau_3(i) || J_\ex \rangle } =- \frac{\langle J_\ex || \displaystyle{\sum_{i=1}^A} \ell(i) \tau_3(i) ||J_\gs \rangle }{ \langle J_\ex || \displaystyle{\sum_{i=1}^A} \sigma(i) \tau_3(i) || J_\gs \rangle }\]
The matrix element ratios $\beta$ and $\eta$ are also used in \cite{PhysRevLett.66.2557,PhysRevD.37.618}.

One can substitute Eq. (\ref{eq:gamma}) into Eq. (\ref{eq:resonant1}), replacing the isovector spin matrix element
with the measured $\gamma$-decay rate.  Note that the square of the reduced nuclear matrix elements are identical for 
emission and absorption, while the associated spin statistical factors are not.  Setting $k_\gamma=\epsilon_a$ yields
\begin{eqnarray}
    \langle \sigma \rangle &&=  \frac{{\pi} }{ \alpha (1+\delta^2)} \frac{\Gamma_\gamma }{ \epsilon_a^3} \frac{q_a }{ \sqrt{2\pi} \sigma_{TH}} ~\times \nonumber \\
    &&~~\frac{2J_\ex +1 }{ 2J_\gs+1}\left[ \frac{g^0_{aNN}\beta +g^3_{aNN} }{ (\mu_0-\frac{1}{2}) \beta +\mu_1 -\eta} \right]^2,
    \label{eq:resonant2}
\end{eqnarray}
where the laboratory decay width $\Gamma_\gamma \equiv \hbar \omega_\gamma$.  

The  reduced matrix element ratios $\eta$ and $\beta$ are taken from \cite{PhysRevLett.66.2557,PhysRevD.37.618}. The experimental information needed to evaluate Eq. (\ref{eq:resonant2}) is taken from ENSDF
and listed in Table \ref{tab:data}. The effective couplings $g_{aNN}^\mathrm{eff}$ are given in terms of $g_{app}$ and $g_{ann}$  for
$^{57}$Fe and $^{23}$Na in Table
\ref{tab:isotopes}.
The thermal width is determined by the temperature $T_8$ of the axion source. Normalizing the $^{23}$Na cross section
to the nominal carbon-burning temperature of $T_8 =10$ (see Table
\ref{tab:data}) and assuming $\epsilon_a \gg m_a$, one obtains 
\begin{align}
    \langle \sigma(T_8) \rangle_{^{23}\mathrm{Na}}~=~ 0.53 \times 10^{-40} \mathrm{cm^2}~ \sqrt{\frac{10}{T_8}} ~\left|\frac{ g_{aNN}^{\mathrm{eff}~^{23}\mathrm{Na}} }{ 10^{-7}}\right|^2 ~.
    \label{eq:RA}
\end{align}
For the KSVZ axion, one can rewrite Eq. (\ref{eq:RA}) in terms of $m_a$ as
\begin{align}
    \langle \sigma(T_8) \rangle_{^{23}\mathrm{Na}} \big| \underset{\text{\tiny{KSVZ}}}{=}& 0.32 \times 10^{-40} \mathrm{cm^2}~ \sqrt{\frac{10}{T_8}} \left(\frac{m_a }{1~\mathrm{eV}} \right)^2 .
    \label{eq:crossKSVZ}
\end{align}

\begin{table}[h!]
	\centering
    \resizebox{\columnwidth}{!}{%
	\begin{tabular}{ccccccc}
		\hline
		& & & & & & \\[-4pt]
		Nucleus & $T_8$ & $\sigma_\mathrm{TH}$\,(eV) & $\Gamma_\gamma$\,(eV) & $\delta$  & $\beta$ & $\eta$ \\[4pt]
		\hline
		&& & & & \\[-4pt]
		$^{23}$Na & 10 & 883.& 4.01$\times 10^{-4}$  &   0.065  &  0.884  & -1.20  \\
        $^{57}$Fe & 0.14 & 2.17 & 4.85$\times 10^{-10}$  &  0.00223   & -1.192    & 0.80  \\[4pt]
		\hline
	\end{tabular}%
    }
	\caption{Parameters entering in the evaluation of $\langle \sigma \rangle$ for $^{23}$Na and $^{57}$Fe. The line widths
    $\sigma_\mathrm{TH}$ are evaluated at the indicated temperature. }
	\label{tab:data}
\end{table}

For comparison, we give the corresponding results for $^{57}$Fe. In this case, we normalize the source temperature to $T_8$=0.14, a value representative of the central core
of the sun. Note that the $^{57}$Fe 14.4 keV state's radiative width $\Gamma_\gamma$
differs substantially from the total width $\Gamma_\mathrm{tot}$, as the decays proceed primarily through internal conversion. The internal conversion 
coefficient of $8.56 \pm 0.26$ was used to extract $\Gamma_\gamma$ from the $\Gamma_\mathrm{tot}$.  The results
corresponding to Eqs. (\ref{eq:RA}) and (\ref{eq:crossKSVZ}) are
\begin{align}
    \langle \sigma(T_8) \rangle_{^{57}\mathrm{Fe}} =& 1.45 \times 10^{-40} \mathrm{cm^2} ~\sqrt{\frac{0.14}{T_8}}~\left|\frac{ g_{aNN}^{\mathrm{eff}~^{57}\mathrm{Fe}} }{ 10^{-7}}\right|^2 \nonumber \\
     \underset{\text{\tiny{KSVZ}}}{=}&  1.47 \times 10^{-42} \mathrm{cm^2} ~\sqrt{\frac{0.14}{T_8}}~\left(\frac{m_a }{1~\mathrm{eV}} \right)^2 
    \label{eq:RAB}
\end{align}
The first of Eqs. (\ref{eq:RAB}) shows that the narrow width of the solar $^{57}$Fe axion line compensates for the weaker
B(M1) value, while the second shows the effect of the suppressed KSVZ coupling to neutrons.

\section{The Axion Flux and Event Rates} 
\label{sec:IV}
Here, we discuss three issues controlling the galactic $^{23}$Na axion flux at Earth:
\begin{enumerate}
\item The expected flux of axions, which is a statistical quantity due to the large number of contributing galactic sources, in the regime where the star is transparent to their emission.
\item The possibility that now is a special time, when the flux is elevated, due to recent arguments that Betelgeuse may be in the late stages of carbon burning.
\item The axion stellar escape probability. We find that the axion opacity is governed by resonant reabsorption on $^{23}$Na.  Consequently, axion production, reabsorption, 
and detection
are all governed by the single coupling $g_{aNN}^\mathrm{eff~^{23}Na}$.
\end{enumerate}

\subsection{The $^{23}$Na flux probability distribution}
The $^{23}$Na flux at Earth is expected to be a statistical quantity, reflecting the contributions of several hundred
massive stars in our galaxy currently burning carbon. If the 440 keV state in $^{23}$Na is thermally
excited, the rate for axion emission can be evaluated in the allowed limit, just as described for
absorption, yielding
\begin{equation}
	{\omega_a \over \omega_\gamma} = {1 \over 2 \pi \alpha} {1 \over 1+ \delta^2} ~{q_a^3 \over \omega_a^3}~\left[ {g^0_{aNN} \beta+ g^3_{aNN} \over (\mu_0-{1 \over 2}) \beta +\mu_1 -\eta} \right]^2.
	\label{eq:rate}
\end{equation}
The axion yield from a single star at a specific time can be computed by integrating over the carbon-burning region, taking into account the mass fraction of $^{23}$Na and thermal population of the
440 keV level, both functions of the radius $r$.  The axion flux at earth from the entire galaxy requires an integration
over a galactic model in which the numbers, masses, and evolutionary stages of the carbon-burning
stars are specified.
The resulting axion flux probability distribution was determined in \cite{HLMR}, where many simulations were performed, each of which distributed the carbon-burning WD and SN progenitors
in time according to a Poisson distribution, with the progenitor type, mass,
and galactic position determined by Monte Carlo sampling over distributions
fitted to astrophysical data.  The results are accurately represented by a split Gaussian/Lorentzian distribution,
\begin{equation}
    P[\phi]=\sqrt{\frac{2}{\pi}} \frac{1 }{ \sigma_1+\sigma_2} \left\{ \begin{array}{cc} e^{-(\phi-\phi_0)^2/2 \sigma_1^2}, & \phi<\phi_0 \\ \left[ 1+\pi \frac{(\phi-\phi_0)^2 }{ 2 \sigma_2^2} \right]^{-1}, & \phi>\phi_0 \end{array} \right.
\end{equation}
where $\sigma_1 = 0.151 \phi_0$ and $\sigma_2=0.492 \phi_0$, where $\phi_0$ is the most probable flux, 
\begin{equation}
    \phi_0 = 0.173 \times 10^6 \left| \frac{ g_{ann}^\mathrm{eff \, ^{23}Na} }{ 10^{-7}} \right|^2\mathrm{cm^{-2}s^{-1}}.
\end{equation}
The distribution is quite skewed, reflecting the underlying statistics.  Because there are many sources 
contributing at any given time, the low-flux side of this distribution is rather sharply bounded, described by
a Gaussian with a relatively small standard deviation $\sigma_1$: an unexpectedly low flux requires an unlikely correlation among many sources, e.g., a void around the Earth in which normally there would be several contributing progenitors.  In contrast, the high-flux side of the distribution is extended, described by a Lorentzian with a relatively large
$\sigma_2$:  a single progenitor 
anomalously close to the Earth can produce an elevated total galactic flux.

The Gaussian behavior of $P[\phi]$ for low $\phi$ is important for experiment, as it limits large fluctuations to 
the low-flux side. In an experiment, the distribution $P[\phi]$ would presumably be combined with detector counting uncertainties to exclude values
of $|g_{aNN}^\mathrm{eff}|$ at a desired confidence level.  For present exploratory
purposes, however, we follow \cite {HLMR} in using a fixed representative flux, the median value
\begin{equation}
    \langle \phi_a^\mathrm{galactic} \rangle = 0.214 \times 10^6 \left| \frac{ g_{ann}^\mathrm{eff \, ^{23}Na} }{ 10^{-7}} \right|^2\mathrm{cm^{-2}s^{-1}}.
    \label{eq:standard}
\end{equation}
This value exceeds $\phi_0$ because the probability that today's flux lies in the Lorentzian portion of $P[\phi]$
is $\sigma_2/(\sigma_1+\sigma_2) \approx$ 0.76.
 
\subsection{Betelgeuse}
Because of the distribution's Lorentzian tail, an elevated flux today would not be highly improbable.
Such a fluctuation would arise if a carbon-burning progenitor were unusually close
to Earth.  A recent paper \cite{Saio} interpreting brightness variations in Betelgeuse concluded that this may be the case today. If this claim is correct, then the galactic $^{23}$Na axion flux at Earth
is almost an order of magnitude larger than the median value $\langle \phi_a^\mathrm{galactic} \rangle$.  The $a ~priori$ probability of an enhancement at or above this level is $\approx$ 0.02.

Betelgeuse is a bright red supergiant, the second brightest star in the constellation Orion.  Its brightness varies,
with a pattern characterized by periods of 2200, 420, 230, and 185 days.  The authors of \cite{Saio} identify
the pulsations as the radial
fundamental mode and its first, second, and third overtones, generated in the envelope of a star in the late
evolutionary stages of carbon burning.  They evolved a series of candidate stellar models in which a linear nonadiabatic 
pulsation analysis was performed, in order to identify more precisely the stellar conditions consistent with the observations.  
They determined that Betelgeuse most likely arose as a $\approx 19M_\odot$ ZAMS progenitor
that subsequently experienced significant mass loss, producing a star of 11--12$M_\odot$ with today's
observed pulsations.  The models providing the best fits to observations place the star at $\sim 300$ yr from the
end of carbon burning, so that both the $^{23}$Na abundance and core temperature are nearly optimal for maximizing
the $^{23}$Na axion flux. We note that these conclusions have been challenged \cite{Molr_2023}.

\begin{figure}
\centering
\includegraphics[scale=0.2]{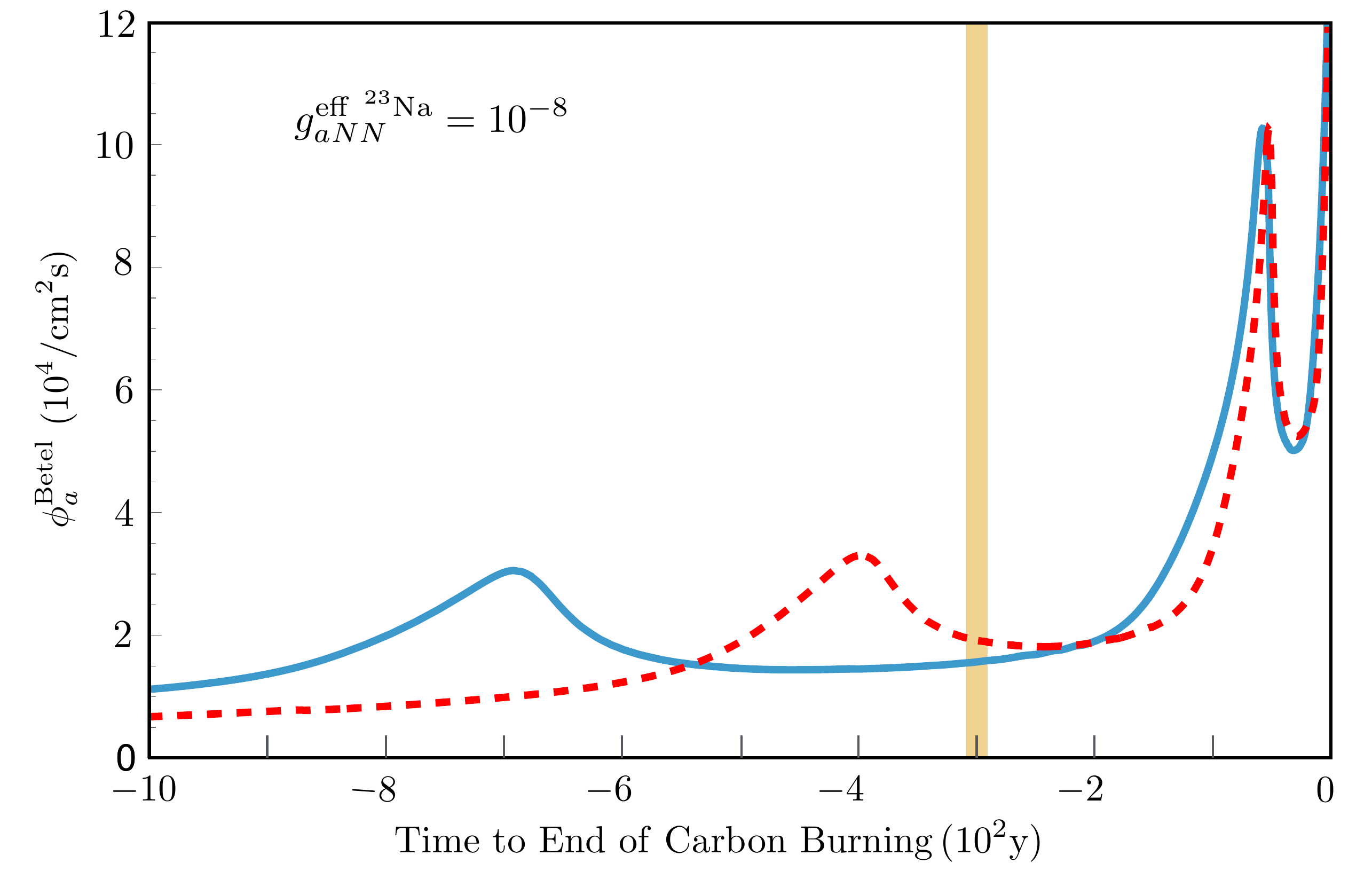}
\caption{The flux expected from Betelgeuse as a function of the time prior to the end carbon burning, for $M$ = 11$M_\odot$ (solid blue) and 12$M_\odot$ (dashed red). The vertical band at -300 yr (yellow shaded) is the current evolutionary stage of Betelgeuse according to the analysis of \cite{Saio}. The end of carbon burning defines $t=0$.  We argue (see text) that an enhanced axion flux due to Betelgeuse should not be considered in experimental analyses setting limits.}
\label{fig:Flux}
\end{figure}

Betelgeuse is sufficiently close that its distance from
Earth can be determined by parallax. We use the distance obtained by the 2020 Solar Mass Ejection Imager (SMEI) of $0.168^{+0.029}_{-0.014}$ kpc \cite{Joyce_2020}, which appears to be the most accurate available.
We estimated the axion flux from Betelgeuse by combining the SMEI distance with simulations using $M=$11 and 12 $M_\odot$
progenitors, evolved with the MESA code \cite{Paxton2011,Paxton2013} through carbon and oxygen burning.
We define the end of carbon burning, $t=0$ yr, as the time when the principal burning products, $^{20}$Ne
and $^{23}$Na, reach their maximum abundances, prior to destruction in subsequent burning cycles.  We then
computed the $^{23}$Na axion production for the 1000-year window leading up to this time.  The results
are shown in Fig. \ref{fig:Flux}.  

The predicted fluxes at $t=-300$y yr are
\begin{equation}
\phi_a^\mathrm{Betel} = \left\{ \begin{array}{l} 1.57 \\ 1.91 \end{array}  \right. \times 10^6  \left| \frac{ g_{ann}^\mathrm{eff \, ^{23}Na} }{ 10^{-7}} \right|^2\mathrm{cm^{-2}s^{-1}}~
\label{eq:Bet}
\end{equation}
where the upper (lower) value corresponds to the choice $M_\odot=11$ ($M_\odot=12$).
As the axion yield remains near
these levels for the last $\approx$ 600 years of carbon burning, the results
are not highly sensitive to the specific time of $-300$ yr identified in \cite{Saio}.

Assuming that Betelgeuse is burning carbon, we can estimate the galactic flux by taking the average of the two values given 
by Eq.~(\ref{eq:Bet}), and combining that with the contribution from the remainder of the galaxy given by Eq.~(\ref{eq:standard}).  We remove the nearest source from each simulation, replacing it with the atypical contribution of Betelgeuse.  This yields
\begin{align}
\langle \phi_a^\mathrm{galactic+Betel}\rangle =   1.91 \times 10^6 \left| \frac{ g_{ann}^\mathrm{eff \, ^{23}Na} }{ 10^{-7}} \right|^2\mathrm{cm^{-2}s^{-1}}  ~~
\label{eq:Bettot}
\end{align}
with 90\% of the total due to Betelgeuse.  

While we have noted the possibility of an enhanced flux, we strongly discourage the use of Eq.~(\ref{eq:Bettot}) in any experiment seeking to set limits. There are other plausible explanations for the brightness variations seen in
Betelgeuse, including one that has recent observational support --- eclipsing by a companion star orbiting deep in the chromosphere of the red giant \cite{dupree2026betelgeusedetectionexpandingwake,howell2025probabledirectimagingdetectionstellar}. Consequently, we adopt the statistical estimate of the flux, Eq.~(\ref{eq:standard}), using this more conservative value in the remainder of our analysis.

If a signal were to be seen in a NaI experiment, however, one would need an experiment with directional sensitivity to confirm or exclude the possibility of
a dominant contribution from Betelgeuse, in order to relate the observed rate to axion couplings

The possibility of an anomalous axion flux from Betelgeuse does not affect the analysis of \cite{HLMR}, where very light $^{23}$Na ALPs are detected through their conversion to 440.2 keV $\gamma$-rays. The axion flux and axion-photon conversion probability in the galactic magnetic field scale 
with distance $d$ as $1/d^2$ and $B_T^2d^2$, respectively.  Thus, while the axion flux from a nearby source is elevated, the $\gamma$-ray conversion probability is reduced proportionately.  To a first approximation, the $\gamma$-ray flux at Earth from a given solid angle is determined by the number of galactic sources within the cone, but not their distances.  

\subsection{The escape probability of $^{23}$Na axions}
Once a $^{23}$Na axion is produced, any interaction within the star that converts it into another particle or degrades its energy will prevent detection in NaI. In contrast to other cases we will discuss later in this paper, the
interactions of interest are quite limited. Because of its 440.2 keV energy, the axion cannot break up nuclei, while
absorption requires a discrete state at the requisite energy.  One readily concludes that the only hadronic process of
importance is reabsorption on the $^{23}$Na along the path from the production point to Earth.  Other processes 
that can contribute include Compton scattering off the electrons and the Primakoff process. Our calculations,
however, show that the Primakoff contribution to the opacity is always negligible while the impact of Compton scattering is modest and limited to a specific coupling window.

The escape probability is given by 
\begin{equation}
    P_\mathrm{esc} [z_0]= \exp\left[-\int_{z_0}^{z_S}~\sum_i n_i(z) ~\sigma_i ~dz \right].
    \label{eq:prob}
\end{equation}
where $n_i$ is the number density of a given axion absorber and $\sigma_i$ is the corresponding cross section.
The integral is performed along the line connecting $z_0$, the point in the star where the axion is produced, to 
the emission point on the stellar surface, denoted as $z_S$.  As any interaction removes the axion from the flux, a sum is performed over all channels $i$, folding the target number density $n_i(z)$ with the corresponding cross section $\sigma_i$.

Resonant absorption on $^{23}$Na converts the axion into a photon.  The cross section is given by Eq. (\ref{eq:RA})
with the replacement
\[  T_8 \equiv T_8^\mathrm{source} \rightarrow  T_8^\mathrm{source} +T_8^\mathrm{sink}. \]
which accounts for thermal motion at the sites of production and reabsorption.   The cross section
for the $a \rightarrow \gamma$ process via Compton scattering on an electron depends on the coupling
$g_{aee}$ (see Eq.~(23) of \cite{PhysRevD.37.618}).  The Primakoff conversion cross section, which
depends on $g_{a \gamma \gamma}$ (see Eq.~(18) of \cite{Caputo:2024oqc}), is
\begin{equation}
\sigma_{a \rightarrow \gamma} = \frac{g_{a\gamma \gamma} \alpha Z^2  }{ 4} \left[ \left( 1 + \frac{ \kappa^2 }{ 4 \epsilon_a^2} \right) ~\mathrm{ln} \left(1 + \frac{4 \epsilon_a^2 }{ \kappa^2} \right) -1 \right],
\end{equation}
where $\kappa$ is the Debye-H\"{u}ckel frequency. This result is valid for $m_a \ll \epsilon_a$. In the later stages of carbon burning, when the abundance of
$^{23}$Na is near its peak value, the cross section for resonant absorption on
$^{23}$Na exceeds that for the Compton scattering by $\approx$ 250, if
one sets $g_{aee}$ to its DFSZ value for $\sin^2{\beta} \equiv 0.5$.  The cross section for Primakoff absorption is negligible.  This hierarchy governs the
opacity physics.

The escaping axion fraction is determined by integrating over the carbon-burning region of a given progenitor,
folding the axion production from each infinitesimal volume $dV$ with the escape probability for that volume
given by Eq. (\ref{eq:prob}), then dividing by the total production.
The calculation is rather insensitive to the choice of progenitor: at sufficiently large $|g_{aNN}^{\mathrm{eff}~^{23}\mathrm{Na}}|$ the mean free path of the axion becomes much smaller than the radius of the carbon-burning region.
At that point, the only axions that can escape the star without interacting are those emitted from the outside edge of the carbon-burning region, and approximately in an outward radial direction. Only these axions can avoid resonant capture on $^{23}$Na.  This region becomes thinner with increasing $|g_{aNN}^{\mathrm{eff}~^{23}\mathrm{Na}}|$, eventually cutting off the flux. To properly evaluate the small surviving fraction of axions, a careful integration over angles for each shell of matter must be performed.

\begin{figure}[h!]
\centering
\includegraphics[scale=0.3]{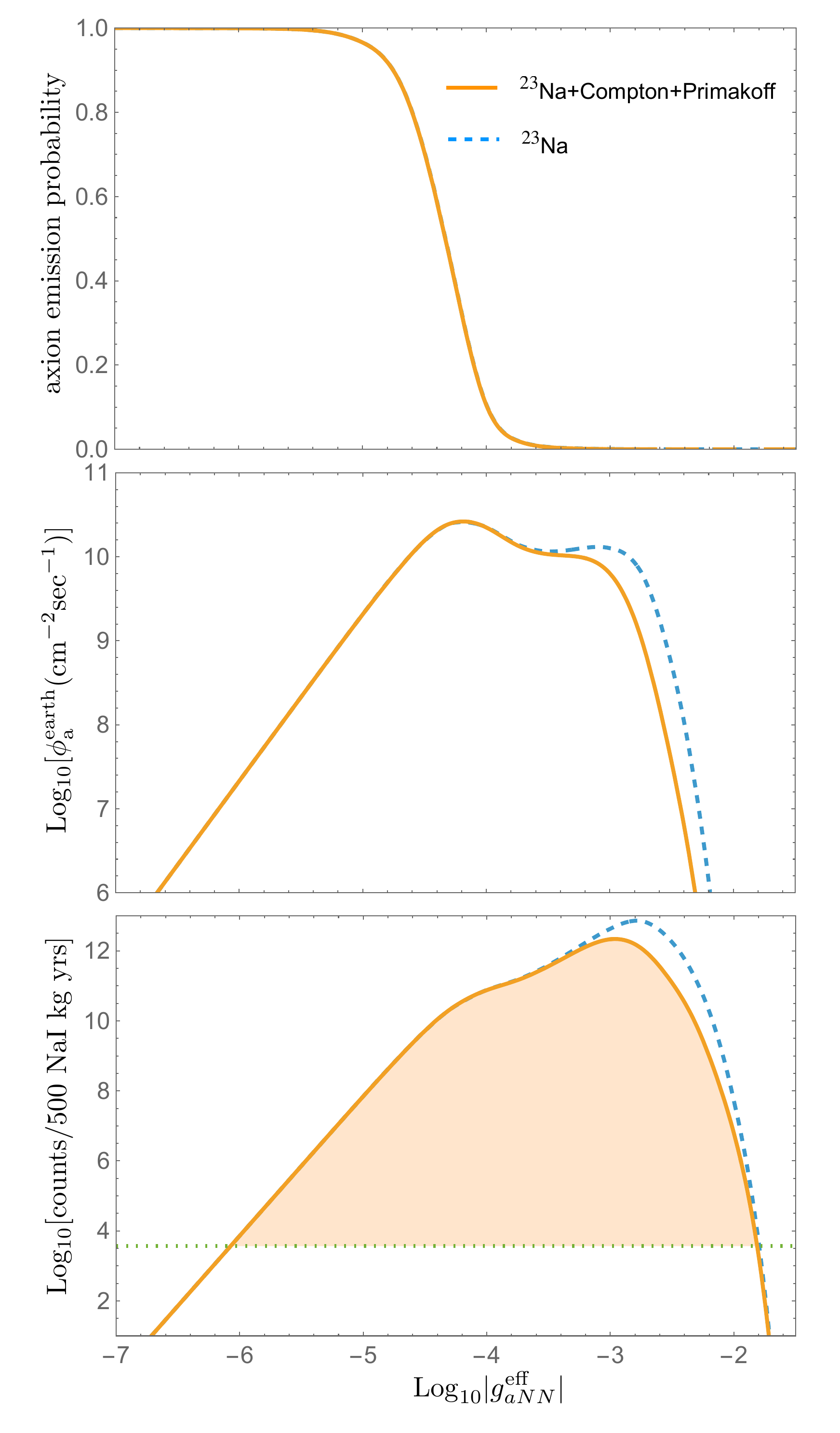}
\caption{The axion emission probability (top), $^{23}$Na axion flux at Earth (middle), and the axion counts in a NaI experiment (bottom) are shown as a function of the axion-nucleon effective coupling, $|g_{aNN}^\mathrm{eff}|$. The emission probability was evaluated for a 11$M_\odot$ star 1000 years prior to the end of carbon burning. In the bottom panel, the
horizontal dotted line shows the axion counting rate that could be detected at
$3\sigma$ given 500 NaI kg-yr of data, assuming that the average background rate is subtracted (see text).  The shaded region is the resulting excluded region. The solid lines result from stellar opacity calculations that include resonant absorption on $^{23}$Na, Compton, and Primakoff scattering, while the dashed lines include only resonant absorption on $^{23}$Na. The figure is reproduced from the companion letter~\cite{HLRR}.}
\label{fig:Count}
\end{figure}

Numerical results are given in Fig. \ref{fig:Count}.  The top panel, computed for a 11$M_\odot$ star
at a point 1000 years prior to the end of carbon burning, shows the
reabsorption probability as a function of $|g_{aNN}^{\mathrm{eff}~^{23}\mathrm{Na}}|$.  Absorption becomes important for couplings
$\gtrsim 10^{-5}$.  The solid line is the result when resonant absorption on $^{23}$Na, Compton scattering, and the Primakoff
process are included, using DFSZ values for $g_{aee}$ and $g_{a\gamma \gamma}$ evaluated for $\sin^2\beta =1/2$.  The dashed
line is the result when only resonant absorption in included.  As the differences cannot be distinguished on a linear plot, only the solid is visible.

The middle panel gives the galactic $^{23}$Na flux at Earth. As the stellar production grows as $|g_{aNN}^{\mathrm{eff}~^{23}\mathrm{Na}}|^2$, a broad maximum in the flux forms between couplings $\sim 10^{-3}-10^{-5}$, with
$\phi_a \approx 10^{10}$/cm$^2$/sec. The flux peaks at $|g_{aNN}^{\mathrm{eff}~^{23}\mathrm{Na}}|\approx 6.5 \times 10^{-5}$ at a value
of $\approx 2.1 \times 10^{10}$ axions/cm$^2$/sec.  The impact of including absorption channels governed by $g_{aee}$ (Compton) and $g_{a \gamma \gamma}$ (Primakoff) are now visible but small; they slightly diminish the sensitivity to very large couplings.

The bottom panel shows counts obtained in NaI after a 500 kg-yr exposure.  As the product of stellar
production and detection scales as $|g_{aNN}^{\mathrm{eff}~^{23}\mathrm{Na}}|^4$, the curves (relative to the middle panel) shift 
somewhat toward higher couplings, and flatten somewhat at lower couplings. The shaded region
indicates the couplings that might be ruled out in a 500 kg-yr experiment, based on the
estimate of the next section that $\sim 10^4$ axion-induced events at 440.2 keV would stand
out above backgrounds by $\gtrsim 3 \sigma$.

\section{Detection in N\MakeLowercase{a}I} 
\label{sec:V}
While the contribution from Compton scattering is visible in the third panel of
Fig. \ref{fig:Count}, it has no impact
on the range of couplings that could be probed in a terrestrial resonant absorption 
experiment.  Thus, the production, stellar attenuation, and
detection of $^{23}$Na axions are governed by a single parameter, $|g_{aNN}^\mathrm{eff~^{23}Na}|$,
provided the ratio $|g_{aee}/g_{aNN}^\mathrm{eff~^{23}Na}|$ is not wildly enhanced relative to QCD axion expectations. Consequently, future $^{23}$Na experiments will constrain $|g_{aNN}^\mathrm{eff~^{23}Na}|$, regardless of other couplings the axion or
ALP might have. 

The axion capture signal in NaI is the 440.2 keV $\gamma$-ray emitted as $^{23}$Na
returns to its ground state. 
The DM community has already developed the low-background NaI(Tl) detector arrays needed to see such a signal.  The DAMA/LIBRA collaboration currently operates a 242 kg array
at Gran Sasso, exploring a low-energy signal they have attributed to the scattering of light WIMPs.  Their results have motivated
other efforts worldwide, as described previously.  Such detectors can also be used for 440.2 keV axion detection
and allow in some cases for WIMP and axion experiments to be carried out simultaneously.  
Such dual operations require a detector instrumented to see signals at two different
gains, so that the DAMA region (high gain) and 440.2 keV axion region (low gain) can both be probed
\cite{reina}.  COSINE-100, for example, has this capability.

We can make a sensitivity estimate by comparing background rates at 440.2 keV with the axion counting rates
already calculated.  The $3\sigma$ statistical uncertainty in the measured background counting rate in the vicinity 
of the 440.2 keV peak can be estimated as
\begin{equation}
   \Delta_\mathrm{bg}^\mathrm{stat} \approx 3 \sqrt{M\, t\,  b\,  \Gamma_\mathrm{FWHM}}
\end{equation}
where $b$ is the background rate per unit time, mass, and energy, $t$ is the integration time,
$M$ is the detector mass, and $\Gamma_\mathrm{FWHM}$ is the detector's resolution. We take as a nominal 
exposure $Mt$ = 500 kg yrs.  We assume a resolution of
$\Gamma_\mathrm{FWHM} \approx$ 22 keV, based on the $\Delta E/E \approx 0.05$ found
in COSINE-100 crystals for energies $\gtrsim 100$ keV \cite{COSINEBG}.  We adopt a background
rate $b$ = 0.37 counts/kg/d/keV, obtained from the average of the backgrounds in five COSINE-100 crystals
measured in their 430-440 and 440-450 keV bins \cite{bins}.
This yields $\Delta_\mathrm{bg}^\mathrm{stat} \approx 3660$ events.  

A search for a statistically significant increase in the rate at a designated energy would typically
employ a maximum likelihood analysis to extract the localized signal from a smoothly varying background.
The shape of the signal would be known, determined by observing similar $\gamma$-rays to determine the detector's resolution.  The analysis would involve a fit to a range of energy bins around 440 keV, under the assumption that the background varies gently on the scale of the signal's width.
If the COSINE-100 background rates in the 430-440 and 440-450 bins are fit to the form $a(1+b \Delta E )$,
$b$ is typically $\sim 10^{-3}$/keV for their crystals, implying a change in the background of about 1\% between consecutive
10 keV bins.  The analysis would determine whether a simple description of this form adequately
describes backgrounds in the vicinity of 440 keV.

Pending such an analysis, we determine a tentative $3 \sigma$ exclusion for $|g_{aNN}^{\mathrm{eff}~^{23}\mathrm{Na}}|$ by requiring a counting rate in excess of $\Delta_\mathrm{bg}^\mathrm{stat}$. This yields
\begin{equation}
    8.4 \times 10^{-7} \lesssim |g_{aNN}^{\mathrm{eff}~^{23}\mathrm{Na}}| \lesssim 1.5 \times 10^{-2},
\end{equation}
as shown in the bottom panel of Fig. \ref{fig:Count}.

While we will discuss the significance of this exclusion in the next section, there is
good motivation for extending the exclusion band modestly, e.g., to a lower boundary
of $\approx (3$--$5) \times 10^{-7}$.  It is unlikely that
background reduction in the 440 keV region has been a major concern to DM NaI collaborations,
given their focus to date on the lower energy DAMA/LIBRA signal, but perhaps the motivation
we have provided here will change this.  In the $^{76}$Ge LEGEND double
beta decay experiment \cite{LEGEND:2017cdu} --- where the background goal is an impressive
$b < 3 \times 10^{-8}$ counts/kg/d/keV --- impurities are removed from the feedstock from which the crystals are grown
by zone refinement \cite{Gradwohl:2020kms}. Zone refinement has been discussed for NaI DM experiments \cite{PhysRevResearch.2.013223,PhysRevD.104.L021302}.  LEGEND also reduces external backgrounds by immersing naked crystals in a liquid Ar active shield
that employs radiopure underground argon \cite{BURLAC2023167943}, with this strategy and others designed to minimize contact between the crystals and
external materials. In principle, cosmogenic $^{22}$Na ($\tau_{1/2}$=2.6 y) and its 511 keV annihilation photons could be
eliminated by using underground sources of Na.  Thus, there may be opportunities for further reducing DM NaI backgrounds
in the 440 keV region of interest.

\section{Axions Constraints in the Turner Window}
\label{sec:VI}
The ``Turner window" is a term used to describe axions with couplings sufficiently
large that they evade the SN1987A cooling bound.  This corresponds roughly to QCD
axions with masses $\gtrsim$ 1 eV.  As we discuss below, the existing constraints in this window come primarily from astrophysics.  These limits are re-examined here, in order to assess their relation to a possible NaI experiment.  

Isospin is important to the discussion: While many exclusion plots for $g_{aNN}$ appear in the literature in 2D --- $g_{aNN}$ as a function of $m_a$ --- the general axion-nucleon parameter space is 3D, a function of $m_a$, $g_{aNN}^0$, and $g^3_{aNN}$.  These parameters are connected only if one
assumes a specific axion model, e.g., a specific QCD axion. For
KSVZ axions, all couplings are determined by the axion mass scale $f_a$
\begin{equation}
\left. \begin{array}{l} g_{app}^\mathrm{KSVZ} \\ g_{ann}^\mathrm{KSVZ} \\ m_a \end{array} \right\} = \frac{m_N }{ f_a} 
\left\{ \begin{array}{c} -0.47 \\ -0.02 \\ 6.07~ \mathrm{MeV} \end{array} \right.  
\label{eq:KSVZ}
\end{equation}
defining a one-dimensional path through the 3D parameter space. For this axion,
given QCD uncertainties, the possibility of $g_{ann}^\mathrm{KSVZ} =0$ (an equal mixture of isovector and isoscalar couplings) is open \cite{Caputo:2024oqc}.  For DFSZ axions,
\begin{equation}
     \left. \begin{array}{l} g_{app}^\mathrm{DFSZ} \\ g_{ann}^\mathrm{DFSZ} \\ m_a \end{array} \right\} = \frac{m_N }{ f_a} 
\left\{ \begin{array}{c} -0.182-0.435 \sin^2{\beta} \\ -0.160+0.414\sin^2{\beta} \\ 6.07~ \mathrm{MeV} \end{array} \right. 
\label{eq:DFSZ}
\end{equation}
The DFSZ parameter space is 2D, determined by $f_a$ and the mixing angle $\beta$, defined as in \cite{Caputo:2024oqc}.  DFSZ nucleon couplings
evolve from nearly isoscalar, to an equal isoscalar/isovector admixture with $g_{aNN}^0 = g_{aNN}^3$, to dominantly isovector as $\sin^2{\beta}$ is
dialed from 0 to 1. 

In our treatment here, we take the view that we are probing an axion or ALP with couplings $g^0_{aNN}$ 
and $g^3_{aNN}$ to nucleons, without further assumptions about the ALP mass, apart from the kinematic requirement that $m_a < 440$ keV.  In the figures below, constraints --- existing as well as those possible with new NaI experiments --- are presented as functions of the two isospin couplings.

Existing astrophysical constraints on Turner-window axions include:
\begin{enumerate}
\item Limits derived from the production of neutrons in the SNO detector, through energetic solar axions breaking up deuterium \cite{PhysRevLett.126.091601}. The solar source is the axion branch in
\begin{equation}
    p+d \rightarrow {^3 \mathrm{He}}+ \left\{ \begin{array}{c} \gamma \\ a \end{array} \right.
\end{equation}
which was first discussed in \cite{RAFFELT1982323}. This reaction produces axions with $\epsilon_a=5.5$ MeV, an energy well above the 2.22 MeV threshold for producing neutrons in the 
SNO heavy-water detector by deuteron breakup, $a+d \rightarrow p + n$. 
This reaction probes $g_{aNN}^3$:
the isoscalar contribution is exceedingly small,
e.g., producing a $\approx 0.01$\% correction at 5 MeV for the
analogous photodisintegration cross section \cite{RUPAK2000405}. The result 
of \cite{PhysRevLett.126.091601} can be expressed as
\begin{equation}
1.9 \times 10^{-5} < |g_{aNN}^3| < 0.95 \times 10^{-3},
\label{eq:SNO}
\end{equation}
where the lower bound reflects the limit of SNO sensitivity and the upper bound the requirement that the axions are 
sufficiently weakly coupled that they can escape the Sun.
\item Limits derived from the detection of $\gamma$'s in the Kamiokande II (KII)
detector, coincident with the neutrino burst from SN1987A.  The $\gamma$'s are produced by
the decay of nuclear excited states following axion inelastic absorption on $^{16}$O.  The limit obtained in \cite{PhysRevLett.65.960} converts to the following constraint on our couplings,
\begin{equation}
    ~~~~~1.3 \times 10^{-6} < \sqrt{g_{app}^2 + g_{ann}^2} < 1.4 \times 10^{-3} ,
    \label{eq:SN1987count}
\end{equation}
though in Sec. \ref{sec:gammas} we describe contributions to the axion opacity that narrow this range considerably (see Fig.~\ref{fig:limits}).

\item Limits obtained from axion cooling of SN1987A, a process that can shorten the time over which supernova burst neutrinos are emitted \cite{PhysRevLett.58.1490,PhysRevLett.58.1494}. For axions to
compete with neutrinos in SN cooling, they
must have couplings to matter not too much in excess of 
the weak scale, so they can promptly escape.  This sets an upper bound on the parameter
region excluded. The neutrino burst seen in SN1987A was more extended than expected,
and this has caused some concern about the robustness of the cooling argument.  For
example, it has been suggested that the
SN1987A explosion might have failed, with the observed neutrinos coming from an accretion disk \cite{PhysRevD.101.123025}, though see 
\cite{Caputo:2024oqc} for a critique of this scenario.  The coupling constraint
from \cite{Caputo:2024oqc} is
\begin{eqnarray}
   1.19 \times 10^{-9} &< \big[(g_{app}+0.435 g_{ann})^2 + \nonumber \\
   &~~~~~~~1.45 g_{ann}^2\big]^{1/2}\lesssim 3 \times 10^{-7}
    \label{eq:cooling}
\end{eqnarray}
where the upper bound represents the onset of trapping.
The authors of \cite{PhysRevD.109.023001} obtained a more aggressive upper bound
\begin{equation} 
3 \times 10^{-7} \rightarrow 2.5 \times 10^{-6}
\label{eq:cooling2}
\end{equation}
For the reasons discussed in Sec. \ref{sec:cooling}, we have used Eq. (\ref{eq:cooling}).
Were we to instead use Eq. (\ref{eq:cooling2}), the upper boundary of the supernova cooling band
would overlap the lower boundary of the NaI band, in figures we will later show (Figs. \ref{fig:KSVZDFSZ},
\ref{fig:gppgnn}, and \ref{fig:g0g3}). This would leave no room for QCD axions or ALPs
in the associated mass range, unless the latter were fine tuned to couple only to neutrons. 
\end{enumerate}

Separate from this study, work is underway to update these and other astrophysical bounds on axions in
the Turner window \cite{Janka}. Below we give preliminary results
relevant to the potential impact of new NaI axion experiments.

\subsection{Solar axion capture in SNO}  
The SNO constraint on axion couplings and the proposed NaI measurements
share two attractive features: First, in both scenarios the axions are produced by
stars in their quiescent burning phases, where the relevant nuclear astrophysics
is grounded in laboratory measurements of cross sections.
Second, both axion production and subsequent detection depend on the same axion
coupling, generating rates that depend on
$|g_{aNN}^3|^4$ and $|g_{aNN}^{\mathrm{eff}~^{23}\mathrm{Na}}|^4$, respectively.

One possible concern about \cite{PhysRevLett.126.091601}, pointed out by the authors, was
the simple estimate made of the $a+d \rightarrow p+n$ cross section.  The deuteron ground 
state was taken
from the solution of a one-body delta-function potential and the 
scattering state was treated as an undistorted s-wave. We were able to cross-check
this result: due to constraints from selection rules
and antisymmetry, the cross section can be taken directly --- that is, without nuclear structure corrections like the ratios $\beta$ and $\eta$ of Eq. (\ref{eq:gamma}) --- from the known M1 contribution to the deuteron photodisintegration cross section \cite{RUPAK2000405,TORNOW20038}. We found a result in reasonably good with with the estimate of \cite{PhysRevLett.126.091601}, and
consequently have made no correction.

The upper bound of the $|g_{aNN}^3|$ exclusion region, 0.95 $\times 10^{-3}$, originates
from the work of \cite{RAFFELT1982323}, where it was recognized that the 5.5 MeV
axions produced in the $p+d$ reaction could be absorbed on solar $^{17}$O
and $^{13}$C, due to their low thresholds for neutron breakup. These isotopes are
produced in the core through CN-cycle burning. This absorption mechanism dominates
over others, including the Compton and Primakoff processes.
As $^{17}$O and $^{13}$C have odd valence neutrons, we expect the breakup cross section to be governed by $|g_{ann}|$ rather than  $|g^3_{aNN}|$.
Consequently, we have revised the the upper bound of \cite{PhysRevLett.126.091601}, yielding
\begin{align}
&1.9 \times 10^{-5} < |g_{aNN}^3|~~~  \mathrm{and}~~ \nonumber \\    &|g^0_{aNN}-g^3_{aNN}|< 0.95 \times 10^{-3}.
\label{eq:SNO1}
\end{align}

\subsection{SN1987A axion-induced $\gamma$'s in Kamiokande II} 
\label{sec:gammas}
For axion couplings somewhat larger than those defined by the onset of trapping in SN1987A, there is still emission from the ``axio-sphere," the region outside core near the point of last
scattering.  This residual flux, interacting in the Kamiokande II (KII) detector, would have excited abnormal-parity transitions in $^{16}$O that then can decay electromagnetically, producing $\gamma$'s above the KII detection threshold.  

The modeling 
of the axion-induced $\gamma$ decay yield involves some challenging nuclear physics. It was recognized decades ago
that $^{16}$O is an unusually complex light nucleus: the ground state consists of entangled spherical and highly deformed shapes,
which mix through tunneling \cite{BROWN1966401}.  The quantum mechanics driving this phenomenon 
is the plunging of a 2s1d-shell $\kappa = \frac{1}{2}$ 
band in energy under deformation. Because of this intruder state and its possible
occupancies, certain 2p2h and 4p4h deformed configurations have energies comparable 
to that of the na\"ive spherical ground state, allowing them to strongly mix --- the phenomenon of shape
coexistence.  This physics has been confirmed in many microscopic calculations,
e.g., the shell-model studies of \cite{PhysRevLett.65.1325,WARBURTON19927}. The same physics impacts
the excited-state spectrum, for example, accounting for the appearance of
a $0^+_2$ state at low excitation energies, dominantly 4p4h in character.

Although other work has been done on KII $\gamma$'s, see for example \cite{PhysRevC.109.015501}, in our view the original study of
Engel, Seckel, and Hayes \cite{PhysRevLett.65.960} is the \textit{only} treatment that adequately addresses the nuclear physics described above.  These authors generated shell-model response functions
using bases that included configurations up to 4 and 5 $\hbar \omega$, respectively, for the positive- and negative-parity states, although with truncations based on selecting
certain SU(3) representations.  The resulting KII $\gamma$ yield, summarized in Fig. 1 of \cite{PhysRevLett.65.960},
shows a response dominated by axion excitation of the 10.96 MeV $J^\pi T= 0^-0$ state. This energy is above
the breakup threshold to $\alpha+^{12}$C of 7.17 MeV, but due to its parity and angular momentum, the state decays 
primarily by two-photon emission.  The authors note that this state's
contribution to the $\gamma$ decay yield, which depends on $|g_{aNN}^0|$ only, is almost entirely responsible for extending the extracted upper bound on $|g_{aNN}^0|$ from $\sim 10^{-5}$ to
$\sim 10^{-3}$. 

Additional $\gamma$'s can be produced by transitions to continuum states above the proton (12.13 MeV)
and neutron (15.66 MeV) emission thresholds, depending on their branching ratios for resonance decay $\Gamma_\gamma/\Gamma_\mathrm{total}$.   This contribution, shown separately in Fig. 1 of \cite{PhysRevLett.65.960}, becomes important for $|g_{aNN}| \lesssim \mathrm{few} \times 10^{-5}$.
The authors note that these $\gamma$'s are produced earlier in the burst, when more energetic
axions are available to drive transitions to the continuum. Details about the isospin dependence of
this contribution to the KII $\gamma$ yield are not given.  As the target has $N=Z$, one 
would expect the isoscalar and isovector axion absorption responses to be similar, for 
$|g_{aNN}^0|\approx |g_{aNN}^3|$.  However, there are a variety of effects that could induce
isospin dependence in the $\gamma$ yield, including the suppression of isoscalar E1 $\gamma$ transitions in this self-conjugate nucleus.  

The $\gamma$ spectrum from \cite{PhysRevLett.65.960} is concentrated at energies of 5-10 MeV, making
the low-energy triggering efficiency of KII important.  The authors take this into account, although
details are not given in \cite{PhysRevLett.65.960}.  The KII electron efficiency was available at the
time, from Fig. 3 of \cite{PhysRevD.38.448}.  We note for comparison that of the 12 KII events attributed to the SN1987A neutrino burst,
the lowest observed electron energy was 6.3$\pm$1.7 MeV.

Our principal concern about this calculation, and others similar to it, is the assumption that
the main axion absorption process is $NNa \rightarrow NN$, so that axions decouple from the star once this
process becomes inefficient. However, in an expanding nucleon-dominated plasma, 
axion absorption from $NNa \rightarrow NN$ scales as $\rho^{5/3}$ due to the need
for spatially correlated nucleons.  In contrast, absorption on nuclei scales as $\rho^{2/3}$ and thus can be important over a far more extended region of the star. During the explosion,
nuclei can be found at radii just a few times that of the neutrinosphere. 
At 0.3 sec after core bounce in the SN1987A model discussed below, the freezeout of $\alpha$'s becomes significant at $r \gtrsim 2\times 10^7$ cm and 
densities $\rho \gtrsim 10^8$ g/cm$^3$,
while iron-group abundances become significant at $r \gtrsim 7 \times 10^7$ and
densities $\gtrsim 10^7$ g/cm$^3$.  In addition, the axions must traverse nearly a solar mass of stellar $^{16}$O before
they can interact with the $^{16}$O in KII.
While the $^{16}$O zone is further from the core, $r \gtrsim 2 \times 10^8$ cm, its impact is enhanced through resonant absorption.
This carves out a deep absorption line in the axion spectrum at 10.96 MeV, depleting the flux needed
to produce $0^-0$ state $\gamma$'s in Kamioka II.  (As noted previously in connection with the 440.2 keV axions, the thermal width of the absorption line is much larger than the red shift of the line centroid.)

The opacity evolves with time, as the star expands and the composition
is altered by the passage of the shock wave, and is sensitive to the breakup
thresholds of the nuclear species responsible for axion absorption.
We have calculated the survival probability of axions launched in or near the axio-sphere using Eq. (\ref{eq:prob}),
but replacing our previous detailed angular integration by a single axion trajectory, as the nuclear
absorption is sufficiently distant from the region of axion production in the core to make this a reasonable approximation. 
We include the four species we expect to play the largest
roles in the absorption: $^4$He, $^{56}$Ni (which we use as representative of the iron group), $^{28}$Si, and
$^{16}$O.  The SN1987A explosion model was provided by the Garching group, a 15$M_\odot$ progenitor
viewed at 0.3, 1.0, 3.0, and 10.0 sec after core bounce \cite{Garching,Garching2}.  The first two times are most relevant, as approximately half of the events of \cite{PhysRevLett.65.960} occur in the first second. Although the Garching simulations were done in 3D, the density and composition are provided as 1D projections onto a representative spherical radius $r$.

Two of the absorbers, $^{28}$Si and $^{56}$Ni, are open-shell nuclei with strong spin responses, for which
we summed the allowed contributions of the $1^+$ $T=0$ and $T=1$ states.  (We also computed the second-forbidden $3^+$ responses, as they
required little extra work.)  For the closed-shell nuclei $^4$He and $^{16}$O, the allowed $1^+$, first-forbidden $0^-$ and $2^-$, and $3^+$ responses were computed. 
The shell-model calculations for $^{28}$Si and $^{56}$Ni were performed with the USDB \cite{USDB} and GXPF1 \cite{GXP1} interactions, respectively, 
including all configurations in the $2s1d$ and $2p1f$ spaces. The $^{16}$O negative-
and positive-parity shell-model diagonalizations were performed in bases that included all
1+3$\hbar \omega$ and 0+2+4$\hbar \omega$ configurations, using the interaction developed in \cite{PhysRevLett.65.1325}.
The $^4$He calculations were done with the Sussex potential \cite{ELLIOTT1968279}, an interaction derived from $NN$ phase shifts,
supplemented by the one-pion-exchange potential, as certain long-distance interactions are not specified in the Sussex fit.  The bases included all configurations through 5$\hbar \omega$. The Sussex
potential is known to reproduce spectra of light nuclei well, provided the potential is scaled by an overall factor, 
consistent with the uncertainty inherent in the 
Sussex fits \cite{PhysRevC.38.2335}. We followed this procedure, adjusting the scale factor so that the centroid of the
low-lying excited states, those below 27 MeV, matched that of experiment.  The scaling value obtained, 1.245, agrees well with that
found by others using similar spaces, e.g., the value of 1.225 obtained in the $4\hbar \omega$ calculations of \cite{PhysRevC.38.2335}.

\begin{figure}
    \centering
    \includegraphics[scale=0.30]{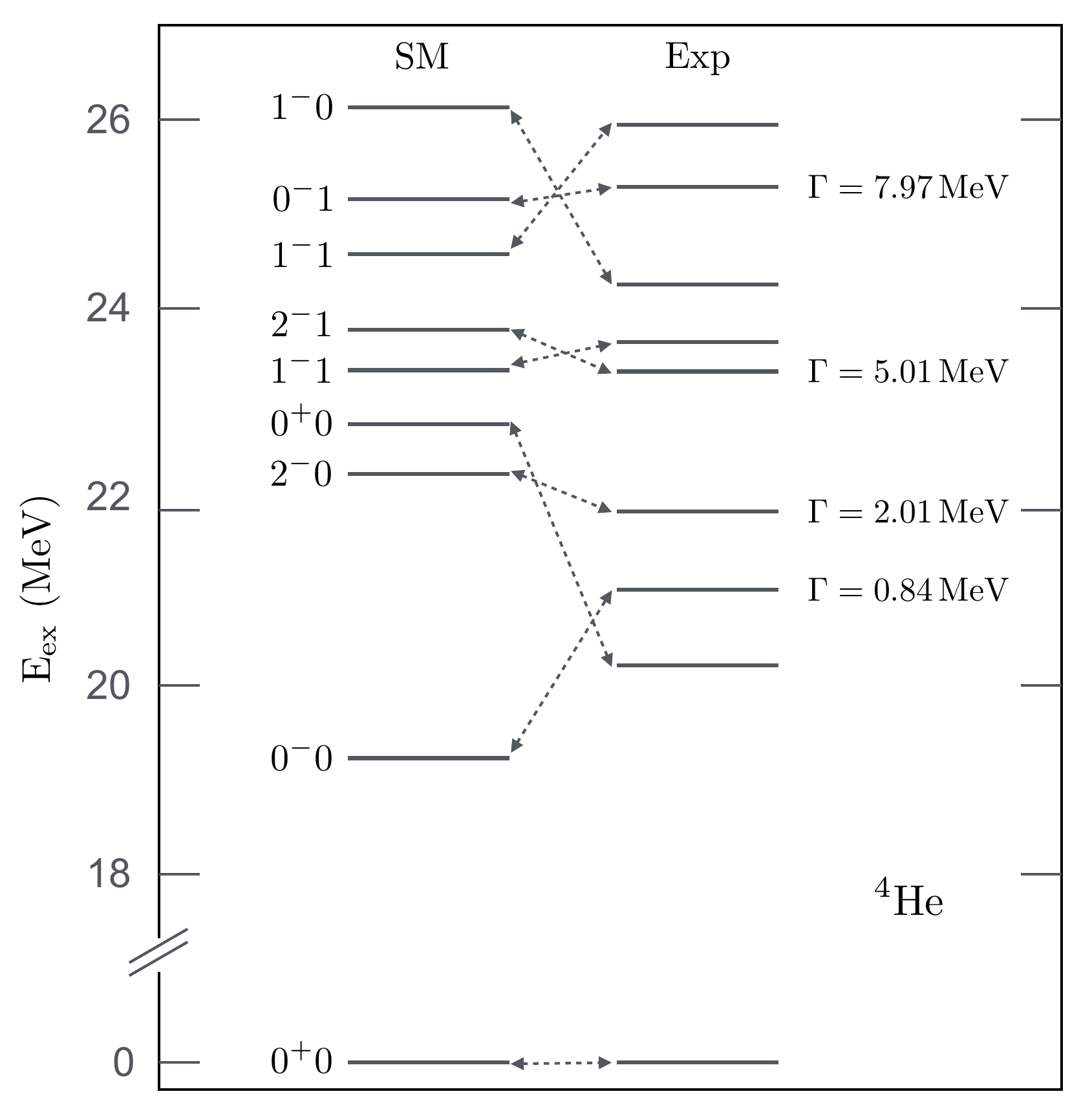}
    \caption{The shell model spectrum for $^4$He,
    computed with a modified Sussex interaction (see text),
    compared to experiment.
    For the four abnormal-parity transitions, the experimental
    energies (and widths) are used in the axion opacity calculation,
    rather than the shell model energies.}
    \label{fig:4He}
\end{figure}

Figure \ref{fig:4He} shows the low-lying Sussex
potential shell model spectrum for $^4$He, the corresponding
experimental spectrum, and the experimental widths of the
first four abnormal-parity states. In the opacity calculations
described below, we use the experimental energies and widths
for these four states.  

\begin{figure}
    \centering
    \includegraphics[scale=0.34]{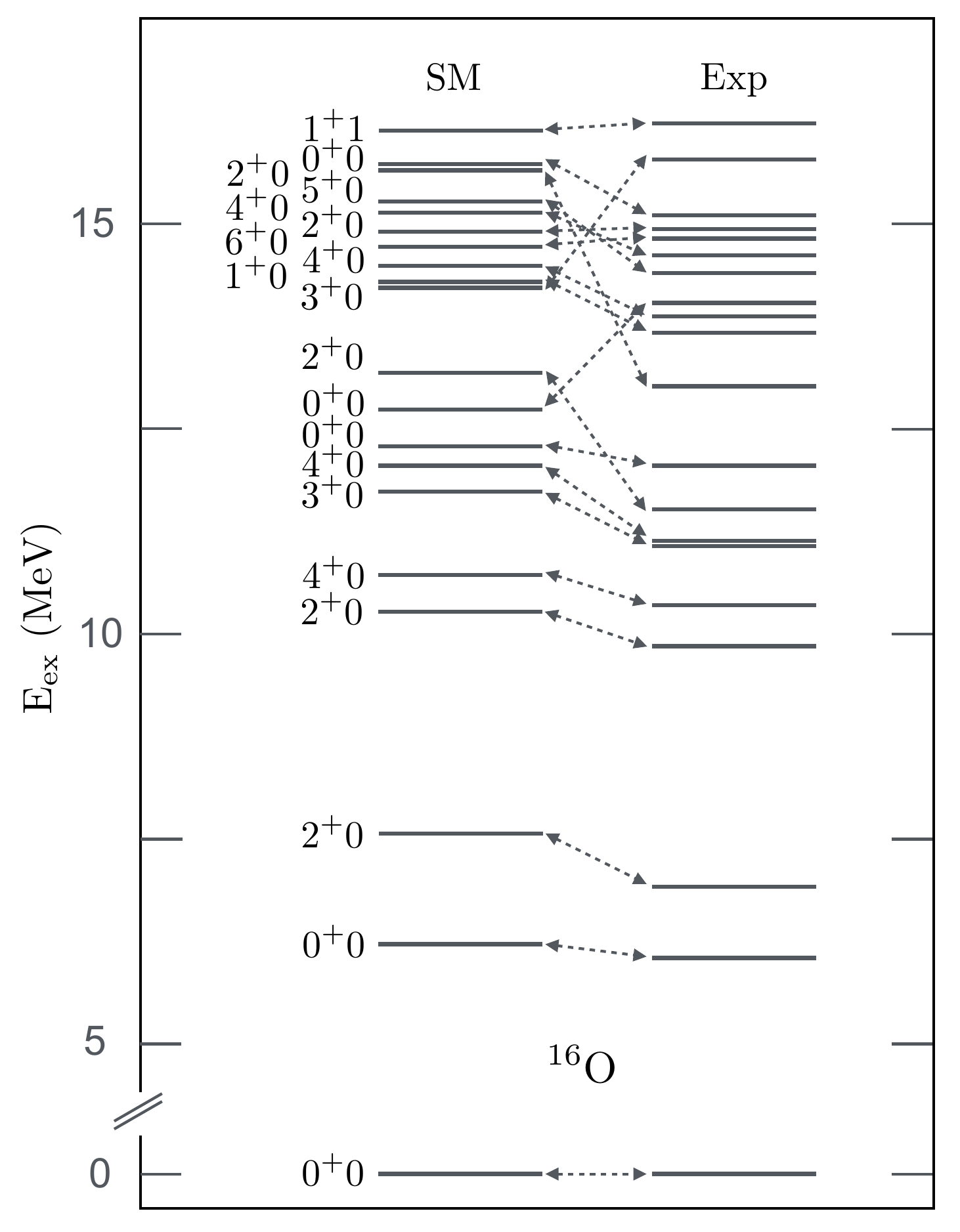}
    \caption{The shell model positive-parity spectrum for $^{16}$O, computed in a 4$\hbar \omega$ space using the interaction developed in \cite{PhysRevLett.65.1325},
    compared to experiment.}
    \label{fig:16O}
\end{figure}

Figure \ref{fig:16O} compares the shell model
$^{16}$O spectrum of even parity states to experiment, up to the first isovector state (the
isospin analog of the $^{16}$N/$^{16}$F ground state). The agreement is excellent, with
states in 1-to-1 correspondence.   The negative-parity states --- see discussion in \cite{PhysRevLett.65.1325} --- are not shown, but are also in good accord with experiment.
The first negative-parity transitions that can be excited by axion absorption
are to the shell-model levels $0^-0$ (12.14 MeV), $2^-0$ (12.60 MeV), $0^-1$ (14.41 MeV), and $2^-1$ (16.00 MeV). The respective experimental energies are 10.96, 8.87, 12.80, and 12.97 MeV.
The shell-model energies are thus displaced by $\approx$ 2.4 MeV on average.  This is expected as the repulsion that otherwise would be generated by mixing with the
$5 \hbar \omega$ spectrum is absent due to the $4\hbar \omega$ truncation of the model space.

As the use of large bases precludes direct summation over the excited
states --- the dimension of the $^{56}$Ni $m$-scheme basis exceeds $10^9$  --- the responses were generated 
by the Lanczos 
moments method, as described in \cite{PhysRevLett.65.1325}.  The number of iterations performed ranged from 150 in $^{56}$Ni to 300 in other cases, which ensures that
Lanczos distributions reproduce at least the first 300 energy moments of the exact shell-model distributions.  
We adapted the 
algorithm to generate the full momentum dependence of nuclear responses, through an expansion in powers of the three-momentum transfer $q^2$ \cite{Janka}.  Away from the energy extrema in the spectral distribution, the shell-model level density typically becomes very high and the Lanczos
eigenstates represent the contributions of many underlying continuum eigenstates. Except in the case of $^4$He,
we extract from these moments a continuous distribution by smearing. A skewed Lorentzian with $\sigma_\mathrm{FWHM} \approx
1.5$ MeV is used, where the skewing preserves normalization by vanishing at threshold.  More details are given in \cite{Janka}.  Physically, the smearing accounts for both the natural widths of
individual resonances and their aggregation into the Lanczos eigenstate representation.  For $^4$He, there is
excellent agreement between the shell model and experimental spectra of resonances up to $\approx 30$ MeV, all of
which have known widths, typically of several MeV.  For these states, we use the measured widths.

\begin{figure*}
\centering
\includegraphics[scale=0.27]{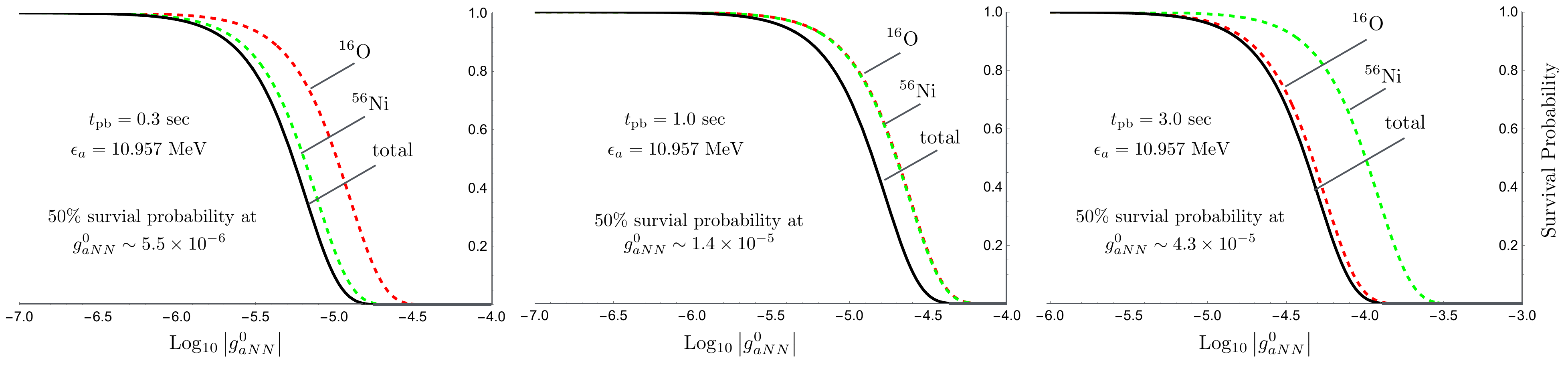}
\caption{The probability that an $\epsilon_a=10.957$ MeV axion --- the energy
of the transition to the first $0^-0$ state in $^{16}$O --- produced at the axio-sphere avoids reabsorption
as it traverses cooler regions of the star with high mass fractions of $^{56}$Ni and $^{16}$O,
as a function of $|g_{aNN}^0|$.
The calculations were done for a 15$M_\odot$ Garching model of SN1987A at the indicated
times post core bounce. We have taken $|g_{aNN}^3|=0$.  If nonzero, the absorption on $^{56}$Ni
will further increase.}
\label{fig:bound}
\end{figure*}

The three panels of Fig. \ref{fig:bound} give the probability that an axion emitted from the axio-sphere 
at times $t_\mathrm{pb}=0.3$, 1.0, and 3.0 sec after core bounce with energy 10.96 MeV --- the energy required for exciting the
transition to the $J^\pi T=0^-0$ state in $^{16}$O  --- avoids reabsorption in SN1987A's outer layers.  In addition to the nuclear processes discussed above, we include the Compton and Primakoff processes, though these are never of any numerical consequence. The important absorbers are $^{56}$Ni, due to its low breakup threshold, and $^{16}$O,
due to resonant absorption, with the former dominating at 0.3 sec and the latter at 3.0 sec, due to the earlier impact of the shock wave on nuclear shells closer to the core. 

The results show that the axion flux is significantly attenuated for couplings $|g_{aNN}^0| \gtrsim 10^{-5}$,
very much reducing the axion parameter space that can be constrained by KII.  This conclusion is conservative:
we have not included the first-forbidden contributions from $^{56}$Ni and we have assumed the axion coupling 
is purely isoscalar. The addition of a nonzero isovector coupling increases the $^{56}$Ni absorption but not the 
probability of KII detection.

\begin{figure*}[h!]
\centering
\includegraphics[scale=0.27]{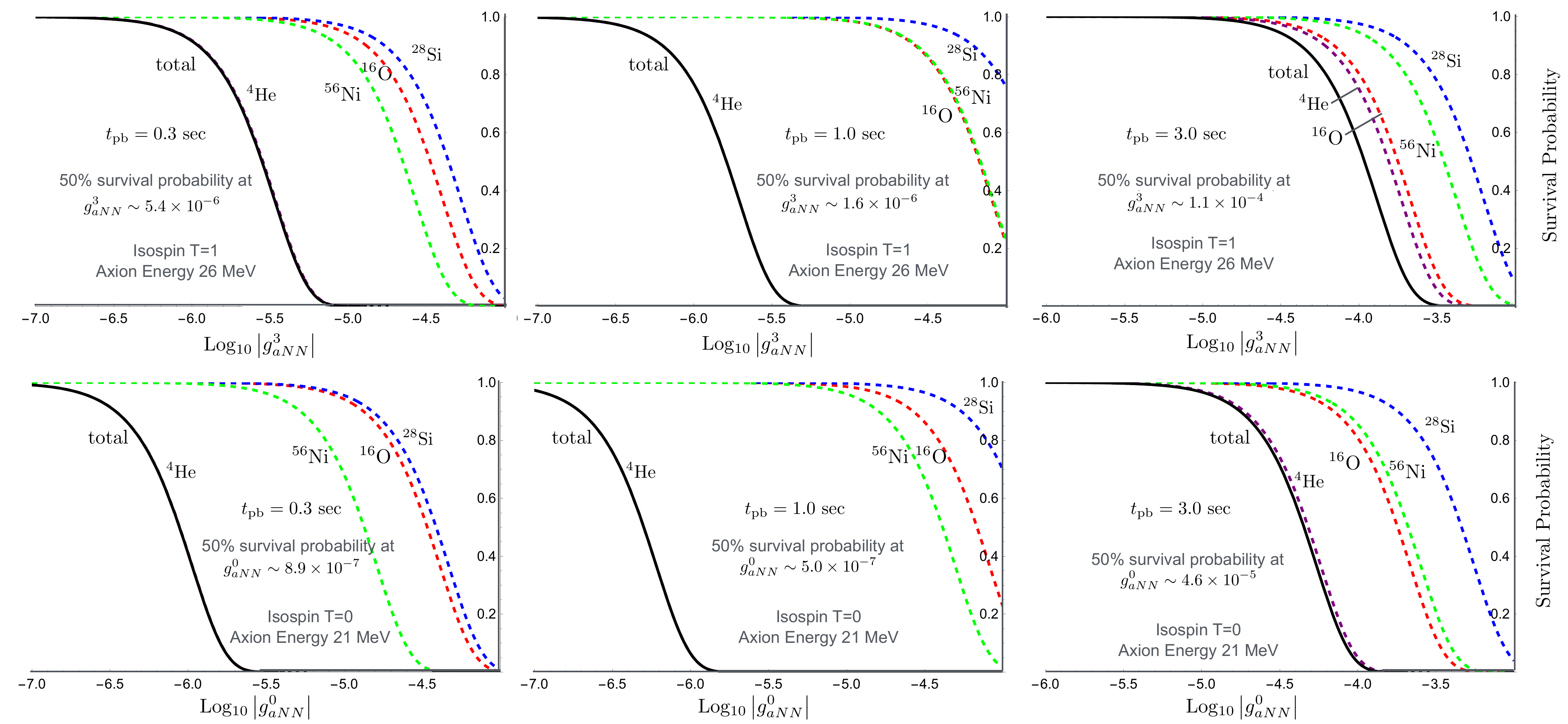}
\caption{As in Fig. \ref{fig:bound}, for either isoscalar or isovector
axion couplings, at two energies, $\epsilon_a=21$ and 26 MeV, that straddle the
$^{16}$O photoabsorption peak. }
\label{fig:cont}
\end{figure*}

Analogous calculations can be performed as a function of time, coupling, 
and axion energy, for energies above particle breakup in $^{16}$O, to
evaluate this contribution to the KII $\gamma$ yield. Representative results are given in Fig. \ref{fig:cont} for two axion energies, $\epsilon_a=21$ and 26 MeV, which straddle the
photoabsorption peak in $^{16}$O \cite{TANNER196445,PhysRevLett.111.122502}, with isospins $T=0$ and 1,
respectively.  (As we include no sources of isospin mixing, the axion
survival probabilities would be multiplicative, assuming both $|g_{aNN}^0|$
and $|g_{aNN}^3|$ are nonzero.) Three features that stand out are
the scale characterizing the onset of strong axion absorption, $|g_{aNN}| \approx 10^{-6}$; the persistence of absorption at approximately this level throughout the first second of the explosion; and the dominance
of absorption on $^4$He, which contributes for $\epsilon_a > 19.82$ MeV.
The processing of $^{56}$Ni to $\alpha$'s during the explosion 
moves absorption strength from lower energies to the $^{16}$O 
giant-resonance region, accounting for the time evolution.

\begin{figure*}
\centering
\includegraphics[scale=0.27]{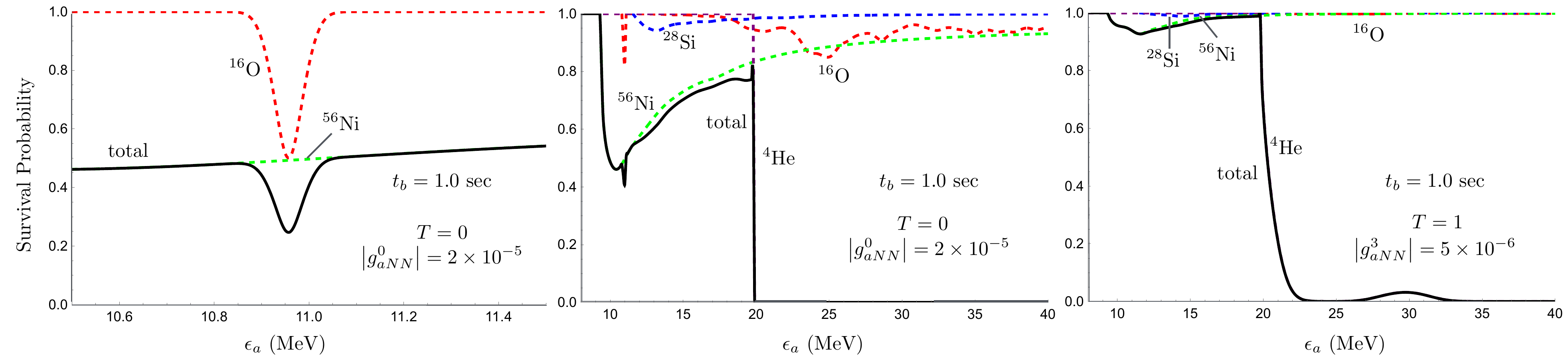}
\caption{The axion survival probability as a function of axion energy $\epsilon_a$
for the indicated $|g^0_{aNN}|$ and $|g_{aNN}^3|$, in the vicinity of the $^{16}\mathrm{O}$ $0^-0$ absorption
line (left) and for the $T=0$ and $T=1$ responses, all shown at one second 
after core bounce.}
\label{fig:energy}
\end{figure*}

In Fig. \ref{fig:energy}, the axion survival probability is plotted as a 
function of $\epsilon_a$ for selected fixed $g_{aNN}$.  The left
panel shows the absorption line that forms in the interval around the 10.96 MeV energy of the $^{16}$O bound $0^-0$ state, while the panels
on the right illustrate the impact of the nuclear breakup thresholds.

The 10.96 MeV state shows that axion detection is inhibited 
because there is a correlation between absorption in the star and
in KII, as the axions interact with $^{16}$O in both cases.
Flux is preferentially removed at the energies where the detector 
is most sensitive. It is likely that such correlations also 
affect continuum contributions, which would further reduce the
KII signal, relative to the statistical calculations done here.
The requirements for a resonance to be important
to the KII $\gamma$ yield are a large cross section for axion
absorption and a favorable branching ratio $\Gamma_\gamma/\Gamma_\mathrm{total}$. As electromagnetic widths seldom
exceed 10 eV, a resonance with at least a 1\% electromagnetic
width would then have $\Gamma_\mathrm{total} \lesssim$ 1 keV
$\ll \sigma_\mathrm{TH}$. Consequently, axion absorption in the star would be resonant, producing an absorption line at the energy needed
for exciting that resonance in KII, just as we showed for the $0^-0$ 10.96 MeV quasi-bound state. This is another reason for anticipating
that a more complete calculation will further increase axion
absorption in the star. Properly accounting for the correlation between opacity and detection is not trivial, as this physics will not be captured in standard statistical approaches. But some improvements could be made by making better
use of measured electromagnetic and total resonance widths, extending the
approach used here for the 10.957 MeV level to other states \cite{Janka}.

Pending such work, we estimate the impact
of the present work on \cite{PhysRevLett.65.960}
using Fig.~1 of that paper, which decomposes the total event rate in KII
into the $0^-0$ contribution and a
particle-emission contribution.  As we lack detailed
information on the time evolution of these events, we
adopt for our probabilities those calculated at $t_\mathrm{pb}=1$ sec
as representative values.  (In \cite{PhysRevLett.65.960} the authors
note that half of the events occur in the first second,
which motivated them to fix the parameters of a schematic model
of the supernova axion emission to the results of a dynamical SN model evaluated
at 1 sec.)  As the probabilities also depend on axion energy, we
evaluated the contribution of states above particle breakup by assigning half of the
strength to 21 MeV and half to 26 MeV, based on the qualitative argument that these
energies span the photoabsorption peak, and thus may be a good 
choice for the spin-dependent giant resonances that are related to the E1 response
in supermultiplet theories of the giant resonances \cite{ERAMZHYAN1986229}.  The variation in probabilities
with energy is modest from the $^4$He breakup threshold to $\sim$ 50 MeV.  It is
stated in \cite{PhysRevLett.65.960} that the particle-emission contribution
is insensitive to $|g_{aNN}^0|$ and $|g_{aNN}^3|$ separately, so we have
assumed equal strengths for the isoscalar and isovector responses.

The results we show in Sec. \ref{sec:results} considerably weaken the constraints
that can be extracted from the absence of SN1987A-associated $\gamma$'s at KII, making
exclusions much more dependent on details of the calculation, as well as the criteria 
for setting an exclusion.  As the opacity evolves with the explosion, in our view a calculation is needed that treats the time dependence of both axion production and absorption. We will discuss these points a bit more in Sec. \ref{sec:results}. Such improvements are among our goals for future work \cite{Janka}.

\subsection{SN1987A cooling}
\label{sec:cooling}
\begin{figure*}[t]
\centering
\includegraphics[scale=0.31]{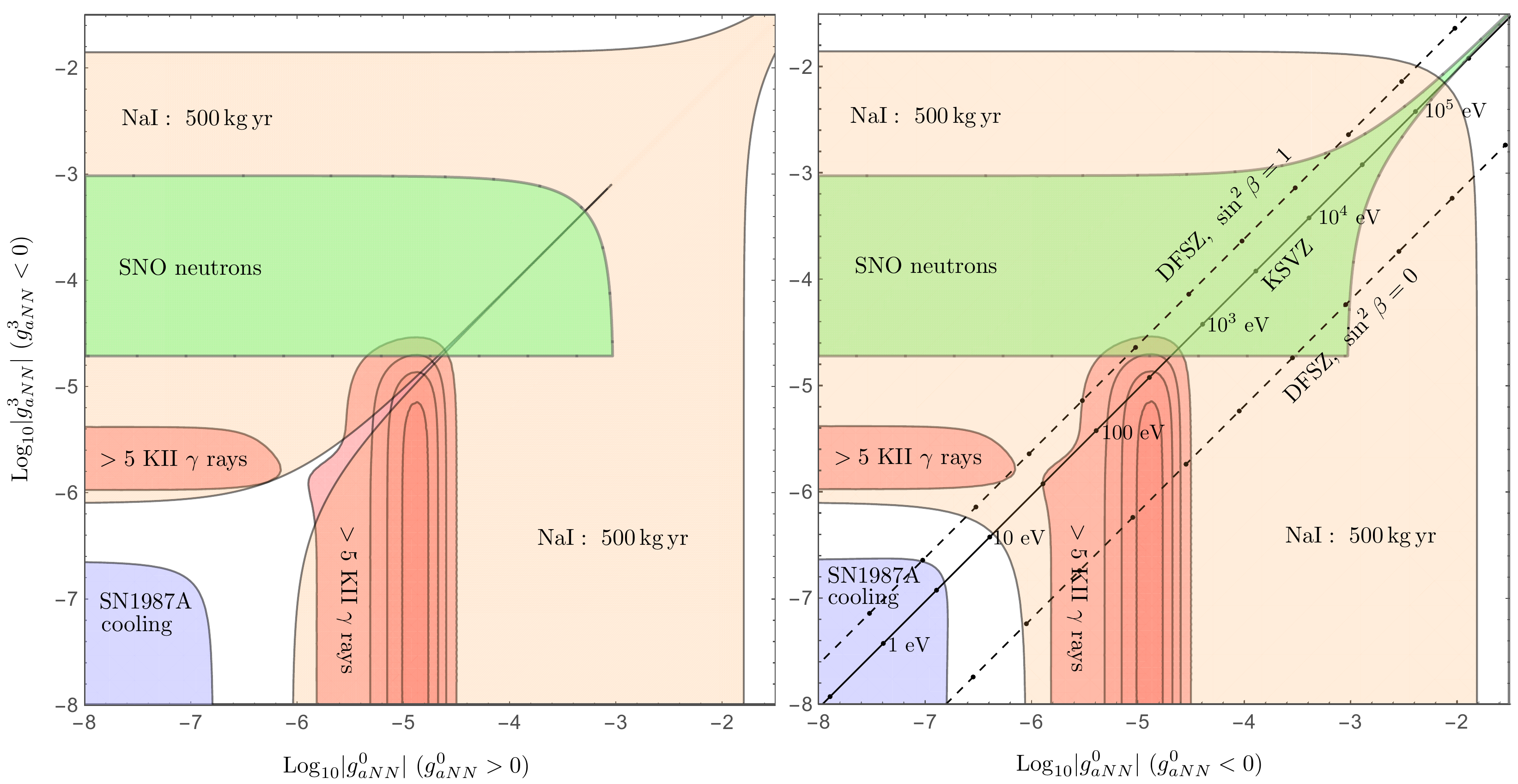}
\caption{Axion exclusions showing the potential impact of a NaI experiment.  The left (right) panel are the limits when $g_{aNN}^0$ and $g_{aNN}^3$ have opposite (same) signs. The DFSZ and KSVZ QCD axion trajectories are shown on the right panel. The contours for the KII photon exclusion correspond to 5, 15, 25, and 35 events above background.  The figure is reproduced from the companion letter~\cite{HLRR}.}.
\label{fig:limits}
\end{figure*}

For the SN1987A cooling bound in Fig.~\ref{fig:limits}, we have used Eq.~(\ref{eq:cooling}). However, as noted in Eq. (\ref{eq:cooling2}), there are recent calculations~\cite{PhysRevD.109.023001} that extend the exclusion to larger couplings by including a harder axion spectrum resulting from the presence of pions deep in the core, probing $|g_{app}|$ up to 2.5 $\times 10^{-6}$. But as we have seen, higher energy axions are also those most effectively absorbed on $\alpha$'s and iron-group elements residing
just outside the axio-sphere. Further discussion is deferred to \cite{Janka}.

\subsection{Other Constraints}
While not utilized here, one other astrophysical bound should be noted: Axion production through thermal excitation of stellar nuclei was first considered in connection with $^{57}$Fe cooling of red giants \cite{PhysRevLett.66.2557}, where the temperature of the degenerate $^4$He core is ideal for such cooling. In contrast to earlier work, which focused on stellar cooling via the Primakoff process \cite{PhysRevD.36.2211,PhysRevD.26.1840,PhysRevD.22.839},
Ref.~\cite{PhysRevLett.66.2557} was based on the metallicity dependence of cooling that would accompany axion emission by $^{57}$Fe.   The result \cite{PhysRevLett.66.2557} 
\begin{equation}
   1.6 \times 10^{-7} < |g^0_{aNN}-0.838 g^3_{aNN}|,
\end{equation}
is significant as it overlaps with the bound established by SN1987A cooling (see Eq. (\ref{eq:cooling})). There is strong motivation for  updating the work \cite{PhysRevLett.66.2557}, including detailed modeling of the transition between the red giant and horizontal branches, when stellar core temperatures peak.

\section{Results and the Potential Impact of N\lowercase{a}I Searches}
\label{sec:results}
\subsection{Constraints in the Isospin Plane}
The main results of this study are summarized in the figures of this section. In Fig.~\ref{fig:limits} we characterize the ALP in terms of its nucleon couplings $g_{aNN}^0$ and $g^3_{aNN}$, and 
its mass is left unspecified.  As the parameter space is defined by 
$\{m_a, ~g_{aNN}^0, ~g_{ann}^3\}$, the adoption of a specific model will determine
a track through the $\{g_{aNN}^0, g_{aNN}^3\}$ plane, parameterized by $m_a$. While the overall sign of the interaction plays no role in observables,
the relative sign does. We arbitrarily fix $g_{aNN}^3$ to be negative,
so that the two panels correspond to $g^0_{aNN} >0$ (left, opposite signs) and $g^0_{aNN}<0$ (right, same signs).

For the QCD axion, either the KSVZ axion of Eqs. (\ref{eq:KSVZ}) or the DFSZ axion of Eqs. (\ref{eq:DFSZ}),
$g_{aNN}^0$ and $g_{aNN}^3$ are both negative, and thus
these axion models are confined to the right panel.  In the case of the KSVZ
axion, there is a single track which can be parameterized in terms of the KSVZ axion mass.
For the DFSZ axion, there are multipole tracks depending on the mixing angle
$\beta$. The limiting tracks, $\sin^2\beta$ = 0 and 1, are
shown, defining a band that must be fully probed to exclude this axion. For  $\sin^2{\beta}=1$, the SNO constraint is quite restrictive, 
while the photon yield from KII is marginal, never exceeding 15 events.
For $\sin^2{\beta}=0$, the relationship between the two experiments is reversed.

The shapes of the various exclusion regions reflect the isospin dependence of axion production and opacity in the stellar source.  For the SNO signal, the dominant axion reabsorption
process within the Sun is breakup of the odd-neutron nuclei $^{17}$O and $^{13}$C. Consequently, the exclusion region in the right panel of Fig. \ref{fig:limits} is extended along the $g^0_{aNN}=g^3_{aNN}$ diagonal, as $^{17}$O and $^{13}$C are ineffective as absorbers for axions that couple dominantly to protons.
In the left panel, the diagonal is the $g^0_{aNN}=-g^3_{aNN}$ neutron limit, where axion reabsorption is maximized.  
Similar physics affects the NaI band, but as the dominant opacity
source is resonant absorption on the odd proton nucleus $^{23}$Na, the
behavior in the two panels is opposite that found for the SNO neutron signal.  In addition, as 
$^{23}$Na is also the stellar axion source, we see a second impact along
the diagonal of the left panel, an indentation in the lower bound of the excluded region,
indicating a loss of sensitivity to small couplings.

As we have detailed in the previous section, the reabsorption of SN1987A bremsstrahlung axions on 
$\alpha$'s, iron group elements, $^{16}$O, and $^{28}$Si substantially weakens
the constraints obtained in \cite{PhysRevLett.65.960}.  The contours of the excluded region,
the red area in Fig. \ref{fig:limits}, correspond to SN1987A photon excesses of 5, 15, 25, and 35. 
In the original work, it was assumed that an excess of five or more photons would constitute a signal, 
though the precise floor was not of great importance, as the counting rate was high throughout most
of the excluded region.  That is now changed: the exclusion area is much
reduced, and the signal within this area is weak.  Along the intersection of this area with the trajectory of the KSVZ axion,  
the most likely number of excess photons is 5-15.  This makes exclusion a delicate exercise.  As we have noted, there are significant modeling uncertainties with the potential to further reduce the signal.  In addition,  the choice of integration time is complicated by the unexpected duration 
of the SN1987A neutrino signal, as the twelve events spanned 12.5 sec. If the axion signal tracks the neutrino signal, one finds from the KII background rate of 0.6 Hz that a 99\% C.L. exclusion requires 8 excess events. This is uncomfortably close to the revised signal we have computed.

The potential impact of a 500 kg-yr NaI experiment is the feature that dominates Fig. \ref{fig:limits}.  The exclusion region extends 
over most of the plane, encompassing both the SNO neutron and KII photon regions.  Like solar $p+d \rightarrow ^3\mathrm{He}+a$ axions,
$^{23}$Na axions are continuously produced during a well-understood phase of quiescent stellar evolution, and thus are always available as an experimental
target of opportunity.

\subsection{NaI Experiments and QCD Axions}
If one is interested in testing a particular model of axions or ALPs, then
presumably the isospin dependence of the axion-nucleon interaction is known,
defining a specific trajectory through the plane of Fig. \ref{fig:limits},
e.g., as shown there for the KSVZ axion. Experimental limits are similar, with
each experiment testing some combination of $g_{app}$ and $g_{ann}$
(or equivalently $g^0_{aNN}$ and $g^3_{aNN}$).  In most of the examples 
we have discussed, this dependence has been a linear 
combination. 

\begin{figure}[t]
\centering
\includegraphics[scale=0.42]{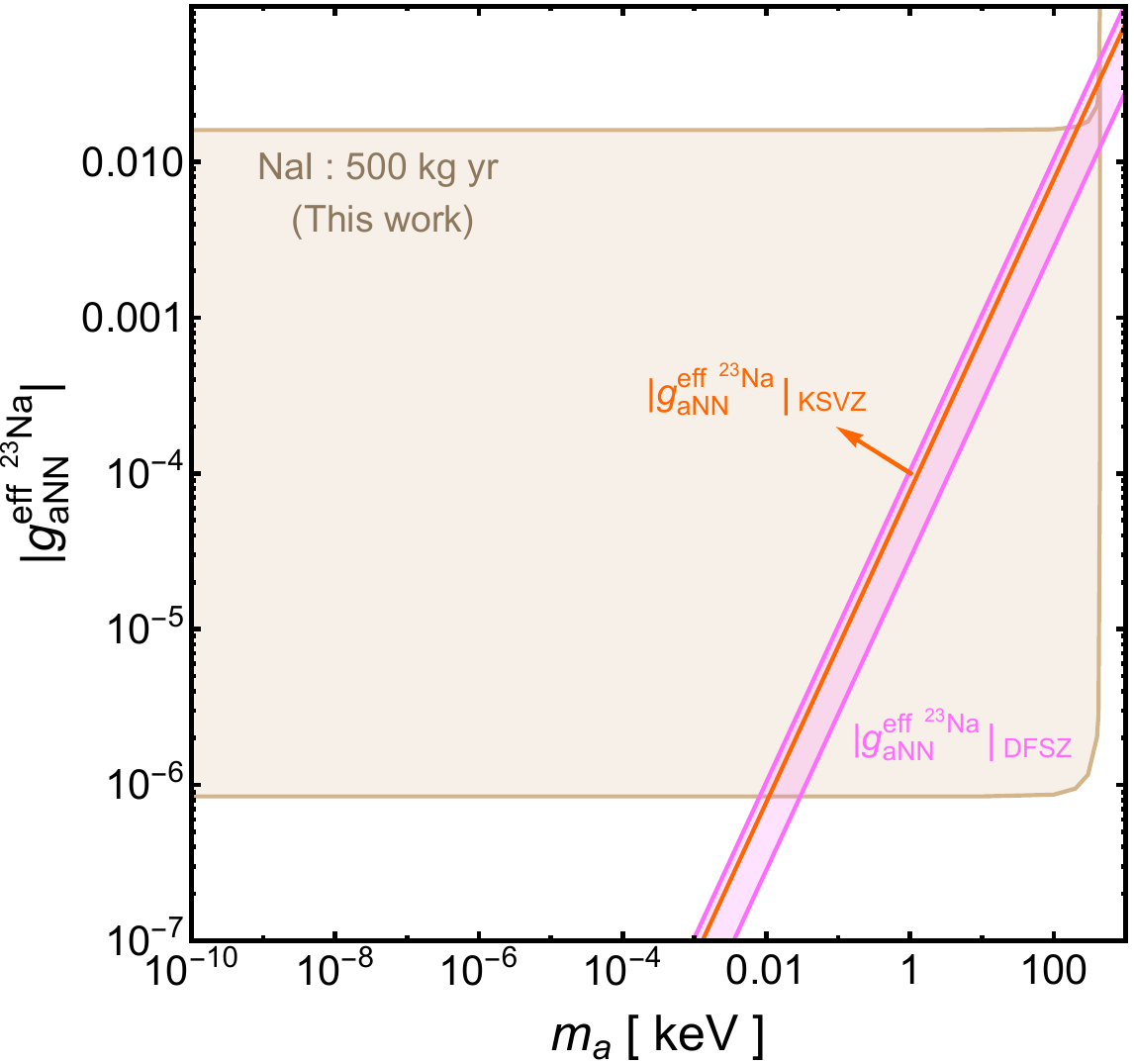}
\caption{NaI limits that could be imposed on the ALP coupling $g_{aNN}^\mathrm{eff~^{23}Na}$ as a function of $m_a$, assuming
500 kg-yr of data and current background levels. The
KSVZ and DFSZ axions would also be excluded where they overlap the shaded area.} 
\label{fig:Na23eff}
\end{figure}

The isospin dependence can be exploited.  First, it allows one to focus on the relevant
1D path through Fig. \ref{fig:limits}, so that a 2D plot can made with
$m_a$ the control parameter.  For NaI this path is naturally defined by
$g_{aNN}^\mathrm{eff~^{23}Na}$. In such a plot the phase-space effects of $m_a$ can be captured. For example, the factor
\begin{equation}
   \left[ 1 - {m_a^2 \over \epsilon_a^2} \right]^{3/2}
\end{equation}
arises due to the $q_a$-dependence of the stellar axion production rate
computed from Eq.~(\ref{eq:rate}). Similar effects arise in the cross section for absorption in NaI and in the opacity calculations, where both allowed and forbidden transitions play a key role. Figure \ref{fig:Na23eff}
gives the resulting ALP constraints as a function of $m_a$: the
rounded upturn in the excluded area is a reflection of the restricted
phase-space near the kinematic endpoint.

A second use of the isospin dependence arises when comparing experiment and theory (or in determining the relationship between experiments).  The natural bases for
describing the 2D isospin space are 
$\{g_{app},g_{ann}\}$ and  $\{g_{aNN}^0,g_{aNN}^3\}$.  Adopting the former,
from Table \ref{tab:data} the sensitivity of NaI experiments to ALP
couplings is given by the vectors
\begin{equation}
\vec{v}^\mathrm{~Na}_\parallel \equiv \{1,-0.062\}~~~~~~~~~~\vec{v}^\mathrm{~Na}_\perp \equiv \{0.062,1 \}
\end{equation}
NaI experiments are maximally sensitive to interactions that are aligned with
$\vec{v}^\mathrm{~Na}_\parallel$ and blind to those aligned with $\vec{v}^\mathrm{~Na}_\perp$.

From Eq.~(\ref{eq:KSVZ}), the corresponding vectors for the KSVZ axion in the same basis are
\begin{equation}
\vec{v}^\mathrm{~KSVZ}_\parallel \equiv \{-0.47,-0.02\}~~~~~~~~~~\vec{v}^\mathrm{~KSVZ}_\perp \equiv \{0.02 ,-0.47 \}
\label{eq:gKSVZ}
\end{equation}
The metric measuring whether an experiment effectively tests a
theory is a normalized dot product, which in the present example is
\begin{equation}
{| \vec{v}_\parallel^\mathrm{~Na} \cdot \vec{v}_\parallel^\mathrm{~KSVZ} | \over \sqrt{ |\vec{v}_\parallel^\mathrm{~Na}|^2 ~|\vec{v}_\parallel^\mathrm{~KSVZ}|^2}} \approx\ 0.995
\end{equation}
For the DFSZ axion, Eq. (\ref{eq:DFSZ}), the vectors depend on the mixing angle
\begin{align}
\vec{v}^\mathrm{~DFSZ}_\parallel[\beta] &\equiv \{-0.182 -0.435 \sin^2{\beta},-0.160+0.414 \sin^2{\beta}\} \nonumber \\\vec{v}^\mathrm{~DFSZ}_\perp[\beta] &\equiv \{~ 0.160-0.414 \sin^2{\beta},-0.182-0.435 \sin^2{\beta} \}
\label{eq:gDFSZ}
\end{align}
We find
\begin{equation}
{| \vec{v}_\parallel^\mathrm{~Na} \cdot \vec{v}_\parallel^\mathrm{~DFSZ}[\beta] | \over \sqrt{ |\vec{v}_\parallel^\mathrm{~Na}|^2 ~|\vec{v}_\parallel^\mathrm{~DFSZ}[\beta]|^2}} \approx\ 0.71-1.00
\end{equation}
with values falling below 0.94 only for small mixing angles $\sin^2{\beta}\lesssim 0.2$. With values near one, NaI experiments
will be very sensitive probes of QCD axions.

This approach is also useful when using multiple
experiments to constrain a given model.  If we take the KSVZ
axion as an example, the relevance of a given experiment
depends on its alignment with $\vec{v}_\parallel^\mathrm{~KSVZ}$. Rather than using
$\{g_{app},g_{ann} \}$, the coefficients associated with the orthogonal isospin operators
$\{ {1 + \tau_3 \over 2},{1-\tau_3 \over 2} \}$,
one can rotate to an isospin operator basis aligned with  Eq. (\ref{eq:KSVZ}), 
with coefficients
\begin{align}
 g_{aNN}^\mathrm{\parallel~KSVZ}&=g_{app}+{2 \over 47}g_{ann} \nonumber \\
g_{aNN}^\mathrm{\perp~KSVZ} &=-{2 \over 47} g_{app} +g_{ann} \underset{KSVZ}{=} 0~.
\label{eq:rotKSVZ}
\end{align}
The second line states that if $g_{app}$ and $g_{ann}$ are replaced by their KSVZ 
values, this linear combination vanishes. Consequently, if an experimental constraint is
given as a function of $g_{app}$ and $g_{ann}$, by rewriting it in terms of $g_{aNN}^\mathrm{\parallel~KSVZ}$ and $g_{aNN}^\mathrm{\perp~KSVZ}$, any dependence
on the latter can be ignored, if the goal is to constrain KSVZ axions.  This also applies to
any ALP with the same isospin couplings as the KSVZ axion. Thus, in combining
experimental limits,
use of $g_{aNN}^\mathrm{\parallel~KSVZ}$ simplifies the task. Experimental observables
with a strong dependence on $g_{aNN}^\mathrm{\parallel~KSVZ}$ will be more restrictive
than those with a weak dependence.

The same procedure can be used with DFSZ axions, though the basis depends on the mixing
angle $\beta$. One finds
\begin{align}
 g_{aNN}^\mathrm{\parallel~DFSZ}&=g_{app}+{0.160-0.414 \sin^2{\beta} \over 0.182 + 0.435 \sin^2{\beta}}~g_{ann} \nonumber \\
g_{aNN}^\mathrm{\perp~DFSZ} &=-{0.160-0.414 \sin^2{\beta} \over 0.182 + 0.435 \sin^2{\beta}} g_{app} +g_{ann} \underset{DFSZ}{=} 0~.
\label{eq:rotDFSZ}
\end{align}

\begin{figure*}[t]
\centering
\includegraphics[scale=0.4]{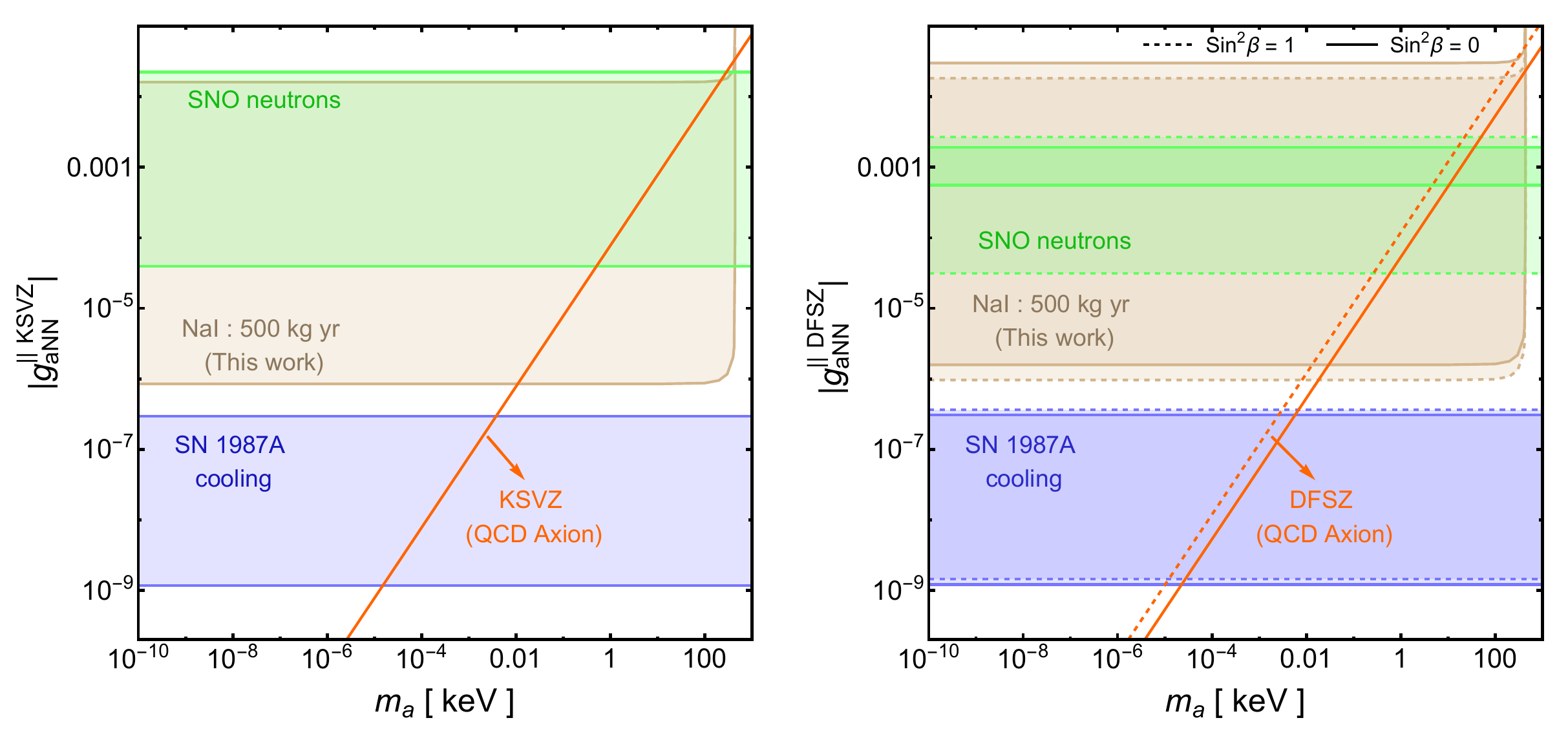}
\caption{Constraints on KSVZ-like (left panel) and DFSZ-like (right panel) ALPs (ALPs with the same isospin couplings as the KSVZ and DFSZ axions).
The KSVZ axion trajectory is the solid line in the left panel, while the
the DFSZ axion trajectories for $\sin^2{\beta}=0,1$ are shown as solid and dashed lines, respectively, on the right.  The SNO neutron and SN1987A cooling exclusions are given, along with the potential exclusion from resonant absorption of galactic axions in NaI. See Eqs.~(\ref{eq:rotKSVZ}) and (\ref{eq:rotDFSZ}) for the definitions
of  $g_{aNN}^\mathrm{\parallel~KSVZ}$ and  $g_{aNN}^\mathrm{\parallel~DFSZ}$.}
\label{fig:KSVZDFSZ}
\end{figure*}

Fig.~\ref{fig:KSVZDFSZ} gives the astrophysical
constraints on ALPs with KSVZ-like or DFSZ-like isospin
couplings. If one specializes to QCD axions, then the
additional information from Eqs.~(\ref{eq:KSVZ}) and
(\ref{eq:DFSZ}) relating the axion mass to its couplings 
is used.  This yields the KSVZ line and DFSZ band
shown in Fig. \ref{fig:KSVZDFSZ}, the latter bounded by
results for the choices $\sin^2{\beta}=0$ and 1. 

It is apparent that NaI experiments would be very helpful
in further constraining the parameter space open to hadronically coupled ALPs.  A gap remains
between the NaI and SN1987A cooling excluded area. In principle, that gap could be closed
by NaI experiments, given some combination of increased mass, 
longer exposures, and lower backgrounds.  As
noted previously, the region is constrained by arguments based on the metallicity dependence
of red giant cooling \cite{PhysRevLett.66.2557}, a study that should be updated.

In Fig.~\ref{fig:KSVZDFSZ} and in Figs.~\ref{fig:gppgnn} and
\ref{fig:g0g3} discussed below, we omit the constraint from 
KII photons, given concerns previously expressed about whether a rigorous  exclusion of axion parameters can be confidently based on this observable.
Had we done otherwise, it is apparent from Fig. \ref{fig:limits} 
that the impact would be modest, affecting a narrow range of  couplings that, for the most part, would also be excluded by anticipated NaI experiments.

\subsection{Constraints on other ALP Couplings}
\begin{figure*}[t]
\centering
\includegraphics[scale=0.4]{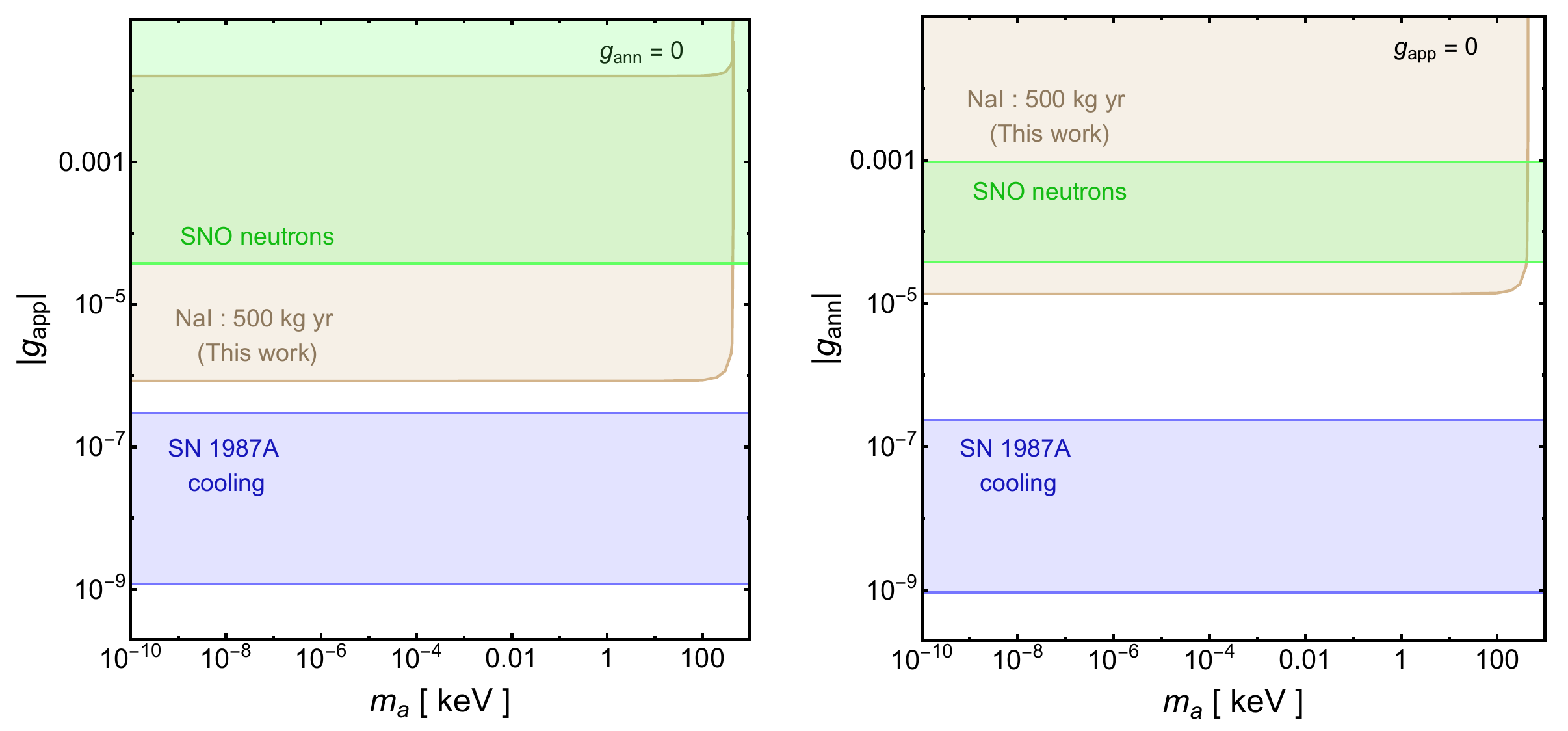}
\caption{Constraints on ALPs (SNO: green shaded, SN1987 cooling: blue shaded, NaI (this work): darker yellow shaded) that couple only to the proton (left panel) or only to the neutron (right panel).}
\label{fig:gppgnn}
\vspace*{.5cm}
\includegraphics[scale=0.4]{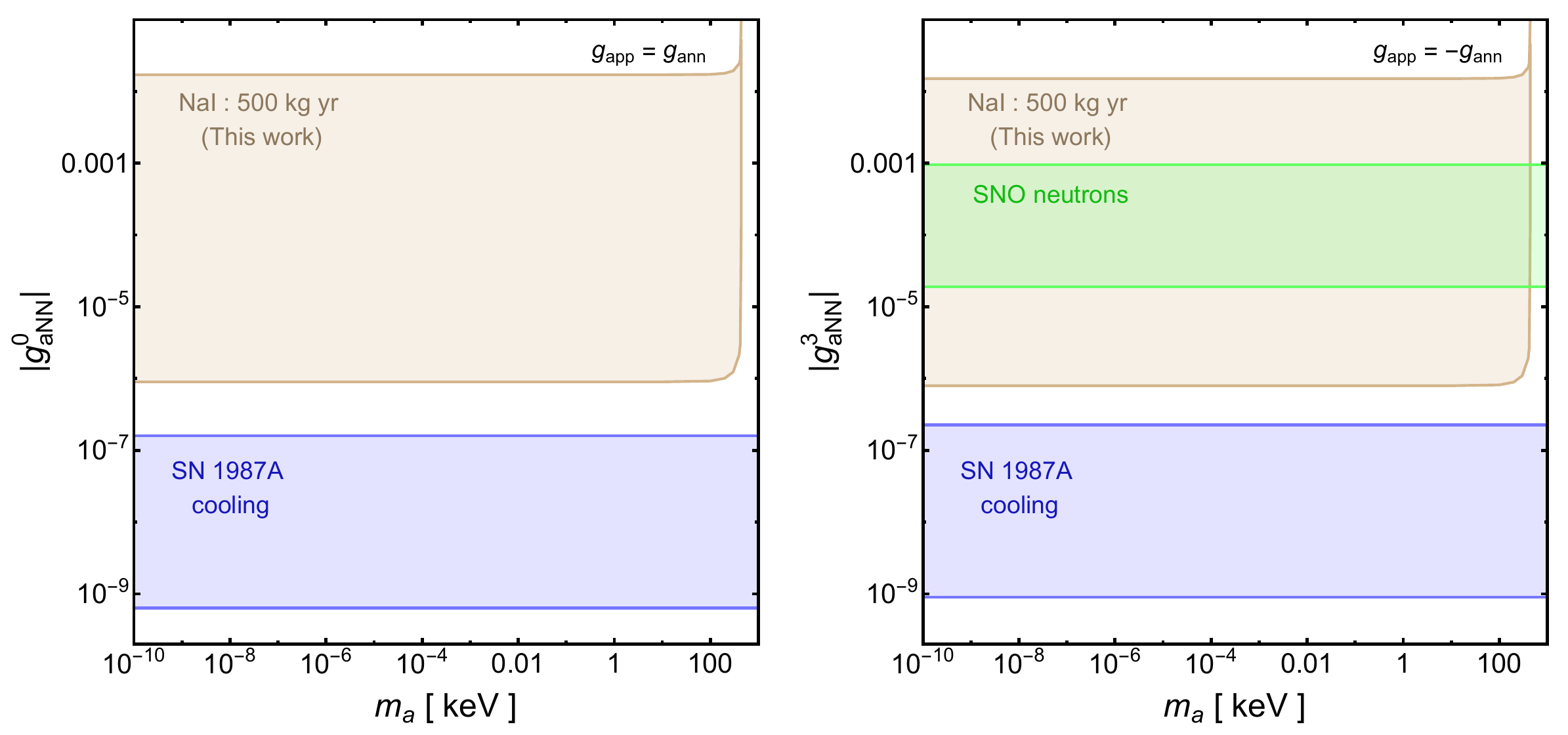}
\caption{As in Fig. \ref{fig:gppgnn}, only for isoscalar (left) and isovector (right) couplings.}
\label{fig:g0g3}
\end{figure*}

The procedure just described can be followed for any ALP
coupling.  While there is an infinity of choices, 
we can illustrate the interplay between NaI and other constraints
by considering the representative cases of
Figs.  \ref{fig:gppgnn} and \ref{fig:g0g3}, ALPs that 
couple just to protons or neutrons, or just to the isoscalar
and isovector combinations $(g_{app}+g_{ann})/2$ and
$(g_{app}-g_{ann})/2$.

It is evident from these figures that the experimental constraints depend sensitively on the assumed ALP framework, shifting as the model couplings are varied. Nonetheless, every panel shows that NaI axion experiments can have significant impact.

While our analysis is restricted to ALPs that interact with the standard model through their couplings to nucleons, once a specific model is introduced, one expects addition interactions to arise.  Constraints arising from, for example, observables connected with $g_{a\gamma \gamma}$ or $g_{aee}$ would need to be added to the results given here.

\section{Summary and Outlook}
\label{sec:summary}
It was recently demonstrated that if ALPs exist with couplings to 
nucleons, carbon-burning stars within the Milky Way will produce a continuous axion line source
at 440 keV. The ALPs, produced by 
stars with masses $\gtrsim 7.5M_\odot$, come from unusual nuclear astrophysics:
such stars can synthesize up to 0.1 $M_\odot$ of $^{23}$Na, then maintain the isotope at temperatures of $T \approx 10^9$ K for periods that can exceed 10,000 yrs.
As odd-A isotopes typically burn at high temperatures, this situation is
very rare.  The production mechanism is the conversion of thermal energy
into ALPs by the repeated M1 electro-excitation and axio-deexcitation of the
440 keV level in $^{23}$Na, as described in \cite{PhysRevLett.66.2557}.

The proposal in this paper to directly search for these axions depends on
a rather remarkable series of serendipitous coincidences. The first is the
nuclear physics and astrophysics --- not only the unusual synthesis of $^{23}$Na
described above, but also the fortunate correspondence between the temperature
of the synthesis and the energy needed to excite the strong (0.23 W.u.) M1
transition to the 440 keV level.  Thus, typically 10\% of the $^{23}$Na resides
in the excited state and is available to emit ALPs.

The second is the fact that the carbon-burning time is sufficiently long
that there are many contributing sources in our galaxy.  As described
previously \cite{HLMR}, this means that the expected flux is a
statistical quantity that can be computed from the galactic inventory of 
massive stars and our knowledge of quiescent nuclear burning. 
Deviations from the mean on the low-flux side are described by a 
Gaussian probability distribution, as a low flux requires a 
conspiracy among the positions and evolutionary times of many sources. 
The ``outliers" that might make today a special time tend to produce
elevated fluxes, as these can come from a single
chance nearby event and are distributed as a Lorentzian.  
(We noted here that some descriptions
of Betelgeuse would make today such an outlier time.)  The important point
for experiment is that the Gaussian
distribution on the low-flux side can be incorporated into the statistics
of an experiment, providing a basis for defining confidence levels on
axion parameter exclusions.

The third fortunate aspect is the existence of NaI detectors: 
resonant absorption at Earth can be used to detect these axions.  Target and
detector are the same, and the detection process is enhanced by the ratio 
of the axion energy to the thermal width of the axion line. Natural sodium
is also 100\% $^{23}$Na.  These properties contrast with those of the 
better known case of $^{57}$Fe, for which isotopic enrichment, the use of thin
targets, and external detectors are required.

Fourth, $^{23}$Na is not only the axion source and detector, but also
dominates any stellar reabsorption. The connection between the underlying
particle physics --- the ALP coupling $|g_{aNN}^\mathrm{eff}|$ --- and the
astro and experimental physics may be uniquely simple.

Finally and perhaps most serendipitously, several detectors capable of 
conducting the search for 440 keV ALPs already exist because of DAMA/LIBRA
and its low-energy signal.  These low-energy WIMP detectors are also 
the natural choice for initial studies of galactic $^{23}$Na ALPs.
In some cases, it might be possible to search simultaneously for the WIMP
and 440 keV ALP signals. The main goal of this 
paper is to bring this possible use of NaI DM detectors to the attention
of the collaborations who developed these detectors.  
As we have shown here, the limits one can 
obtain from such experiments extend over large regions
of the $\{g_{aNN}^0,~g_{aNN}^3\}$ ALP parameter space corresponding to mass scales that,
for KSVZ and DFSZ axions, begin at a few eV.  This region resides just above that 
excluded by SN1987A cooling arguments.  

In this paper, we have examined some of the existing astrophysical constraints
that probe similar couplings. For neutrons generated in SNO by the
interactions of 5.5 MeV solar axions from the $p+d$ reaction, we
have noted some small updates that can be made, but the effects on the
existing exclusion region will be small. Almost all of this region can
be covered in future NaI experiments with exposures well below the 500 kg-yr
target discussed here.  We also re-examined the production of SN1987A-related
$\gamma$'s from axion reactions on $^{16}$O in KII, finding that 
the inclusion of axion reabsorption on $\alpha$'s, $^{56}$Ni, $^{16}$O, and
$^{28}$Si at and near the SN shock front significantly attenuates the axion
signal.  For lower axion energies, including axions that excite the 10.96 
MeV $0^-0$ level that dominates the signal found in \cite{PhysRevLett.65.960}, the important processes are resonant absorption on $^{16}$O and breakup 
of $^{56}$Ni.  The absorption on $\alpha$'s, which dominates above the $^4$He particle
breakup threshold, is extremely effective in removing higher energy axions from the flux.  Our estimate of the impact on the KII counting
rate was based on postprocessing the results of \cite{PhysRevLett.65.960}.
As the effects are quite sensitive to both nuclear thresholds (and thus the 
spectrum of axion absorption strength) and to explosion dynamics (the opacity
evolves due to expansion and shock-wave associated burning), improvements are
needed. The calculations we performed were based on limited dynamical
information (snapshots of a Garching group SN1987A model at 0.3, 1.0, and 3.0
sec after core bounce) and a large-basis shell-model calculation that 
reproduces spectral information and weak rates very well and is quite
similar (but not identical) to that used in \cite{PhysRevLett.65.960}.

Both of these calculations are being improved \cite{Janka}.  In addition,
a third constraint that we intend to reexamine comes from
the anomalous metallicity dependence of red
giant cooling due to axion emission from $^{57}$Fe \cite{PhysRevLett.66.2557}. This observable 
has the sensitivity to probe most of the ``white region"
of Fig. \ref{fig:limits} --- couplings $|g_{aNN}| \sim 10^{-6}-10^{-7}$
between the SN1987A cooling limit of \cite{Caputo:2024oqc} and those that
are or can be established from KII $\gamma$'s and new NaI experiments,
respectively. We are particularly interested in improving the original calculation \cite{PhysRevLett.66.2557} by following the transition between the red giant and horizontal branches
in detail, using stellar evolution simulations, as we anticipate the largest effects
due to nuclear cooling could come at this time, when the core temperature peaks.

Finally, experimentalists involved in current
DM NaI experiments will need to check the na\"ive signal/background estimates
made here.  As background reduction in the 440 keV region likely has not been  a priority in the existing experiments, this issue could be considered. We note that resonant absorption on $^{23}$Na will take place in any sodium-bearing material.  Thus, NaI detectors do not also exhaust the experimental possibilities.

\section*{Acknowledgements}
We are very grateful to Thomas Janka, Daniel Kresse, and Annop Wongwathanarat for making available output from the Garching group's SN1987A simulations, and to Reina Maruyama, Gyunho Yo, Eunju Jeon, Karlheinz Langanke, and Jerry Miller for helpful discussions.  WH acknowledges support by the US Department of Energy under grants DE-SC0004658, DE-SC0023663, and DE-AC02-05CH11231, the National Science Foundation under cooperative agreement 2020275, and the Heising-Simons Foundation under award 00F1C. This work was carried out under the auspices of the National Nuclear Security Administration of the U.S. Department of Energy at Los Alamos National Laboratory under Contract No. 89233218CNA000001. ER acknowledges support by the Laboratory Directed Research and Development program of Los Alamos National Laboratory under project numbers 20251163PRD3 and 20260043DR. AR acknowledges support from the Natural Sciences and
Engineering Research Council of Canada (NSERC) and
Arthur B. McDonald Canadian Astroparticle Physics Research Institute. Research at Perimeter Institute is supported by the Government of Canada through the Department of Innovation, Science, and Economic Development, and by the Province of Ontario.

\bibliographystyle{apsrev4-2}
\bibliography{Na23_new.bib}

\begin{thebibliography}{73}%
\makeatletter
\providecommand \@ifxundefined [1]{%
 \@ifx{#1\undefined}
}%
\providecommand \@ifnum [1]{%
 \ifnum #1\expandafter \@firstoftwo
 \else \expandafter \@secondoftwo
 \fi
}%
\providecommand \@ifx [1]{%
 \ifx #1\expandafter \@firstoftwo
 \else \expandafter \@secondoftwo
 \fi
}%
\providecommand \natexlab [1]{#1}%
\providecommand \enquote  [1]{``#1''}%
\providecommand \bibnamefont  [1]{#1}%
\providecommand \bibfnamefont [1]{#1}%
\providecommand \citenamefont [1]{#1}%
\providecommand \href@noop [0]{\@secondoftwo}%
\providecommand \href [0]{\begingroup \@sanitize@url \@href}%
\providecommand \@href[1]{\@@startlink{#1}\@@href}%
\providecommand \@@href[1]{\endgroup#1\@@endlink}%
\providecommand \@sanitize@url [0]{\catcode `\\12\catcode `\$12\catcode
  `\&12\catcode `\#12\catcode `\^12\catcode `\_12\catcode `\%12\relax}%
\providecommand \@@startlink[1]{}%
\providecommand \@@endlink[0]{}%
\providecommand \url  [0]{\begingroup\@sanitize@url \@url }%
\providecommand \@url [1]{\endgroup\@href {#1}{\urlprefix }}%
\providecommand \urlprefix  [0]{URL }%
\providecommand \Eprint [0]{\href }%
\providecommand \doibase [0]{https://doi.org/}%
\providecommand \selectlanguage [0]{\@gobble}%
\providecommand \bibinfo  [0]{\@secondoftwo}%
\providecommand \bibfield  [0]{\@secondoftwo}%
\providecommand \translation [1]{[#1]}%
\providecommand \BibitemOpen [0]{}%
\providecommand \bibitemStop [0]{}%
\providecommand \bibitemNoStop [0]{.\EOS\space}%
\providecommand \EOS [0]{\spacefactor3000\relax}%
\providecommand \BibitemShut  [1]{\csname bibitem#1\endcsname}%
\let\auto@bib@innerbib\@empty
\bibitem [{\citenamefont {Haxton}\ \emph
  {et~al.}(2026{\natexlab{a}})\citenamefont {Haxton}, \citenamefont {Liu},
  \citenamefont {Ray},\ and\ \citenamefont {Rule}}]{HLRR}%
  \BibitemOpen
  \bibfield  {author} {\bibinfo {author} {\bibfnamefont {W.~C.}\ \bibnamefont
  {Haxton}}, \bibinfo {author} {\bibfnamefont {X.}~\bibnamefont {Liu}},
  \bibinfo {author} {\bibfnamefont {A.}~\bibnamefont {Ray}},\ and\ \bibinfo
  {author} {\bibfnamefont {E.}~\bibnamefont {Rule}},\ }\href@noop {} {\
  (\bibinfo {year} {2026}{\natexlab{a}})},\ \Eprint
  {https://arxiv.org/abs/2603.16998} {arXiv:2603.16998 [hep-ph]} \BibitemShut
  {NoStop}%
\bibitem [{\citenamefont {Haxton}\ \emph {et~al.}(2025)\citenamefont {Haxton},
  \citenamefont {Liu}, \citenamefont {McCutcheon},\ and\ \citenamefont
  {Ray}}]{HLMR}%
  \BibitemOpen
  \bibfield  {author} {\bibinfo {author} {\bibfnamefont {W.~C.}\ \bibnamefont
  {Haxton}}, \bibinfo {author} {\bibfnamefont {X.}~\bibnamefont {Liu}},
  \bibinfo {author} {\bibfnamefont {A.}~\bibnamefont {McCutcheon}},\ and\
  \bibinfo {author} {\bibfnamefont {A.}~\bibnamefont {Ray}},\ }\href
  {https://doi.org/10.1103/n3v9-2wml} {\bibfield  {journal} {\bibinfo
  {journal} {Phys. Rev. Lett.}\ }\textbf {\bibinfo {volume} {135}},\ \bibinfo
  {pages} {222701} (\bibinfo {year} {2025})}\BibitemShut {NoStop}%
\bibitem [{\citenamefont {Haxton}\ and\ \citenamefont
  {Lee}(1991)}]{PhysRevLett.66.2557}%
  \BibitemOpen
  \bibfield  {author} {\bibinfo {author} {\bibfnamefont {W.~C.}\ \bibnamefont
  {Haxton}}\ and\ \bibinfo {author} {\bibfnamefont {K.~Y.}\ \bibnamefont
  {Lee}},\ }\href {https://doi.org/10.1103/PhysRevLett.66.2557} {\bibfield
  {journal} {\bibinfo  {journal} {Phys. Rev. Lett.}\ }\textbf {\bibinfo
  {volume} {66}},\ \bibinfo {pages} {2557} (\bibinfo {year}
  {1991})}\BibitemShut {NoStop}%
\bibitem [{\citenamefont {Peccei}\ and\ \citenamefont
  {Quinn}(1977)}]{PhysRevLett.38.1440}%
  \BibitemOpen
  \bibfield  {author} {\bibinfo {author} {\bibfnamefont {R.~D.}\ \bibnamefont
  {Peccei}}\ and\ \bibinfo {author} {\bibfnamefont {H.~R.}\ \bibnamefont
  {Quinn}},\ }\href {https://doi.org/10.1103/PhysRevLett.38.1440} {\bibfield
  {journal} {\bibinfo  {journal} {Phys. Rev. Lett.}\ }\textbf {\bibinfo
  {volume} {38}},\ \bibinfo {pages} {1440} (\bibinfo {year}
  {1977})}\BibitemShut {NoStop}%
\bibitem [{\citenamefont {Weinberg}(1978)}]{PhysRevLett.40.223}%
  \BibitemOpen
  \bibfield  {author} {\bibinfo {author} {\bibfnamefont {S.}~\bibnamefont
  {Weinberg}},\ }\href {https://doi.org/10.1103/PhysRevLett.40.223} {\bibfield
  {journal} {\bibinfo  {journal} {Phys. Rev. Lett.}\ }\textbf {\bibinfo
  {volume} {40}},\ \bibinfo {pages} {223} (\bibinfo {year} {1978})}\BibitemShut
  {NoStop}%
\bibitem [{\citenamefont {Wilczek}(1978)}]{PhysRevLett.40.279}%
  \BibitemOpen
  \bibfield  {author} {\bibinfo {author} {\bibfnamefont {F.}~\bibnamefont
  {Wilczek}},\ }\href {https://doi.org/10.1103/PhysRevLett.40.279} {\bibfield
  {journal} {\bibinfo  {journal} {Phys. Rev. Lett.}\ }\textbf {\bibinfo
  {volume} {40}},\ \bibinfo {pages} {279} (\bibinfo {year} {1978})}\BibitemShut
  {NoStop}%
\bibitem [{\citenamefont {Caputo}\ and\ \citenamefont
  {Raffelt}(2024)}]{Caputo:2024oqc}%
  \BibitemOpen
  \bibfield  {author} {\bibinfo {author} {\bibfnamefont {A.}~\bibnamefont
  {Caputo}}\ and\ \bibinfo {author} {\bibfnamefont {G.}~\bibnamefont
  {Raffelt}},\ }\href {https://doi.org/10.22323/1.454.0041} {\bibfield
  {journal} {\bibinfo  {journal} {PoS}\ }\textbf {\bibinfo {volume}
  {COSMICWISPers}},\ \bibinfo {pages} {041} (\bibinfo {year} {2024})},\ \Eprint
  {https://arxiv.org/abs/2401.13728} {arXiv:2401.13728 [hep-ph]} \BibitemShut
  {NoStop}%
\bibitem [{\citenamefont {Moriyama}(1995)}]{PhysRevLett.75.3222}%
  \BibitemOpen
  \bibfield  {author} {\bibinfo {author} {\bibfnamefont {S.}~\bibnamefont
  {Moriyama}},\ }\href {https://doi.org/10.1103/PhysRevLett.75.3222} {\bibfield
   {journal} {\bibinfo  {journal} {Phys. Rev. Lett.}\ }\textbf {\bibinfo
  {volume} {75}},\ \bibinfo {pages} {3222} (\bibinfo {year}
  {1995})}\BibitemShut {NoStop}%
\bibitem [{\citenamefont {Krcmar}\ \emph {et~al.}(1998)\citenamefont {Krcmar},
  \citenamefont {Krecak}, \citenamefont {Stipcevic}, \citenamefont {Ljubicic},\
  and\ \citenamefont {Bradley}}]{KRCMAR199838}%
  \BibitemOpen
  \bibfield  {author} {\bibinfo {author} {\bibfnamefont {M.}~\bibnamefont
  {Krcmar}}, \bibinfo {author} {\bibfnamefont {Z.}~\bibnamefont {Krecak}},
  \bibinfo {author} {\bibfnamefont {M.}~\bibnamefont {Stipcevic}}, \bibinfo
  {author} {\bibfnamefont {A.}~\bibnamefont {Ljubicic}},\ and\ \bibinfo
  {author} {\bibfnamefont {D.~A.}\ \bibnamefont {Bradley}},\ }\href
  {https://doi.org/https://doi.org/10.1016/S0370-2693(98)01231-3} {\bibfield
  {journal} {\bibinfo  {journal} {Phys. Letts. B}\ }\textbf {\bibinfo {volume}
  {442}},\ \bibinfo {pages} {38} (\bibinfo {year} {1998})}\BibitemShut
  {NoStop}%
\bibitem [{\citenamefont {Namba}(2007)}]{NAMBA2007398}%
  \BibitemOpen
  \bibfield  {author} {\bibinfo {author} {\bibfnamefont {T.}~\bibnamefont
  {Namba}},\ }\href
  {https://doi.org/https://doi.org/10.1016/j.physletb.2007.01.005} {\bibfield
  {journal} {\bibinfo  {journal} {Physics Letters B}\ }\textbf {\bibinfo
  {volume} {645}},\ \bibinfo {pages} {398} (\bibinfo {year}
  {2007})}\BibitemShut {NoStop}%
\bibitem [{\citenamefont {Derbin}\ \emph {et~al.}(2011)\citenamefont {Derbin},
  \citenamefont {Muratova}, \citenamefont {Semenov},\ and\ \citenamefont
  {Unzhakov}}]{Derbin}%
  \BibitemOpen
  \bibfield  {author} {\bibinfo {author} {\bibfnamefont {A.~V.}\ \bibnamefont
  {Derbin}}, \bibinfo {author} {\bibfnamefont {V.~N.}\ \bibnamefont
  {Muratova}}, \bibinfo {author} {\bibfnamefont {D.~A.}\ \bibnamefont
  {Semenov}},\ and\ \bibinfo {author} {\bibfnamefont {E.~V.}\ \bibnamefont
  {Unzhakov}},\ }\href
  {https://doi.org/https://doi.org/10.1134/S1063778811040041} {\bibfield
  {journal} {\bibinfo  {journal} {Physics of Atomic Nuclei}\ }\textbf {\bibinfo
  {volume} {74}},\ \bibinfo {pages} {596} (\bibinfo {year} {2011})}\BibitemShut
  {NoStop}%
\bibitem [{\citenamefont {collaboration (S. Andriamonje~et
  al.)}(2009)}]{CASTcollaboration_2009}%
  \BibitemOpen
  \bibfield  {author} {\bibinfo {author} {\bibfnamefont {C.}~\bibnamefont
  {collaboration (S. Andriamonje~et al.)}},\ }\href
  {https://doi.org/10.1088/1475-7516/2009/12/002} {\bibfield  {journal}
  {\bibinfo  {journal} {Journal of Cosmology and Astroparticle Physics}\
  }\textbf {\bibinfo {volume} {2009}}\bibinfo  {number} { (12)},\ \bibinfo
  {pages} {002}}\BibitemShut {NoStop}%
\bibitem [{\citenamefont {Alessandria}\ \emph {et~al.}(2013)\citenamefont
  {Alessandria} \emph {et~al.}}]{CUORE:2012ymr}%
  \BibitemOpen
\bibfield  {number} {  }\bibfield  {author} {\bibinfo {author} {\bibfnamefont
  {F.}~\bibnamefont {Alessandria}} \emph {et~al.} (\bibinfo {collaboration}
  {CUORE}),\ }\href {https://doi.org/10.1088/1475-7516/2013/05/007} {\bibfield
  {journal} {\bibinfo  {journal} {JCAP}\ }\textbf {\bibinfo {volume} {05}},\
  \bibinfo {pages} {007}},\ \Eprint {https://arxiv.org/abs/1209.2800}
  {arXiv:1209.2800 [hep-ex]} \BibitemShut {NoStop}%
\bibitem [{\citenamefont {Aprile}\ \emph {et~al.}(2020)\citenamefont {Aprile}
  \emph {et~al.}}]{XENON:2020rca}%
  \BibitemOpen
  \bibfield  {author} {\bibinfo {author} {\bibfnamefont {E.}~\bibnamefont
  {Aprile}} \emph {et~al.} (\bibinfo {collaboration} {XENON}),\ }\href
  {https://doi.org/10.1103/PhysRevD.102.072004} {\bibfield  {journal} {\bibinfo
   {journal} {Phys. Rev. D}\ }\textbf {\bibinfo {volume} {102}},\ \bibinfo
  {pages} {072004} (\bibinfo {year} {2020})},\ \Eprint
  {https://arxiv.org/abs/2006.09721} {arXiv:2006.09721 [hep-ex]} \BibitemShut
  {NoStop}%
\bibitem [{\citenamefont {Di~Luzio}\ \emph {et~al.}(2022)\citenamefont
  {Di~Luzio} \emph {et~al.}}]{DiLuzio:2021qct}%
  \BibitemOpen
  \bibfield  {author} {\bibinfo {author} {\bibfnamefont {L.}~\bibnamefont
  {Di~Luzio}} \emph {et~al.},\ }\href
  {https://doi.org/10.1140/epjc/s10052-022-10061-1} {\bibfield  {journal}
  {\bibinfo  {journal} {Eur. Phys. J. C}\ }\textbf {\bibinfo {volume} {82}},\
  \bibinfo {pages} {120} (\bibinfo {year} {2022})},\ \Eprint
  {https://arxiv.org/abs/2111.06407} {arXiv:2111.06407 [hep-ph]} \BibitemShut
  {NoStop}%
\bibitem [{\citenamefont {Fleury}\ \emph {et~al.}(2023)\citenamefont {Fleury},
  \citenamefont {Caiazzo},\ and\ \citenamefont {Heyl}}]{Fleury:2022plh}%
  \BibitemOpen
  \bibfield  {author} {\bibinfo {author} {\bibfnamefont {L.}~\bibnamefont
  {Fleury}}, \bibinfo {author} {\bibfnamefont {I.}~\bibnamefont {Caiazzo}},\
  and\ \bibinfo {author} {\bibfnamefont {J.}~\bibnamefont {Heyl}},\ }\href
  {https://doi.org/10.1103/PhysRevD.107.L101303} {\bibfield  {journal}
  {\bibinfo  {journal} {Phys. Rev. D}\ }\textbf {\bibinfo {volume} {107}},\
  \bibinfo {pages} {L101303} (\bibinfo {year} {2023})},\ \Eprint
  {https://arxiv.org/abs/2208.00405} {arXiv:2208.00405 [astro-ph.HE]}
  \BibitemShut {NoStop}%
\bibitem [{\citenamefont {Ning}\ \emph {et~al.}(2026)\citenamefont {Ning},
  \citenamefont {Ray},\ and\ \citenamefont {Safdi}}]{Ning:2025kyu}%
  \BibitemOpen
  \bibfield  {author} {\bibinfo {author} {\bibfnamefont {O.}~\bibnamefont
  {Ning}}, \bibinfo {author} {\bibfnamefont {A.}~\bibnamefont {Ray}},\ and\
  \bibinfo {author} {\bibfnamefont {B.~R.}\ \bibnamefont {Safdi}},\ }\href
  {https://doi.org/10.1103/xyjw-9bw7} {\bibfield  {journal} {\bibinfo
  {journal} {Phys. Rev. D}\ }\textbf {\bibinfo {volume} {113}},\ \bibinfo
  {pages} {035010} (\bibinfo {year} {2026})},\ \Eprint
  {https://arxiv.org/abs/2509.03569} {arXiv:2509.03569 [hep-ph]} \BibitemShut
  {NoStop}%
\bibitem [{\citenamefont {Kim}(1979)}]{PhysRevLett.43.103}%
  \BibitemOpen
  \bibfield  {author} {\bibinfo {author} {\bibfnamefont {J.~E.}\ \bibnamefont
  {Kim}},\ }\href {https://doi.org/10.1103/PhysRevLett.43.103} {\bibfield
  {journal} {\bibinfo  {journal} {Phys. Rev. Lett.}\ }\textbf {\bibinfo
  {volume} {43}},\ \bibinfo {pages} {103} (\bibinfo {year} {1979})}\BibitemShut
  {NoStop}%
\bibitem [{\citenamefont {Shifman}\ \emph {et~al.}(1980)\citenamefont
  {Shifman}, \citenamefont {Vainshtein},\ and\ \citenamefont
  {Zakharov}}]{SHIFMAN1980493}%
  \BibitemOpen
  \bibfield  {author} {\bibinfo {author} {\bibfnamefont {M.}~\bibnamefont
  {Shifman}}, \bibinfo {author} {\bibfnamefont {A.}~\bibnamefont
  {Vainshtein}},\ and\ \bibinfo {author} {\bibfnamefont {V.}~\bibnamefont
  {Zakharov}},\ }\href
  {https://doi.org/https://doi.org/10.1016/0550-3213(80)90209-6} {\bibfield
  {journal} {\bibinfo  {journal} {Nuclear Physics B}\ }\textbf {\bibinfo
  {volume} {166}},\ \bibinfo {pages} {493} (\bibinfo {year}
  {1980})}\BibitemShut {NoStop}%
\bibitem [{\citenamefont {Dine}\ \emph {et~al.}(1981)\citenamefont {Dine},
  \citenamefont {Fischler},\ and\ \citenamefont {Srednicki}}]{DINE1981199}%
  \BibitemOpen
  \bibfield  {author} {\bibinfo {author} {\bibfnamefont {M.}~\bibnamefont
  {Dine}}, \bibinfo {author} {\bibfnamefont {W.}~\bibnamefont {Fischler}},\
  and\ \bibinfo {author} {\bibfnamefont {M.}~\bibnamefont {Srednicki}},\ }\href
  {https://doi.org/https://doi.org/10.1016/0370-2693(81)90590-6} {\bibfield
  {journal} {\bibinfo  {journal} {Physics Letters B}\ }\textbf {\bibinfo
  {volume} {104}},\ \bibinfo {pages} {199} (\bibinfo {year}
  {1981})}\BibitemShut {NoStop}%
\bibitem [{\citenamefont {Zhitnitskii}(1980)}]{osti_7063072}%
  \BibitemOpen
  \bibfield  {author} {\bibinfo {author} {\bibfnamefont {A.~P.}\ \bibnamefont
  {Zhitnitskii}},\ }\href {https://www.osti.gov/biblio/7063072} {\bibfield
  {journal} {\bibinfo  {journal} {Sov. J. Nucl. Phys. (Engl. Transl.); (United
  States)}\ }\textbf {\bibinfo {volume} {31:2}} (\bibinfo {year}
  {1980})}\BibitemShut {NoStop}%
\bibitem [{\citenamefont {Tomsick}\ \emph {et~al.}(2023)\citenamefont {Tomsick}
  \emph {et~al.}}]{Tomsick:2023aue}%
  \BibitemOpen
  \bibfield  {author} {\bibinfo {author} {\bibfnamefont {J.~A.}\ \bibnamefont
  {Tomsick}} \emph {et~al.},\ }\href {https://doi.org/10.22323/1.444.0745}
  {\bibfield  {journal} {\bibinfo  {journal} {PoS}\ }\textbf {\bibinfo {volume}
  {ICRC2023}},\ \bibinfo {pages} {745} (\bibinfo {year} {2023})},\ \Eprint
  {https://arxiv.org/abs/2308.12362} {arXiv:2308.12362 [astro-ph.HE]}
  \BibitemShut {NoStop}%
\bibitem [{\citenamefont {Avignone~III}\ \emph {et~al.}(1988)\citenamefont
  {Avignone~III}, \citenamefont {Baktash}, \citenamefont {Barker},
  \citenamefont {Calaprice}, \citenamefont {Dunford}, \citenamefont {Haxton},
  \citenamefont {Kahana}, \citenamefont {Kouzes}, \citenamefont {Miley},\ and\
  \citenamefont {Moltz}}]{PhysRevD.37.618}%
  \BibitemOpen
  \bibfield  {author} {\bibinfo {author} {\bibfnamefont {F.~T.}\ \bibnamefont
  {Avignone~III}}, \bibinfo {author} {\bibfnamefont {C.}~\bibnamefont
  {Baktash}}, \bibinfo {author} {\bibfnamefont {W.~C.}\ \bibnamefont {Barker}},
  \bibinfo {author} {\bibfnamefont {F.~P.}\ \bibnamefont {Calaprice}}, \bibinfo
  {author} {\bibfnamefont {R.~W.}\ \bibnamefont {Dunford}}, \bibinfo {author}
  {\bibfnamefont {W.~C.}\ \bibnamefont {Haxton}}, \bibinfo {author}
  {\bibfnamefont {D.}~\bibnamefont {Kahana}}, \bibinfo {author} {\bibfnamefont
  {R.~T.}\ \bibnamefont {Kouzes}}, \bibinfo {author} {\bibfnamefont {H.~S.}\
  \bibnamefont {Miley}},\ and\ \bibinfo {author} {\bibfnamefont {D.~M.}\
  \bibnamefont {Moltz}},\ }\href {https://doi.org/10.1103/PhysRevD.37.618}
  {\bibfield  {journal} {\bibinfo  {journal} {Phys. Rev. D}\ }\textbf {\bibinfo
  {volume} {37}},\ \bibinfo {pages} {618} (\bibinfo {year} {1988})}\BibitemShut
  {NoStop}%
\bibitem [{\citenamefont {Saio}\ \emph {et~al.}(2023)\citenamefont {Saio},
  \citenamefont {Nandal}, \citenamefont {Meynet},\ and\ \citenamefont
  {Ekst\"{o}m}}]{Saio}%
  \BibitemOpen
  \bibfield  {author} {\bibinfo {author} {\bibfnamefont {H.}~\bibnamefont
  {Saio}}, \bibinfo {author} {\bibfnamefont {D.}~\bibnamefont {Nandal}},
  \bibinfo {author} {\bibfnamefont {G.}~\bibnamefont {Meynet}},\ and\ \bibinfo
  {author} {\bibfnamefont {S.}~\bibnamefont {Ekst\"{o}m}},\ }\href
  {https://doi.org/10.1093/mnras/stad2949} {\bibfield  {journal} {\bibinfo
  {journal} {MNRAS}\ }\textbf {\bibinfo {volume} {526}},\ \bibinfo {pages}
  {2765} (\bibinfo {year} {2023})}\BibitemShut {NoStop}%
\bibitem [{\citenamefont {Maruyama}()}]{reina}%
  \BibitemOpen
  \bibfield  {author} {\bibinfo {author} {\bibfnamefont {R.}~\bibnamefont
  {Maruyama}},\ }\href@noop {} {}\bibinfo {howpublished} {private
  communication}\BibitemShut {NoStop}%
\bibitem [{\citenamefont {Amar\'e}\ \emph {et~al.}(2025)\citenamefont
  {Amar\'e}, \citenamefont {Apilluelo}, \citenamefont {Cebri\'an},
  \citenamefont {Cintas}, \citenamefont {Coarasa}, \citenamefont {Garc\'{\i}a},
  \citenamefont {Mart\'{\i}nez}, \citenamefont {Ortigoza}, \citenamefont
  {de~Sol\'orzano}, \citenamefont {Pardo}, \citenamefont {Puimed\'on},
  \citenamefont {Sarsa},\ and\ \citenamefont {Seoane}}]{ntnl-zrn9}%
  \BibitemOpen
  \bibfield  {author} {\bibinfo {author} {\bibfnamefont {J.}~\bibnamefont
  {Amar\'e}}, \bibinfo {author} {\bibfnamefont {J.}~\bibnamefont {Apilluelo}},
  \bibinfo {author} {\bibfnamefont {S.}~\bibnamefont {Cebri\'an}}, \bibinfo
  {author} {\bibfnamefont {D.}~\bibnamefont {Cintas}}, \bibinfo {author}
  {\bibfnamefont {I.}~\bibnamefont {Coarasa}}, \bibinfo {author} {\bibfnamefont
  {E.}~\bibnamefont {Garc\'{\i}a}}, \bibinfo {author} {\bibfnamefont
  {M.}~\bibnamefont {Mart\'{\i}nez}}, \bibinfo {author} {\bibfnamefont
  {Y.}~\bibnamefont {Ortigoza}}, \bibinfo {author} {\bibfnamefont {A.~O.}\
  \bibnamefont {de~Sol\'orzano}}, \bibinfo {author} {\bibfnamefont
  {T.}~\bibnamefont {Pardo}}, \bibinfo {author} {\bibfnamefont
  {J.}~\bibnamefont {Puimed\'on}}, \bibinfo {author} {\bibfnamefont {M.~L.}\
  \bibnamefont {Sarsa}},\ and\ \bibinfo {author} {\bibfnamefont
  {C.}~\bibnamefont {Seoane}},\ }\href {https://doi.org/10.1103/ntnl-zrn9}
  {\bibfield  {journal} {\bibinfo  {journal} {Phys. Rev. Lett.}\ }\textbf
  {\bibinfo {volume} {135}},\ \bibinfo {pages} {051001} (\bibinfo {year}
  {2025})}\BibitemShut {NoStop}%
\bibitem [{\citenamefont {Yu}\ \emph {et~al.}(2025{\natexlab{a}})\citenamefont
  {Yu} \emph {et~al.}}]{yu2025limitswimpdarkmatter}%
  \BibitemOpen
  \bibfield  {author} {\bibinfo {author} {\bibfnamefont {G.~H.}\ \bibnamefont
  {Yu}} \emph {et~al.} (\bibinfo {collaboration} {COSINE-100}),\ }\href
  {https://doi.org/10.1103/77p2-rnbw} {\bibfield  {journal} {\bibinfo
  {journal} {Phys. Rev. Lett.}\ }\textbf {\bibinfo {volume} {135}},\ \bibinfo
  {pages} {231001} (\bibinfo {year} {2025}{\natexlab{a}})},\ \Eprint
  {https://arxiv.org/abs/2501.13665} {arXiv:2501.13665 [hep-ex]} \BibitemShut
  {NoStop}%
\bibitem [{\citenamefont {{Bernabei}}\ \emph {et~al.}(2021)\citenamefont
  {{Bernabei}}, \citenamefont {{Belli}}, \citenamefont {{Caracciolo}},
  \citenamefont {{Cerulli}}, \citenamefont {{Merlo}}, \citenamefont
  {{Cappella}}, \citenamefont {{d'Angelo}}, \citenamefont {{Incicchitti}},
  \citenamefont {{Dai}}, \citenamefont {{Ma}}, \citenamefont {{Sheng}},
  \citenamefont {{Montecchia}},\ and\ \citenamefont
  {{Ye}}}]{2021arXiv211004734B}%
  \BibitemOpen
  \bibfield  {author} {\bibinfo {author} {\bibfnamefont {R.}~\bibnamefont
  {{Bernabei}}}, \bibinfo {author} {\bibfnamefont {P.}~\bibnamefont {{Belli}}},
  \bibinfo {author} {\bibfnamefont {V.}~\bibnamefont {{Caracciolo}}}, \bibinfo
  {author} {\bibfnamefont {R.}~\bibnamefont {{Cerulli}}}, \bibinfo {author}
  {\bibfnamefont {V.}~\bibnamefont {{Merlo}}}, \bibinfo {author} {\bibfnamefont
  {F.}~\bibnamefont {{Cappella}}}, \bibinfo {author} {\bibfnamefont
  {A.}~\bibnamefont {{d'Angelo}}}, \bibinfo {author} {\bibfnamefont
  {A.}~\bibnamefont {{Incicchitti}}}, \bibinfo {author} {\bibfnamefont {C.~J.}\
  \bibnamefont {{Dai}}}, \bibinfo {author} {\bibfnamefont {X.~H.}\ \bibnamefont
  {{Ma}}}, \bibinfo {author} {\bibfnamefont {X.~D.}\ \bibnamefont {{Sheng}}},
  \bibinfo {author} {\bibfnamefont {F.}~\bibnamefont {{Montecchia}}},\ and\
  \bibinfo {author} {\bibfnamefont {Z.~P.}\ \bibnamefont {{Ye}}},\ }\href
  {https://doi.org/10.48550/arXiv.2110.04734} {\bibfield  {journal} {\bibinfo
  {journal} {arXiv e-prints}\ ,\ \bibinfo {eid} {arXiv:2110.04734}} (\bibinfo
  {year} {2021})},\ \Eprint {https://arxiv.org/abs/2110.04734}
  {arXiv:2110.04734 [hep-ph]} \BibitemShut {NoStop}%
\bibitem [{\citenamefont {{Kim}}(2015)}]{kim2015statuskimsnaiexperiment}%
  \BibitemOpen
  \bibfield  {author} {\bibinfo {author} {\bibfnamefont {K.}~\bibnamefont
  {{Kim}}},\ }\href {https://doi.org/10.48550/arXiv.1511.00023} {\bibfield
  {journal} {\bibinfo  {journal} {arXiv e-prints}\ ,\ \bibinfo {eid}
  {arXiv:1511.00023}} (\bibinfo {year} {2015})},\ \Eprint
  {https://arxiv.org/abs/1511.00023} {arXiv:1511.00023 [physics.ins-det]}
  \BibitemShut {NoStop}%
\bibitem [{\citenamefont {Alner}\ \emph {et~al.}(2005)\citenamefont {Alner},
  \citenamefont {Araújo}, \citenamefont {Arnison}, \citenamefont {Barton},
  \citenamefont {Bewick}, \citenamefont {Bungau}, \citenamefont {Camanzi},
  \citenamefont {Carson}, \citenamefont {Davidge}, \citenamefont {Davies},
  \citenamefont {Davies}, \citenamefont {Daw}, \citenamefont {Dawson},
  \citenamefont {Duffy}, \citenamefont {Durkin}, \citenamefont {Gamble},
  \citenamefont {Hart}, \citenamefont {Hollingworth}, \citenamefont {Homer},
  \citenamefont {Howard}, \citenamefont {Ivaniouchenkov}, \citenamefont
  {Jones}, \citenamefont {Joshi}, \citenamefont {Kirkpatrick}, \citenamefont
  {Kudryavtsev}, \citenamefont {Lawson}, \citenamefont {Lebedenko},
  \citenamefont {Lehner}, \citenamefont {Lewin}, \citenamefont {Lightfoot},
  \citenamefont {Liubarsky}, \citenamefont {Lüscher}, \citenamefont
  {McMillan}, \citenamefont {Morgan}, \citenamefont {Murphy}, \citenamefont
  {Nicklin}, \citenamefont {Nickolls}, \citenamefont {Paling}, \citenamefont
  {Preece}, \citenamefont {Quenby}, \citenamefont {Roberts}, \citenamefont
  {Robinson}, \citenamefont {Smith}, \citenamefont {Smith}, \citenamefont
  {Spooner}, \citenamefont {Sumner}, \citenamefont {Tovey},\ and\ \citenamefont
  {Tziaferi}}]{Alner_2005}%
  \BibitemOpen
  \bibfield  {author} {\bibinfo {author} {\bibfnamefont {G.}~\bibnamefont
  {Alner}}, \bibinfo {author} {\bibfnamefont {H.}~\bibnamefont {Araújo}},
  \bibinfo {author} {\bibfnamefont {G.}~\bibnamefont {Arnison}}, \bibinfo
  {author} {\bibfnamefont {J.}~\bibnamefont {Barton}}, \bibinfo {author}
  {\bibfnamefont {A.}~\bibnamefont {Bewick}}, \bibinfo {author} {\bibfnamefont
  {C.}~\bibnamefont {Bungau}}, \bibinfo {author} {\bibfnamefont
  {B.}~\bibnamefont {Camanzi}}, \bibinfo {author} {\bibfnamefont
  {M.}~\bibnamefont {Carson}}, \bibinfo {author} {\bibfnamefont
  {D.}~\bibnamefont {Davidge}}, \bibinfo {author} {\bibfnamefont
  {G.}~\bibnamefont {Davies}}, \bibinfo {author} {\bibfnamefont
  {J.}~\bibnamefont {Davies}}, \bibinfo {author} {\bibfnamefont
  {E.}~\bibnamefont {Daw}}, \bibinfo {author} {\bibfnamefont {J.}~\bibnamefont
  {Dawson}}, \bibinfo {author} {\bibfnamefont {C.}~\bibnamefont {Duffy}},
  \bibinfo {author} {\bibfnamefont {T.}~\bibnamefont {Durkin}}, \bibinfo
  {author} {\bibfnamefont {T.}~\bibnamefont {Gamble}}, \bibinfo {author}
  {\bibfnamefont {S.}~\bibnamefont {Hart}}, \bibinfo {author} {\bibfnamefont
  {R.}~\bibnamefont {Hollingworth}}, \bibinfo {author} {\bibfnamefont
  {G.}~\bibnamefont {Homer}}, \bibinfo {author} {\bibfnamefont
  {A.}~\bibnamefont {Howard}}, \bibinfo {author} {\bibfnamefont
  {I.}~\bibnamefont {Ivaniouchenkov}}, \bibinfo {author} {\bibfnamefont
  {W.}~\bibnamefont {Jones}}, \bibinfo {author} {\bibfnamefont
  {M.}~\bibnamefont {Joshi}}, \bibinfo {author} {\bibfnamefont
  {J.}~\bibnamefont {Kirkpatrick}}, \bibinfo {author} {\bibfnamefont
  {V.}~\bibnamefont {Kudryavtsev}}, \bibinfo {author} {\bibfnamefont
  {T.}~\bibnamefont {Lawson}}, \bibinfo {author} {\bibfnamefont
  {V.}~\bibnamefont {Lebedenko}}, \bibinfo {author} {\bibfnamefont
  {M.}~\bibnamefont {Lehner}}, \bibinfo {author} {\bibfnamefont
  {J.}~\bibnamefont {Lewin}}, \bibinfo {author} {\bibfnamefont
  {P.}~\bibnamefont {Lightfoot}}, \bibinfo {author} {\bibfnamefont
  {I.}~\bibnamefont {Liubarsky}}, \bibinfo {author} {\bibfnamefont
  {R.}~\bibnamefont {Lüscher}}, \bibinfo {author} {\bibfnamefont
  {J.}~\bibnamefont {McMillan}}, \bibinfo {author} {\bibfnamefont
  {B.}~\bibnamefont {Morgan}}, \bibinfo {author} {\bibfnamefont
  {A.}~\bibnamefont {Murphy}}, \bibinfo {author} {\bibfnamefont
  {G.}~\bibnamefont {Nicklin}}, \bibinfo {author} {\bibfnamefont
  {A.}~\bibnamefont {Nickolls}}, \bibinfo {author} {\bibfnamefont
  {S.}~\bibnamefont {Paling}}, \bibinfo {author} {\bibfnamefont
  {R.}~\bibnamefont {Preece}}, \bibinfo {author} {\bibfnamefont
  {J.}~\bibnamefont {Quenby}}, \bibinfo {author} {\bibfnamefont
  {J.}~\bibnamefont {Roberts}}, \bibinfo {author} {\bibfnamefont
  {M.}~\bibnamefont {Robinson}}, \bibinfo {author} {\bibfnamefont
  {N.}~\bibnamefont {Smith}}, \bibinfo {author} {\bibfnamefont
  {P.}~\bibnamefont {Smith}}, \bibinfo {author} {\bibfnamefont
  {N.}~\bibnamefont {Spooner}}, \bibinfo {author} {\bibfnamefont
  {T.}~\bibnamefont {Sumner}}, \bibinfo {author} {\bibfnamefont
  {D.}~\bibnamefont {Tovey}},\ and\ \bibinfo {author} {\bibfnamefont
  {E.}~\bibnamefont {Tziaferi}},\ }\href
  {https://doi.org/10.1016/j.physletb.2000.09.001} {\bibfield  {journal}
  {\bibinfo  {journal} {Physics Letters B}\ }\textbf {\bibinfo {volume}
  {616}},\ \bibinfo {pages} {17–24} (\bibinfo {year} {2005})}\BibitemShut
  {NoStop}%
\bibitem [{\citenamefont {Choi}\ \emph {et~al.}(2025)\citenamefont {Choi},
  \citenamefont {Ha}, \citenamefont {Jeon}, \citenamefont {Kim}, \citenamefont
  {Kim}, \citenamefont {Kim}, \citenamefont {Kim}, \citenamefont {Kim},
  \citenamefont {Ko}, \citenamefont {Koh}, \citenamefont {Lee}, \citenamefont
  {Lee}, \citenamefont {Lee}, \citenamefont {Lee}, \citenamefont {Lee},
  \citenamefont {Oh},\ and\ \citenamefont {Park}}]{PhysRevLett.134.021802}%
  \BibitemOpen
  \bibfield  {author} {\bibinfo {author} {\bibfnamefont {J.~J.}\ \bibnamefont
  {Choi}}, \bibinfo {author} {\bibfnamefont {C.}~\bibnamefont {Ha}}, \bibinfo
  {author} {\bibfnamefont {E.~J.}\ \bibnamefont {Jeon}}, \bibinfo {author}
  {\bibfnamefont {J.~Y.}\ \bibnamefont {Kim}}, \bibinfo {author} {\bibfnamefont
  {K.~W.}\ \bibnamefont {Kim}}, \bibinfo {author} {\bibfnamefont {S.~H.}\
  \bibnamefont {Kim}}, \bibinfo {author} {\bibfnamefont {S.~K.}\ \bibnamefont
  {Kim}}, \bibinfo {author} {\bibfnamefont {Y.~D.}\ \bibnamefont {Kim}},
  \bibinfo {author} {\bibfnamefont {Y.~J.}\ \bibnamefont {Ko}}, \bibinfo
  {author} {\bibfnamefont {B.~C.}\ \bibnamefont {Koh}}, \bibinfo {author}
  {\bibfnamefont {S.~H.}\ \bibnamefont {Lee}}, \bibinfo {author} {\bibfnamefont
  {I.~S.}\ \bibnamefont {Lee}}, \bibinfo {author} {\bibfnamefont
  {H.}~\bibnamefont {Lee}}, \bibinfo {author} {\bibfnamefont {H.~S.}\
  \bibnamefont {Lee}}, \bibinfo {author} {\bibfnamefont {J.~S.}\ \bibnamefont
  {Lee}}, \bibinfo {author} {\bibfnamefont {Y.~M.}\ \bibnamefont {Oh}},\ and\
  \bibinfo {author} {\bibfnamefont {B.~J.}\ \bibnamefont {Park}} (\bibinfo
  {collaboration} {NEON Collaboration}),\ }\href
  {https://doi.org/10.1103/PhysRevLett.134.021802} {\bibfield  {journal}
  {\bibinfo  {journal} {Phys. Rev. Lett.}\ }\textbf {\bibinfo {volume} {134}},\
  \bibinfo {pages} {021802} (\bibinfo {year} {2025})}\BibitemShut {NoStop}%
\bibitem [{\citenamefont {Milligan}\ \emph {et~al.}(2025)\citenamefont
  {Milligan}, \citenamefont {Urquijo}, \citenamefont {Barberio}, \citenamefont
  {Bashu}, \citenamefont {Bignell}, \citenamefont {Bolognino}, \citenamefont
  {Chhun}, \citenamefont {Dastgiri}, \citenamefont {Fruth}, \citenamefont {Fu},
  \citenamefont {Hill}, \citenamefont {Hua}, \citenamefont {James},
  \citenamefont {Janssens}, \citenamefont {Kapoor}, \citenamefont {Lane},
  \citenamefont {Leaver}, \citenamefont {McGee}, \citenamefont {McKie},
  \citenamefont {McKenzie}, \citenamefont {McNamara}, \citenamefont
  {Melbourne}, \citenamefont {Mews}, \citenamefont {Ng}, \citenamefont {Rule},
  \citenamefont {Slavkovská}, \citenamefont {Stanley}, \citenamefont
  {Stuchbery}, \citenamefont {Suerfu}, \citenamefont {Taylor}, \citenamefont
  {Tempra}, \citenamefont {Tunningley}, \citenamefont {Williams}, \citenamefont
  {Zhong},\ and\ \citenamefont {Zurowski}}]{Milligan_2025}%
  \BibitemOpen
  \bibfield  {author} {\bibinfo {author} {\bibfnamefont {L.}~\bibnamefont
  {Milligan}}, \bibinfo {author} {\bibfnamefont {P.}~\bibnamefont {Urquijo}},
  \bibinfo {author} {\bibfnamefont {E.}~\bibnamefont {Barberio}}, \bibinfo
  {author} {\bibfnamefont {V.}~\bibnamefont {Bashu}}, \bibinfo {author}
  {\bibfnamefont {L.}~\bibnamefont {Bignell}}, \bibinfo {author} {\bibfnamefont
  {I.}~\bibnamefont {Bolognino}}, \bibinfo {author} {\bibfnamefont
  {S.}~\bibnamefont {Chhun}}, \bibinfo {author} {\bibfnamefont
  {F.}~\bibnamefont {Dastgiri}}, \bibinfo {author} {\bibfnamefont
  {T.}~\bibnamefont {Fruth}}, \bibinfo {author} {\bibfnamefont
  {G.}~\bibnamefont {Fu}}, \bibinfo {author} {\bibfnamefont {G.}~\bibnamefont
  {Hill}}, \bibinfo {author} {\bibfnamefont {Y.}~\bibnamefont {Hua}}, \bibinfo
  {author} {\bibfnamefont {R.}~\bibnamefont {James}}, \bibinfo {author}
  {\bibfnamefont {K.}~\bibnamefont {Janssens}}, \bibinfo {author}
  {\bibfnamefont {S.}~\bibnamefont {Kapoor}}, \bibinfo {author} {\bibfnamefont
  {G.}~\bibnamefont {Lane}}, \bibinfo {author} {\bibfnamefont {K.}~\bibnamefont
  {Leaver}}, \bibinfo {author} {\bibfnamefont {P.}~\bibnamefont {McGee}},
  \bibinfo {author} {\bibfnamefont {L.}~\bibnamefont {McKie}}, \bibinfo
  {author} {\bibfnamefont {J.}~\bibnamefont {McKenzie}}, \bibinfo {author}
  {\bibfnamefont {P.}~\bibnamefont {McNamara}}, \bibinfo {author}
  {\bibfnamefont {W.}~\bibnamefont {Melbourne}}, \bibinfo {author}
  {\bibfnamefont {M.}~\bibnamefont {Mews}}, \bibinfo {author} {\bibfnamefont
  {W.}~\bibnamefont {Ng}}, \bibinfo {author} {\bibfnamefont {K.}~\bibnamefont
  {Rule}}, \bibinfo {author} {\bibfnamefont {Z.}~\bibnamefont {Slavkovská}},
  \bibinfo {author} {\bibfnamefont {O.}~\bibnamefont {Stanley}}, \bibinfo
  {author} {\bibfnamefont {A.}~\bibnamefont {Stuchbery}}, \bibinfo {author}
  {\bibfnamefont {B.}~\bibnamefont {Suerfu}}, \bibinfo {author} {\bibfnamefont
  {G.}~\bibnamefont {Taylor}}, \bibinfo {author} {\bibfnamefont
  {D.}~\bibnamefont {Tempra}}, \bibinfo {author} {\bibfnamefont
  {T.}~\bibnamefont {Tunningley}}, \bibinfo {author} {\bibfnamefont
  {A.}~\bibnamefont {Williams}}, \bibinfo {author} {\bibfnamefont
  {Y.}~\bibnamefont {Zhong}},\ and\ \bibinfo {author} {\bibfnamefont
  {M.}~\bibnamefont {Zurowski}},\ }\href
  {https://doi.org/10.1088/1748-0221/20/07/p07049} {\bibfield  {journal}
  {\bibinfo  {journal} {Journal of Instrumentation}\ }\textbf {\bibinfo
  {volume} {20}}\bibinfo  {number} { (07)},\ \bibinfo {pages}
  {P07049}}\BibitemShut {NoStop}%
\bibitem [{\citenamefont {Donnelly}\ and\ \citenamefont
  {Haxton}(1979)}]{DONNELLY1979103}%
  \BibitemOpen
\bibfield  {number} {  }\bibfield  {author} {\bibinfo {author} {\bibfnamefont
  {T.}~\bibnamefont {Donnelly}}\ and\ \bibinfo {author} {\bibfnamefont
  {W.}~\bibnamefont {Haxton}},\ }\href
  {https://doi.org/https://doi.org/10.1016/0092-640X(79)90003-2} {\bibfield
  {journal} {\bibinfo  {journal} {Atomic Data and Nuclear Data Tables}\
  }\textbf {\bibinfo {volume} {23}},\ \bibinfo {pages} {103} (\bibinfo {year}
  {1979})}\BibitemShut {NoStop}%
\bibitem [{\citenamefont {Moln\'ar}\ \emph {et~al.}(2023)\citenamefont
  {Moln\'ar}, \citenamefont {Joyce},\ and\ \citenamefont {Leung}}]{Molr_2023}%
  \BibitemOpen
  \bibfield  {author} {\bibinfo {author} {\bibfnamefont {L.}~\bibnamefont
  {Moln\'ar}}, \bibinfo {author} {\bibfnamefont {M.}~\bibnamefont {Joyce}},\
  and\ \bibinfo {author} {\bibfnamefont {S.-C.}\ \bibnamefont {Leung}},\ }\href
  {https://doi.org/10.3847/2515-5172/acdb7a} {\bibfield  {journal} {\bibinfo
  {journal} {Research Notes of the AAS}\ }\textbf {\bibinfo {volume} {7}},\
  \bibinfo {pages} {119} (\bibinfo {year} {2023})}\BibitemShut {NoStop}%
\bibitem [{\citenamefont {Joyce}\ \emph {et~al.}(2020)\citenamefont {Joyce},
  \citenamefont {Leung}, \citenamefont {Moln\'ar}, \citenamefont {Ireland},
  \citenamefont {Kobayashi},\ and\ \citenamefont {Nomoto}}]{Joyce_2020}%
  \BibitemOpen
  \bibfield  {author} {\bibinfo {author} {\bibfnamefont {M.}~\bibnamefont
  {Joyce}}, \bibinfo {author} {\bibfnamefont {S.-C.}\ \bibnamefont {Leung}},
  \bibinfo {author} {\bibfnamefont {L.}~\bibnamefont {Moln\'ar}}, \bibinfo
  {author} {\bibfnamefont {M.}~\bibnamefont {Ireland}}, \bibinfo {author}
  {\bibfnamefont {C.}~\bibnamefont {Kobayashi}},\ and\ \bibinfo {author}
  {\bibfnamefont {K.}~\bibnamefont {Nomoto}},\ }\href
  {https://doi.org/10.3847/1538-4357/abb8db} {\bibfield  {journal} {\bibinfo
  {journal} {The Astrophysical Journal}\ }\textbf {\bibinfo {volume} {902}},\
  \bibinfo {pages} {63} (\bibinfo {year} {2020})}\BibitemShut {NoStop}%
\bibitem [{\citenamefont {{Paxton}}\ \emph {et~al.}(2011)\citenamefont
  {{Paxton}}, \citenamefont {{Bildsten}}, \citenamefont {{Dotter}},
  \citenamefont {{Herwig}}, \citenamefont {{Lesaffre}},\ and\ \citenamefont
  {{Timmes}}}]{Paxton2011}%
  \BibitemOpen
  \bibfield  {author} {\bibinfo {author} {\bibfnamefont {B.}~\bibnamefont
  {{Paxton}}}, \bibinfo {author} {\bibfnamefont {L.}~\bibnamefont
  {{Bildsten}}}, \bibinfo {author} {\bibfnamefont {A.}~\bibnamefont
  {{Dotter}}}, \bibinfo {author} {\bibfnamefont {F.}~\bibnamefont {{Herwig}}},
  \bibinfo {author} {\bibfnamefont {P.}~\bibnamefont {{Lesaffre}}},\ and\
  \bibinfo {author} {\bibfnamefont {F.}~\bibnamefont {{Timmes}}},\ }\href
  {https://doi.org/10.1088/0067-0049/192/1/3} {\bibfield  {journal} {\bibinfo
  {journal} {Astrophys. J. Suppl.}\ }\textbf {\bibinfo {volume} {192}},\
  \bibinfo {eid} {3} (\bibinfo {year} {2011})},\ \Eprint
  {https://arxiv.org/abs/1009.1622} {arXiv:1009.1622 [astro-ph.SR]}
  \BibitemShut {NoStop}%
\bibitem [{\citenamefont {Paxton}\ \emph {et~al.}(2013)\citenamefont {Paxton}
  \emph {et~al.}}]{Paxton2013}%
  \BibitemOpen
  \bibfield  {author} {\bibinfo {author} {\bibfnamefont {B.}~\bibnamefont
  {Paxton}} \emph {et~al.},\ }\href {https://doi.org/10.1088/0067-0049/208/1/4}
  {\bibfield  {journal} {\bibinfo  {journal} {Astrophys. J. Suppl.}\ }\textbf
  {\bibinfo {volume} {208}},\ \bibinfo {eid} {4} (\bibinfo {year} {2013})},\
  \Eprint {https://arxiv.org/abs/1301.0319} {arXiv:1301.0319 [astro-ph.SR]}
  \BibitemShut {NoStop}%
\bibitem [{\citenamefont {{Dupree}}\ \emph {et~al.}(2026)\citenamefont
  {{Dupree}}, \citenamefont {{Cristofari}}, \citenamefont {{MacLeod}},\ and\
  \citenamefont {{Kravchenko}}}]{dupree2026betelgeusedetectionexpandingwake}%
  \BibitemOpen
  \bibfield  {author} {\bibinfo {author} {\bibfnamefont {A.~K.}\ \bibnamefont
  {{Dupree}}}, \bibinfo {author} {\bibfnamefont {P.~I.}\ \bibnamefont
  {{Cristofari}}}, \bibinfo {author} {\bibfnamefont {M.}~\bibnamefont
  {{MacLeod}}},\ and\ \bibinfo {author} {\bibfnamefont {K.}~\bibnamefont
  {{Kravchenko}}},\ }\href {https://doi.org/10.3847/1538-4357/ae2ed5}
  {\bibfield  {journal} {\bibinfo  {journal} {The Astrophysical Journal}\
  }\textbf {\bibinfo {volume} {998}},\ \bibinfo {eid} {50} (\bibinfo {year}
  {2026})},\ \Eprint {https://arxiv.org/abs/2601.00470} {arXiv:2601.00470
  [astro-ph.SR]} \BibitemShut {NoStop}%
\bibitem [{\citenamefont {{Howell}}\ \emph {et~al.}(2025)\citenamefont
  {{Howell}}, \citenamefont {{Ciardi}}, \citenamefont {{Clark}}, \citenamefont
  {{Hope}}, \citenamefont {{Littlefield}},\ and\ \citenamefont
  {{Furlan}}}]{howell2025probabledirectimagingdetectionstellar}%
  \BibitemOpen
  \bibfield  {author} {\bibinfo {author} {\bibfnamefont {S.~B.}\ \bibnamefont
  {{Howell}}}, \bibinfo {author} {\bibfnamefont {D.~R.}\ \bibnamefont
  {{Ciardi}}}, \bibinfo {author} {\bibfnamefont {C.~A.}\ \bibnamefont
  {{Clark}}}, \bibinfo {author} {\bibfnamefont {D.~A.}\ \bibnamefont {{Hope}}},
  \bibinfo {author} {\bibfnamefont {C.}~\bibnamefont {{Littlefield}}},\ and\
  \bibinfo {author} {\bibfnamefont {E.}~\bibnamefont {{Furlan}}},\ }\href
  {https://doi.org/10.3847/2041-8213/adeaaf} {\bibfield  {journal} {\bibinfo
  {journal} {The Astrophysical Journal Letters}\ }\textbf {\bibinfo {volume}
  {988}},\ \bibinfo {eid} {L47} (\bibinfo {year} {2025})},\ \Eprint
  {https://arxiv.org/abs/2507.15749} {arXiv:2507.15749 [astro-ph.SR]}
  \BibitemShut {NoStop}%
\bibitem [{\citenamefont {Yu}\ \emph {et~al.}(2025{\natexlab{b}})\citenamefont
  {Yu} \emph {et~al.}}]{COSINEBG}%
  \BibitemOpen
  \bibfield  {author} {\bibinfo {author} {\bibfnamefont {G.~H.}\ \bibnamefont
  {Yu}} \emph {et~al.} (\bibinfo {collaboration} {COSINE-100}),\ }\href
  {https://doi.org/10.1140/epjc/s10052-025-13775-0} {\bibfield  {journal}
  {\bibinfo  {journal} {Eur. Phys. J. C}\ }\textbf {\bibinfo {volume} {85}},\
  \bibinfo {pages} {32} (\bibinfo {year} {2025}{\natexlab{b}})},\ \Eprint
  {https://arxiv.org/abs/2408.09806} {arXiv:2408.09806 [astro-ph.IM]}
  \BibitemShut {NoStop}%
\bibitem [{\citenamefont {Yu}\ and\ \citenamefont {Jeon}()}]{bins}%
  \BibitemOpen
  \bibfield  {author} {\bibinfo {author} {\bibfnamefont {G.}~\bibnamefont
  {Yu}}\ and\ \bibinfo {author} {\bibfnamefont {E.}~\bibnamefont {Jeon}},\
  }\href@noop {} {}\bibinfo {howpublished} {private communication}\BibitemShut
  {NoStop}%
\bibitem [{\citenamefont {Abgrall}\ \emph {et~al.}(2017)\citenamefont {Abgrall}
  \emph {et~al.}}]{LEGEND:2017cdu}%
  \BibitemOpen
  \bibfield  {author} {\bibinfo {author} {\bibfnamefont {N.}~\bibnamefont
  {Abgrall}} \emph {et~al.} (\bibinfo {collaboration} {LEGEND}),\ }\href
  {https://doi.org/10.1063/1.5007652} {\bibfield  {journal} {\bibinfo
  {journal} {AIP Conf. Proc.}\ }\textbf {\bibinfo {volume} {1894}},\ \bibinfo
  {pages} {020027} (\bibinfo {year} {2017})},\ \Eprint
  {https://arxiv.org/abs/1709.01980} {arXiv:1709.01980 [physics.ins-det]}
  \BibitemShut {NoStop}%
\bibitem [{\citenamefont {Gradwohl}\ \emph {et~al.}(2020)\citenamefont
  {Gradwohl}, \citenamefont {Moras}, \citenamefont {Janicsk{\'o}-Cs{\'a}thy},
  \citenamefont {Sch{\"o}nert},\ and\ \citenamefont
  {Sumathi}}]{Gradwohl:2020kms}%
  \BibitemOpen
  \bibfield  {author} {\bibinfo {author} {\bibfnamefont {K.-P.}\ \bibnamefont
  {Gradwohl}}, \bibinfo {author} {\bibfnamefont {O.}~\bibnamefont {Moras}},
  \bibinfo {author} {\bibfnamefont {J.}~\bibnamefont
  {Janicsk{\'o}-Cs{\'a}thy}}, \bibinfo {author} {\bibfnamefont
  {S.}~\bibnamefont {Sch{\"o}nert}},\ and\ \bibinfo {author} {\bibfnamefont
  {R.~R.}\ \bibnamefont {Sumathi}},\ }\href
  {https://doi.org/10.1088/1748-0221/15/12/P12010} {\bibfield  {journal}
  {\bibinfo  {journal} {JINST}\ }\textbf {\bibinfo {volume} {15}}\bibfield
  {number} {\bibinfo  {number} { (12)},\ \bibinfo {pages} {P12010}},\ }\Eprint
  {https://arxiv.org/abs/2009.07585} {arXiv:2009.07585 [physics.ins-det]}
  \BibitemShut {NoStop}%
\bibitem [{\citenamefont {Suerfu}\ \emph {et~al.}(2020)\citenamefont {Suerfu},
  \citenamefont {Wada}, \citenamefont {Peloso}, \citenamefont {Souza},
  \citenamefont {Calaprice}, \citenamefont {Tower},\ and\ \citenamefont
  {Ciampi}}]{PhysRevResearch.2.013223}%
  \BibitemOpen
  \bibfield  {author} {\bibinfo {author} {\bibfnamefont {B.}~\bibnamefont
  {Suerfu}}, \bibinfo {author} {\bibfnamefont {M.}~\bibnamefont {Wada}},
  \bibinfo {author} {\bibfnamefont {W.}~\bibnamefont {Peloso}}, \bibinfo
  {author} {\bibfnamefont {M.}~\bibnamefont {Souza}}, \bibinfo {author}
  {\bibfnamefont {F.}~\bibnamefont {Calaprice}}, \bibinfo {author}
  {\bibfnamefont {J.}~\bibnamefont {Tower}},\ and\ \bibinfo {author}
  {\bibfnamefont {G.}~\bibnamefont {Ciampi}},\ }\href
  {https://doi.org/10.1103/PhysRevResearch.2.013223} {\bibfield  {journal}
  {\bibinfo  {journal} {Phys. Rev. Res.}\ }\textbf {\bibinfo {volume} {2}},\
  \bibinfo {pages} {013223} (\bibinfo {year} {2020})}\BibitemShut {NoStop}%
\bibitem [{\citenamefont {Calaprice}\ \emph {et~al.}(2021)\citenamefont
  {Calaprice}, \citenamefont {Copello}, \citenamefont {Dafinei}, \citenamefont
  {D'Angelo}, \citenamefont {D'Imperio}, \citenamefont {Di~Carlo},
  \citenamefont {Diemoz}, \citenamefont {Di~Giacinto}, \citenamefont
  {Di~Ludovico}, \citenamefont {Ianni}, \citenamefont {Iannone}, \citenamefont
  {Marchegiani}, \citenamefont {Mariani}, \citenamefont {Milana}, \citenamefont
  {Nisi}, \citenamefont {Nuti}, \citenamefont {Orlandi}, \citenamefont
  {Pettinacci}, \citenamefont {Pietrofaccia}, \citenamefont {Rahatlou},
  \citenamefont {Souza}, \citenamefont {Suerfu}, \citenamefont {Tomei},
  \citenamefont {Vignoli}, \citenamefont {Wada},\ and\ \citenamefont
  {Zani}}]{PhysRevD.104.L021302}%
  \BibitemOpen
  \bibfield  {author} {\bibinfo {author} {\bibfnamefont {F.}~\bibnamefont
  {Calaprice}}, \bibinfo {author} {\bibfnamefont {S.}~\bibnamefont {Copello}},
  \bibinfo {author} {\bibfnamefont {I.}~\bibnamefont {Dafinei}}, \bibinfo
  {author} {\bibfnamefont {D.}~\bibnamefont {D'Angelo}}, \bibinfo {author}
  {\bibfnamefont {G.}~\bibnamefont {D'Imperio}}, \bibinfo {author}
  {\bibfnamefont {G.}~\bibnamefont {Di~Carlo}}, \bibinfo {author}
  {\bibfnamefont {M.}~\bibnamefont {Diemoz}}, \bibinfo {author} {\bibfnamefont
  {A.}~\bibnamefont {Di~Giacinto}}, \bibinfo {author} {\bibfnamefont
  {A.}~\bibnamefont {Di~Ludovico}}, \bibinfo {author} {\bibfnamefont
  {A.}~\bibnamefont {Ianni}}, \bibinfo {author} {\bibfnamefont
  {M.}~\bibnamefont {Iannone}}, \bibinfo {author} {\bibfnamefont
  {F.}~\bibnamefont {Marchegiani}}, \bibinfo {author} {\bibfnamefont
  {A.}~\bibnamefont {Mariani}}, \bibinfo {author} {\bibfnamefont
  {S.}~\bibnamefont {Milana}}, \bibinfo {author} {\bibfnamefont
  {S.}~\bibnamefont {Nisi}}, \bibinfo {author} {\bibfnamefont {F.}~\bibnamefont
  {Nuti}}, \bibinfo {author} {\bibfnamefont {D.}~\bibnamefont {Orlandi}},
  \bibinfo {author} {\bibfnamefont {V.}~\bibnamefont {Pettinacci}}, \bibinfo
  {author} {\bibfnamefont {L.}~\bibnamefont {Pietrofaccia}}, \bibinfo {author}
  {\bibfnamefont {S.}~\bibnamefont {Rahatlou}}, \bibinfo {author}
  {\bibfnamefont {M.}~\bibnamefont {Souza}}, \bibinfo {author} {\bibfnamefont
  {B.}~\bibnamefont {Suerfu}}, \bibinfo {author} {\bibfnamefont
  {C.}~\bibnamefont {Tomei}}, \bibinfo {author} {\bibfnamefont
  {C.}~\bibnamefont {Vignoli}}, \bibinfo {author} {\bibfnamefont
  {M.}~\bibnamefont {Wada}},\ and\ \bibinfo {author} {\bibfnamefont
  {A.}~\bibnamefont {Zani}},\ }\href
  {https://doi.org/10.1103/PhysRevD.104.L021302} {\bibfield  {journal}
  {\bibinfo  {journal} {Phys. Rev. D}\ }\textbf {\bibinfo {volume} {104}},\
  \bibinfo {pages} {L021302} (\bibinfo {year} {2021})}\BibitemShut {NoStop}%
\bibitem [{\citenamefont {Burlac}\ and\ \citenamefont
  {Salamanna}(2023)}]{BURLAC2023167943}%
  \BibitemOpen
  \bibfield  {author} {\bibinfo {author} {\bibfnamefont {N.}~\bibnamefont
  {Burlac}}\ and\ \bibinfo {author} {\bibfnamefont {G.}~\bibnamefont
  {Salamanna}} (\bibinfo {collaboration} {LEGEND}),\ }\href
  {https://doi.org/10.1016/j.nima.2022.167943} {\bibfield  {journal} {\bibinfo
  {journal} {Nucl. Instrum. Meth. A}\ }\textbf {\bibinfo {volume} {1048}},\
  \bibinfo {pages} {167943} (\bibinfo {year} {2023})}\BibitemShut {NoStop}%
\bibitem [{\citenamefont {Bhusal}\ \emph {et~al.}(2021)\citenamefont {Bhusal},
  \citenamefont {Houston},\ and\ \citenamefont {Li}}]{PhysRevLett.126.091601}%
  \BibitemOpen
  \bibfield  {author} {\bibinfo {author} {\bibfnamefont {A.}~\bibnamefont
  {Bhusal}}, \bibinfo {author} {\bibfnamefont {N.}~\bibnamefont {Houston}},\
  and\ \bibinfo {author} {\bibfnamefont {T.}~\bibnamefont {Li}},\ }\href
  {https://doi.org/10.1103/PhysRevLett.126.091601} {\bibfield  {journal}
  {\bibinfo  {journal} {Phys. Rev. Lett.}\ }\textbf {\bibinfo {volume} {126}},\
  \bibinfo {pages} {091601} (\bibinfo {year} {2021})}\BibitemShut {NoStop}%
\bibitem [{\citenamefont {Raffelt}\ and\ \citenamefont
  {Stodolsky}(1982)}]{RAFFELT1982323}%
  \BibitemOpen
  \bibfield  {author} {\bibinfo {author} {\bibfnamefont {G.}~\bibnamefont
  {Raffelt}}\ and\ \bibinfo {author} {\bibfnamefont {L.}~\bibnamefont
  {Stodolsky}},\ }\href
  {https://doi.org/https://doi.org/10.1016/0370-2693(82)90680-3} {\bibfield
  {journal} {\bibinfo  {journal} {Physics Letters B}\ }\textbf {\bibinfo
  {volume} {119}},\ \bibinfo {pages} {323} (\bibinfo {year}
  {1982})}\BibitemShut {NoStop}%
\bibitem [{\citenamefont {Rupak}(2000)}]{RUPAK2000405}%
  \BibitemOpen
  \bibfield  {author} {\bibinfo {author} {\bibfnamefont {G.}~\bibnamefont
  {Rupak}},\ }\href
  {https://doi.org/https://doi.org/10.1016/S0375-9474(00)00323-7} {\bibfield
  {journal} {\bibinfo  {journal} {Nuclear Physics A}\ }\textbf {\bibinfo
  {volume} {678}},\ \bibinfo {pages} {405} (\bibinfo {year}
  {2000})}\BibitemShut {NoStop}%
\bibitem [{\citenamefont {Engel}\ \emph {et~al.}(1990)\citenamefont {Engel},
  \citenamefont {Seckel},\ and\ \citenamefont {Hayes}}]{PhysRevLett.65.960}%
  \BibitemOpen
  \bibfield  {author} {\bibinfo {author} {\bibfnamefont {J.}~\bibnamefont
  {Engel}}, \bibinfo {author} {\bibfnamefont {D.}~\bibnamefont {Seckel}},\ and\
  \bibinfo {author} {\bibfnamefont {A.~C.}\ \bibnamefont {Hayes}},\ }\href
  {https://doi.org/10.1103/PhysRevLett.65.960} {\bibfield  {journal} {\bibinfo
  {journal} {Phys. Rev. Lett.}\ }\textbf {\bibinfo {volume} {65}},\ \bibinfo
  {pages} {960} (\bibinfo {year} {1990})}\BibitemShut {NoStop}%
\bibitem [{\citenamefont {Hirata}\ \emph {et~al.}(1987)\citenamefont {Hirata},
  \citenamefont {Kajita}, \citenamefont {Koshiba}, \citenamefont {Nakahata},
  \citenamefont {Oyama}, \citenamefont {Sato}, \citenamefont {Suzuki},
  \citenamefont {Takita}, \citenamefont {Totsuka}, \citenamefont {Kifune},
  \citenamefont {Suda}, \citenamefont {Takahashi}, \citenamefont {Tanimori},
  \citenamefont {Miyano}, \citenamefont {Yamada}, \citenamefont {Beier},
  \citenamefont {Feldscher}, \citenamefont {Kim}, \citenamefont {Mann},
  \citenamefont {Newcomer}, \citenamefont {Van}, \citenamefont {Zhang},\ and\
  \citenamefont {Cortez}}]{PhysRevLett.58.1490}%
  \BibitemOpen
  \bibfield  {author} {\bibinfo {author} {\bibfnamefont {K.}~\bibnamefont
  {Hirata}}, \bibinfo {author} {\bibfnamefont {T.}~\bibnamefont {Kajita}},
  \bibinfo {author} {\bibfnamefont {M.}~\bibnamefont {Koshiba}}, \bibinfo
  {author} {\bibfnamefont {M.}~\bibnamefont {Nakahata}}, \bibinfo {author}
  {\bibfnamefont {Y.}~\bibnamefont {Oyama}}, \bibinfo {author} {\bibfnamefont
  {N.}~\bibnamefont {Sato}}, \bibinfo {author} {\bibfnamefont {A.}~\bibnamefont
  {Suzuki}}, \bibinfo {author} {\bibfnamefont {M.}~\bibnamefont {Takita}},
  \bibinfo {author} {\bibfnamefont {Y.}~\bibnamefont {Totsuka}}, \bibinfo
  {author} {\bibfnamefont {T.}~\bibnamefont {Kifune}}, \bibinfo {author}
  {\bibfnamefont {T.}~\bibnamefont {Suda}}, \bibinfo {author} {\bibfnamefont
  {K.}~\bibnamefont {Takahashi}}, \bibinfo {author} {\bibfnamefont
  {T.}~\bibnamefont {Tanimori}}, \bibinfo {author} {\bibfnamefont
  {K.}~\bibnamefont {Miyano}}, \bibinfo {author} {\bibfnamefont
  {M.}~\bibnamefont {Yamada}}, \bibinfo {author} {\bibfnamefont {E.~W.}\
  \bibnamefont {Beier}}, \bibinfo {author} {\bibfnamefont {L.~R.}\ \bibnamefont
  {Feldscher}}, \bibinfo {author} {\bibfnamefont {S.~B.}\ \bibnamefont {Kim}},
  \bibinfo {author} {\bibfnamefont {A.~K.}\ \bibnamefont {Mann}}, \bibinfo
  {author} {\bibfnamefont {F.~M.}\ \bibnamefont {Newcomer}}, \bibinfo {author}
  {\bibfnamefont {R.}~\bibnamefont {Van}}, \bibinfo {author} {\bibfnamefont
  {W.}~\bibnamefont {Zhang}},\ and\ \bibinfo {author} {\bibfnamefont {B.~G.}\
  \bibnamefont {Cortez}},\ }\href {https://doi.org/10.1103/PhysRevLett.58.1490}
  {\bibfield  {journal} {\bibinfo  {journal} {Phys. Rev. Lett.}\ }\textbf
  {\bibinfo {volume} {58}},\ \bibinfo {pages} {1490} (\bibinfo {year}
  {1987})}\BibitemShut {NoStop}%
\bibitem [{\citenamefont {Bionta}\ \emph {et~al.}(1987)\citenamefont {Bionta},
  \citenamefont {Blewitt}, \citenamefont {Bratton}, \citenamefont {Casper},
  \citenamefont {Ciocio}, \citenamefont {Claus}, \citenamefont {Cortez},
  \citenamefont {Crouch}, \citenamefont {Dye}, \citenamefont {Errede},
  \citenamefont {Foster}, \citenamefont {Gajewski}, \citenamefont {Ganezer},
  \citenamefont {Goldhaber}, \citenamefont {Haines}, \citenamefont {Jones},
  \citenamefont {Kielczewska}, \citenamefont {Kropp}, \citenamefont {Learned},
  \citenamefont {LoSecco}, \citenamefont {Matthews}, \citenamefont {Miller},
  \citenamefont {Mudan}, \citenamefont {Park}, \citenamefont {Price},
  \citenamefont {Reines}, \citenamefont {Schultz}, \citenamefont {Seidel},
  \citenamefont {Shumard}, \citenamefont {Sinclair}, \citenamefont {Sobel},
  \citenamefont {Stone}, \citenamefont {Sulak}, \citenamefont {Svoboda},
  \citenamefont {Thornton}, \citenamefont {van~der Velde},\ and\ \citenamefont
  {Wuest}}]{PhysRevLett.58.1494}%
  \BibitemOpen
  \bibfield  {author} {\bibinfo {author} {\bibfnamefont {R.~M.}\ \bibnamefont
  {Bionta}}, \bibinfo {author} {\bibfnamefont {G.}~\bibnamefont {Blewitt}},
  \bibinfo {author} {\bibfnamefont {C.~B.}\ \bibnamefont {Bratton}}, \bibinfo
  {author} {\bibfnamefont {D.}~\bibnamefont {Casper}}, \bibinfo {author}
  {\bibfnamefont {A.}~\bibnamefont {Ciocio}}, \bibinfo {author} {\bibfnamefont
  {R.}~\bibnamefont {Claus}}, \bibinfo {author} {\bibfnamefont
  {B.}~\bibnamefont {Cortez}}, \bibinfo {author} {\bibfnamefont
  {M.}~\bibnamefont {Crouch}}, \bibinfo {author} {\bibfnamefont {S.~T.}\
  \bibnamefont {Dye}}, \bibinfo {author} {\bibfnamefont {S.}~\bibnamefont
  {Errede}}, \bibinfo {author} {\bibfnamefont {G.~W.}\ \bibnamefont {Foster}},
  \bibinfo {author} {\bibfnamefont {W.}~\bibnamefont {Gajewski}}, \bibinfo
  {author} {\bibfnamefont {K.~S.}\ \bibnamefont {Ganezer}}, \bibinfo {author}
  {\bibfnamefont {M.}~\bibnamefont {Goldhaber}}, \bibinfo {author}
  {\bibfnamefont {T.~J.}\ \bibnamefont {Haines}}, \bibinfo {author}
  {\bibfnamefont {T.~W.}\ \bibnamefont {Jones}}, \bibinfo {author}
  {\bibfnamefont {D.}~\bibnamefont {Kielczewska}}, \bibinfo {author}
  {\bibfnamefont {W.~R.}\ \bibnamefont {Kropp}}, \bibinfo {author}
  {\bibfnamefont {J.~G.}\ \bibnamefont {Learned}}, \bibinfo {author}
  {\bibfnamefont {J.~M.}\ \bibnamefont {LoSecco}}, \bibinfo {author}
  {\bibfnamefont {J.}~\bibnamefont {Matthews}}, \bibinfo {author}
  {\bibfnamefont {R.}~\bibnamefont {Miller}}, \bibinfo {author} {\bibfnamefont
  {M.~S.}\ \bibnamefont {Mudan}}, \bibinfo {author} {\bibfnamefont {H.~S.}\
  \bibnamefont {Park}}, \bibinfo {author} {\bibfnamefont {L.~R.}\ \bibnamefont
  {Price}}, \bibinfo {author} {\bibfnamefont {F.}~\bibnamefont {Reines}},
  \bibinfo {author} {\bibfnamefont {J.}~\bibnamefont {Schultz}}, \bibinfo
  {author} {\bibfnamefont {S.}~\bibnamefont {Seidel}}, \bibinfo {author}
  {\bibfnamefont {E.}~\bibnamefont {Shumard}}, \bibinfo {author} {\bibfnamefont
  {D.}~\bibnamefont {Sinclair}}, \bibinfo {author} {\bibfnamefont {H.~W.}\
  \bibnamefont {Sobel}}, \bibinfo {author} {\bibfnamefont {J.~L.}\ \bibnamefont
  {Stone}}, \bibinfo {author} {\bibfnamefont {L.~R.}\ \bibnamefont {Sulak}},
  \bibinfo {author} {\bibfnamefont {R.}~\bibnamefont {Svoboda}}, \bibinfo
  {author} {\bibfnamefont {G.}~\bibnamefont {Thornton}}, \bibinfo {author}
  {\bibfnamefont {J.~C.}\ \bibnamefont {van~der Velde}},\ and\ \bibinfo
  {author} {\bibfnamefont {C.}~\bibnamefont {Wuest}},\ }\href
  {https://doi.org/10.1103/PhysRevLett.58.1494} {\bibfield  {journal} {\bibinfo
   {journal} {Phys. Rev. Lett.}\ }\textbf {\bibinfo {volume} {58}},\ \bibinfo
  {pages} {1494} (\bibinfo {year} {1987})}\BibitemShut {NoStop}%
\bibitem [{\citenamefont {Bar}\ \emph {et~al.}(2020)\citenamefont {Bar},
  \citenamefont {Blum},\ and\ \citenamefont {D'Amico}}]{PhysRevD.101.123025}%
  \BibitemOpen
  \bibfield  {author} {\bibinfo {author} {\bibfnamefont {N.}~\bibnamefont
  {Bar}}, \bibinfo {author} {\bibfnamefont {K.}~\bibnamefont {Blum}},\ and\
  \bibinfo {author} {\bibfnamefont {G.}~\bibnamefont {D'Amico}},\ }\href
  {https://doi.org/10.1103/PhysRevD.101.123025} {\bibfield  {journal} {\bibinfo
   {journal} {Phys. Rev. D}\ }\textbf {\bibinfo {volume} {101}},\ \bibinfo
  {pages} {123025} (\bibinfo {year} {2020})}\BibitemShut {NoStop}%
\bibitem [{\citenamefont {Lella}\ \emph {et~al.}(2024)\citenamefont {Lella},
  \citenamefont {Carenza}, \citenamefont {Co'}, \citenamefont {Lucente},
  \citenamefont {Giannotti}, \citenamefont {Mirizzi},\ and\ \citenamefont
  {Rauscher}}]{PhysRevD.109.023001}%
  \BibitemOpen
  \bibfield  {author} {\bibinfo {author} {\bibfnamefont {A.}~\bibnamefont
  {Lella}}, \bibinfo {author} {\bibfnamefont {P.}~\bibnamefont {Carenza}},
  \bibinfo {author} {\bibfnamefont {G.}~\bibnamefont {Co'}}, \bibinfo {author}
  {\bibfnamefont {G.}~\bibnamefont {Lucente}}, \bibinfo {author} {\bibfnamefont
  {M.}~\bibnamefont {Giannotti}}, \bibinfo {author} {\bibfnamefont
  {A.}~\bibnamefont {Mirizzi}},\ and\ \bibinfo {author} {\bibfnamefont
  {T.}~\bibnamefont {Rauscher}},\ }\href
  {https://doi.org/10.1103/PhysRevD.109.023001} {\bibfield  {journal} {\bibinfo
   {journal} {Phys. Rev. D}\ }\textbf {\bibinfo {volume} {109}},\ \bibinfo
  {pages} {023001} (\bibinfo {year} {2024})}\BibitemShut {NoStop}%
\bibitem [{\citenamefont {Haxton}\ \emph
  {et~al.}(2026{\natexlab{b}})\citenamefont {Haxton}, \citenamefont {Janka}
  \emph {et~al.}}]{Janka}%
  \BibitemOpen
  \bibfield  {author} {\bibinfo {author} {\bibfnamefont {W.~C.}\ \bibnamefont
  {Haxton}}, \bibinfo {author} {\bibfnamefont {H.~T.}\ \bibnamefont {Janka}},
  \emph {et~al.},\ }\href@noop {} {} (\bibinfo {year} {2026}{\natexlab{b}}),\
  \bibinfo {note} {work in progress}\BibitemShut {NoStop}%
\bibitem [{\citenamefont {Tornow}\ \emph {et~al.}(2003)\citenamefont {Tornow},
  \citenamefont {Czakon}, \citenamefont {Howell}, \citenamefont {Hutcheson},
  \citenamefont {Kelley}, \citenamefont {Litvinenko}, \citenamefont
  {Mikhailov}, \citenamefont {Pinayev}, \citenamefont {Weisel},\ and\
  \citenamefont {Wita\l{}a}}]{TORNOW20038}%
  \BibitemOpen
  \bibfield  {author} {\bibinfo {author} {\bibfnamefont {W.}~\bibnamefont
  {Tornow}}, \bibinfo {author} {\bibfnamefont {N.}~\bibnamefont {Czakon}},
  \bibinfo {author} {\bibfnamefont {C.}~\bibnamefont {Howell}}, \bibinfo
  {author} {\bibfnamefont {A.}~\bibnamefont {Hutcheson}}, \bibinfo {author}
  {\bibfnamefont {J.}~\bibnamefont {Kelley}}, \bibinfo {author} {\bibfnamefont
  {V.}~\bibnamefont {Litvinenko}}, \bibinfo {author} {\bibfnamefont
  {S.}~\bibnamefont {Mikhailov}}, \bibinfo {author} {\bibfnamefont
  {I.}~\bibnamefont {Pinayev}}, \bibinfo {author} {\bibfnamefont
  {G.}~\bibnamefont {Weisel}},\ and\ \bibinfo {author} {\bibfnamefont
  {H.}~\bibnamefont {Wita\l{}a}},\ }\href
  {https://doi.org/https://doi.org/10.1016/j.physletb.2003.08.078} {\bibfield
  {journal} {\bibinfo  {journal} {Physics Letters B}\ }\textbf {\bibinfo
  {volume} {574}},\ \bibinfo {pages} {8} (\bibinfo {year} {2003})}\BibitemShut
  {NoStop}%
\bibitem [{\citenamefont {Brown}\ and\ \citenamefont
  {Green}(1966)}]{BROWN1966401}%
  \BibitemOpen
  \bibfield  {author} {\bibinfo {author} {\bibfnamefont {G.}~\bibnamefont
  {Brown}}\ and\ \bibinfo {author} {\bibfnamefont {A.}~\bibnamefont {Green}},\
  }\href {https://doi.org/https://doi.org/10.1016/0029-5582(66)90771-1}
  {\bibfield  {journal} {\bibinfo  {journal} {Nuclear Physics}\ }\textbf
  {\bibinfo {volume} {75}},\ \bibinfo {pages} {401} (\bibinfo {year}
  {1966})}\BibitemShut {NoStop}%
\bibitem [{\citenamefont {Haxton}\ and\ \citenamefont
  {Johnson}(1990)}]{PhysRevLett.65.1325}%
  \BibitemOpen
  \bibfield  {author} {\bibinfo {author} {\bibfnamefont {W.~C.}\ \bibnamefont
  {Haxton}}\ and\ \bibinfo {author} {\bibfnamefont {C.}~\bibnamefont
  {Johnson}},\ }\href {https://doi.org/10.1103/PhysRevLett.65.1325} {\bibfield
  {journal} {\bibinfo  {journal} {Phys. Rev. Lett.}\ }\textbf {\bibinfo
  {volume} {65}},\ \bibinfo {pages} {1325} (\bibinfo {year}
  {1990})}\BibitemShut {NoStop}%
\bibitem [{\citenamefont {Warburton}\ \emph {et~al.}(1992)\citenamefont
  {Warburton}, \citenamefont {Brown},\ and\ \citenamefont
  {Millener}}]{WARBURTON19927}%
  \BibitemOpen
  \bibfield  {author} {\bibinfo {author} {\bibfnamefont {E.}~\bibnamefont
  {Warburton}}, \bibinfo {author} {\bibfnamefont {B.}~\bibnamefont {Brown}},\
  and\ \bibinfo {author} {\bibfnamefont {D.}~\bibnamefont {Millener}},\ }\href
  {https://doi.org/https://doi.org/10.1016/0370-2693(92)91472-L} {\bibfield
  {journal} {\bibinfo  {journal} {Physics Letters B}\ }\textbf {\bibinfo
  {volume} {293}},\ \bibinfo {pages} {7} (\bibinfo {year} {1992})}\BibitemShut
  {NoStop}%
\bibitem [{\citenamefont {Carenza}\ \emph {et~al.}(2024)\citenamefont
  {Carenza}, \citenamefont {Co'}, \citenamefont {Giannotti}, \citenamefont
  {Lella}, \citenamefont {Lucente}, \citenamefont {Mirizzi},\ and\
  \citenamefont {Rauscher}}]{PhysRevC.109.015501}%
  \BibitemOpen
  \bibfield  {author} {\bibinfo {author} {\bibfnamefont {P.}~\bibnamefont
  {Carenza}}, \bibinfo {author} {\bibfnamefont {G.}~\bibnamefont {Co'}},
  \bibinfo {author} {\bibfnamefont {M.}~\bibnamefont {Giannotti}}, \bibinfo
  {author} {\bibfnamefont {A.}~\bibnamefont {Lella}}, \bibinfo {author}
  {\bibfnamefont {G.}~\bibnamefont {Lucente}}, \bibinfo {author} {\bibfnamefont
  {A.}~\bibnamefont {Mirizzi}},\ and\ \bibinfo {author} {\bibfnamefont
  {T.}~\bibnamefont {Rauscher}},\ }\href
  {https://doi.org/10.1103/PhysRevC.109.015501} {\bibfield  {journal} {\bibinfo
   {journal} {Phys. Rev. C}\ }\textbf {\bibinfo {volume} {109}},\ \bibinfo
  {pages} {015501} (\bibinfo {year} {2024})}\BibitemShut {NoStop}%
\bibitem [{\citenamefont {Hirata}\ \emph {et~al.}(1988)\citenamefont {Hirata},
  \citenamefont {Kajita}, \citenamefont {Koshiba}, \citenamefont {Nakahata},
  \citenamefont {Oyama}, \citenamefont {Sato}, \citenamefont {Suzuki},
  \citenamefont {Takita}, \citenamefont {Totsuka}, \citenamefont {Kifune},
  \citenamefont {Suda}, \citenamefont {Takahashi}, \citenamefont {Tanimori},
  \citenamefont {Miyano}, \citenamefont {Yamada}, \citenamefont {Beier},
  \citenamefont {Feldscher}, \citenamefont {Frati}, \citenamefont {Kim},
  \citenamefont {Mann}, \citenamefont {Newcomer}, \citenamefont {Van~Berg},
  \citenamefont {Zhang},\ and\ \citenamefont {Cortez}}]{PhysRevD.38.448}%
  \BibitemOpen
  \bibfield  {author} {\bibinfo {author} {\bibfnamefont {K.~S.}\ \bibnamefont
  {Hirata}}, \bibinfo {author} {\bibfnamefont {T.}~\bibnamefont {Kajita}},
  \bibinfo {author} {\bibfnamefont {M.}~\bibnamefont {Koshiba}}, \bibinfo
  {author} {\bibfnamefont {M.}~\bibnamefont {Nakahata}}, \bibinfo {author}
  {\bibfnamefont {Y.}~\bibnamefont {Oyama}}, \bibinfo {author} {\bibfnamefont
  {N.}~\bibnamefont {Sato}}, \bibinfo {author} {\bibfnamefont {A.}~\bibnamefont
  {Suzuki}}, \bibinfo {author} {\bibfnamefont {M.}~\bibnamefont {Takita}},
  \bibinfo {author} {\bibfnamefont {Y.}~\bibnamefont {Totsuka}}, \bibinfo
  {author} {\bibfnamefont {T.}~\bibnamefont {Kifune}}, \bibinfo {author}
  {\bibfnamefont {T.}~\bibnamefont {Suda}}, \bibinfo {author} {\bibfnamefont
  {K.}~\bibnamefont {Takahashi}}, \bibinfo {author} {\bibfnamefont
  {T.}~\bibnamefont {Tanimori}}, \bibinfo {author} {\bibfnamefont
  {K.}~\bibnamefont {Miyano}}, \bibinfo {author} {\bibfnamefont
  {M.}~\bibnamefont {Yamada}}, \bibinfo {author} {\bibfnamefont {E.~W.}\
  \bibnamefont {Beier}}, \bibinfo {author} {\bibfnamefont {L.~R.}\ \bibnamefont
  {Feldscher}}, \bibinfo {author} {\bibfnamefont {W.}~\bibnamefont {Frati}},
  \bibinfo {author} {\bibfnamefont {S.~B.}\ \bibnamefont {Kim}}, \bibinfo
  {author} {\bibfnamefont {A.~K.}\ \bibnamefont {Mann}}, \bibinfo {author}
  {\bibfnamefont {F.~M.}\ \bibnamefont {Newcomer}}, \bibinfo {author}
  {\bibfnamefont {R.}~\bibnamefont {Van~Berg}}, \bibinfo {author}
  {\bibfnamefont {W.}~\bibnamefont {Zhang}},\ and\ \bibinfo {author}
  {\bibfnamefont {B.~G.}\ \bibnamefont {Cortez}},\ }\href
  {https://doi.org/10.1103/PhysRevD.38.448} {\bibfield  {journal} {\bibinfo
  {journal} {Phys. Rev. D}\ }\textbf {\bibinfo {volume} {38}},\ \bibinfo
  {pages} {448} (\bibinfo {year} {1988})}\BibitemShut {NoStop}%
\bibitem [{\citenamefont {{Utrobin, V. P.}}\ \emph {et~al.}(2019)\citenamefont
  {{Utrobin, V. P.}}, \citenamefont {{Wongwathanarat, A.}}, \citenamefont
  {{Janka, H.-Th.}}, \citenamefont {{M{\"u}ller, E.}}, \citenamefont {{Ertl,
  T.}},\ and\ \citenamefont {{Woosley, S. E.}}}]{Garching}%
  \BibitemOpen
  \bibfield  {author} {\bibinfo {author} {\bibnamefont {{Utrobin, V. P.}}},
  \bibinfo {author} {\bibnamefont {{Wongwathanarat, A.}}}, \bibinfo {author}
  {\bibnamefont {{Janka, H.-Th.}}}, \bibinfo {author} {\bibnamefont
  {{M{\"u}ller, E.}}}, \bibinfo {author} {\bibnamefont {{Ertl, T.}}},\ and\
  \bibinfo {author} {\bibnamefont {{Woosley, S. E.}}},\ }\href
  {https://doi.org/10.1051/0004-6361/201834976} {\bibfield  {journal} {\bibinfo
   {journal} {A\&A}\ }\textbf {\bibinfo {volume} {624}},\ \bibinfo {pages}
  {A116} (\bibinfo {year} {2019})}\BibitemShut {NoStop}%
\bibitem [{\citenamefont {{Utrobin}}\ \emph {et~al.}(2015)\citenamefont
  {{Utrobin}}, \citenamefont {{Wongwathanarat}}, \citenamefont {{Janka}},\ and\
  \citenamefont {{M{\"u}ller}}}]{Garching2}%
  \BibitemOpen
  \bibfield  {author} {\bibinfo {author} {\bibfnamefont {V.~P.}\ \bibnamefont
  {{Utrobin}}}, \bibinfo {author} {\bibfnamefont {A.}~\bibnamefont
  {{Wongwathanarat}}}, \bibinfo {author} {\bibfnamefont {H.-T.}\ \bibnamefont
  {{Janka}}},\ and\ \bibinfo {author} {\bibfnamefont {E.}~\bibnamefont
  {{M{\"u}ller}}},\ }\href {https://doi.org/10.1051/0004-6361/201425513}
  {\bibfield  {journal} {\bibinfo  {journal} {A\&A}\ }\textbf {\bibinfo
  {volume} {581}},\ \bibinfo {eid} {A40} (\bibinfo {year} {2015})},\ \Eprint
  {https://arxiv.org/abs/1412.4122} {arXiv:1412.4122 [astro-ph.SR]}
  \BibitemShut {NoStop}%
\bibitem [{\citenamefont {Brown}\ and\ \citenamefont {Ricter}(2006)}]{USDB}%
  \BibitemOpen
  \bibfield  {author} {\bibinfo {author} {\bibfnamefont {B.~A.}\ \bibnamefont
  {Brown}}\ and\ \bibinfo {author} {\bibfnamefont {W.~A.}\ \bibnamefont
  {Ricter}},\ }\href {https://doi.org/10.1103.PhysRevC.74.034315} {\bibfield
  {journal} {\bibinfo  {journal} {Phys. Rev. C}\ }\textbf {\bibinfo {volume}
  {74}},\ \bibinfo {pages} {034315} (\bibinfo {year} {2006})}\BibitemShut
  {NoStop}%
\bibitem [{\citenamefont {Honma}\ \emph {et~al.}(2004)\citenamefont {Honma},
  \citenamefont {Otsuka}, \citenamefont {Brown},\ and\ \citenamefont
  {Mizusaki}}]{GXP1}%
  \BibitemOpen
  \bibfield  {author} {\bibinfo {author} {\bibfnamefont {M.}~\bibnamefont
  {Honma}}, \bibinfo {author} {\bibfnamefont {T.}~\bibnamefont {Otsuka}},
  \bibinfo {author} {\bibfnamefont {B.~A.}\ \bibnamefont {Brown}},\ and\
  \bibinfo {author} {\bibfnamefont {T.}~\bibnamefont {Mizusaki}},\ }\href
  {https://doi.org/10.1103.PhysRevC.69.034335} {\bibfield  {journal} {\bibinfo
  {journal} {Phys. Rev. C}\ }\textbf {\bibinfo {volume} {69}},\ \bibinfo
  {pages} {034335} (\bibinfo {year} {2004})}\BibitemShut {NoStop}%
\bibitem [{\citenamefont {Elliott}\ and\ \citenamefont
  {Jackson}(1968)}]{ELLIOTT1968279}%
  \BibitemOpen
  \bibfield  {author} {\bibinfo {author} {\bibfnamefont {J.}~\bibnamefont
  {Elliott}}\ and\ \bibinfo {author} {\bibfnamefont {A.}~\bibnamefont
  {Jackson}},\ }\href
  {https://doi.org/https://doi.org/10.1016/0375-9474(68)90420-X} {\bibfield
  {journal} {\bibinfo  {journal} {Nuclear Physics A}\ }\textbf {\bibinfo
  {volume} {121}},\ \bibinfo {pages} {279} (\bibinfo {year}
  {1968})}\BibitemShut {NoStop}%
\bibitem [{\citenamefont {Ceuleneer}\ \emph {et~al.}(1988)\citenamefont
  {Ceuleneer}, \citenamefont {Vandepeutte},\ and\ \citenamefont
  {Semay}}]{PhysRevC.38.2335}%
  \BibitemOpen
  \bibfield  {author} {\bibinfo {author} {\bibfnamefont {R.}~\bibnamefont
  {Ceuleneer}}, \bibinfo {author} {\bibfnamefont {P.}~\bibnamefont
  {Vandepeutte}},\ and\ \bibinfo {author} {\bibfnamefont {C.}~\bibnamefont
  {Semay}},\ }\href {https://doi.org/10.1103/PhysRevC.38.2335} {\bibfield
  {journal} {\bibinfo  {journal} {Phys. Rev. C}\ }\textbf {\bibinfo {volume}
  {38}},\ \bibinfo {pages} {2335} (\bibinfo {year} {1988})}\BibitemShut
  {NoStop}%
\bibitem [{\citenamefont {Tanner}\ \emph {et~al.}(1964)\citenamefont {Tanner},
  \citenamefont {Thomas},\ and\ \citenamefont {Earle}}]{TANNER196445}%
  \BibitemOpen
  \bibfield  {author} {\bibinfo {author} {\bibfnamefont {N.}~\bibnamefont
  {Tanner}}, \bibinfo {author} {\bibfnamefont {G.}~\bibnamefont {Thomas}},\
  and\ \bibinfo {author} {\bibfnamefont {E.}~\bibnamefont {Earle}},\ }\href
  {https://doi.org/https://doi.org/10.1016/0029-5582(64)90674-1} {\bibfield
  {journal} {\bibinfo  {journal} {Nuclear Physics}\ }\textbf {\bibinfo {volume}
  {52}},\ \bibinfo {pages} {45} (\bibinfo {year} {1964})}\BibitemShut {NoStop}%
\bibitem [{\citenamefont {Bacca}\ \emph {et~al.}(2013)\citenamefont {Bacca},
  \citenamefont {Barnea}, \citenamefont {Hagen}, \citenamefont {Orlandini},\
  and\ \citenamefont {Papenbrock}}]{PhysRevLett.111.122502}%
  \BibitemOpen
  \bibfield  {author} {\bibinfo {author} {\bibfnamefont {S.}~\bibnamefont
  {Bacca}}, \bibinfo {author} {\bibfnamefont {N.}~\bibnamefont {Barnea}},
  \bibinfo {author} {\bibfnamefont {G.}~\bibnamefont {Hagen}}, \bibinfo
  {author} {\bibfnamefont {G.}~\bibnamefont {Orlandini}},\ and\ \bibinfo
  {author} {\bibfnamefont {T.}~\bibnamefont {Papenbrock}},\ }\href
  {https://doi.org/10.1103/PhysRevLett.111.122502} {\bibfield  {journal}
  {\bibinfo  {journal} {Phys. Rev. Lett.}\ }\textbf {\bibinfo {volume} {111}},\
  \bibinfo {pages} {122502} (\bibinfo {year} {2013})}\BibitemShut {NoStop}%
\bibitem [{\citenamefont {Eramzhyan}\ \emph {et~al.}(1986)\citenamefont
  {Eramzhyan}, \citenamefont {Ishkhanov}, \citenamefont {Kapitonov},\ and\
  \citenamefont {Neudatchin}}]{ERAMZHYAN1986229}%
  \BibitemOpen
  \bibfield  {author} {\bibinfo {author} {\bibfnamefont {R.}~\bibnamefont
  {Eramzhyan}}, \bibinfo {author} {\bibfnamefont {B.}~\bibnamefont
  {Ishkhanov}}, \bibinfo {author} {\bibfnamefont {I.}~\bibnamefont
  {Kapitonov}},\ and\ \bibinfo {author} {\bibfnamefont {V.}~\bibnamefont
  {Neudatchin}},\ }\href
  {https://doi.org/https://doi.org/10.1016/0370-1573(86)90136-5} {\bibfield
  {journal} {\bibinfo  {journal} {Physics Reports}\ }\textbf {\bibinfo {volume}
  {136}},\ \bibinfo {pages} {229} (\bibinfo {year} {1986})}\BibitemShut
  {NoStop}%
\bibitem [{\citenamefont {Raffelt}\ and\ \citenamefont
  {Dearborn}(1987)}]{PhysRevD.36.2211}%
  \BibitemOpen
  \bibfield  {author} {\bibinfo {author} {\bibfnamefont {G.~G.}\ \bibnamefont
  {Raffelt}}\ and\ \bibinfo {author} {\bibfnamefont {D.~S.~P.}\ \bibnamefont
  {Dearborn}},\ }\href {https://doi.org/10.1103/PhysRevD.36.2211} {\bibfield
  {journal} {\bibinfo  {journal} {Phys. Rev. D}\ }\textbf {\bibinfo {volume}
  {36}},\ \bibinfo {pages} {2211} (\bibinfo {year} {1987})}\BibitemShut
  {NoStop}%
\bibitem [{\citenamefont {Fukugita}\ \emph {et~al.}(1982)\citenamefont
  {Fukugita}, \citenamefont {Watamura},\ and\ \citenamefont
  {Yoshimura}}]{PhysRevD.26.1840}%
  \BibitemOpen
  \bibfield  {author} {\bibinfo {author} {\bibfnamefont {M.}~\bibnamefont
  {Fukugita}}, \bibinfo {author} {\bibfnamefont {S.}~\bibnamefont {Watamura}},\
  and\ \bibinfo {author} {\bibfnamefont {M.}~\bibnamefont {Yoshimura}},\ }\href
  {https://doi.org/10.1103/PhysRevD.26.1840} {\bibfield  {journal} {\bibinfo
  {journal} {Phys. Rev. D}\ }\textbf {\bibinfo {volume} {26}},\ \bibinfo
  {pages} {1840} (\bibinfo {year} {1982})}\BibitemShut {NoStop}%
\bibitem [{\citenamefont {Dicus}\ \emph {et~al.}(1980)\citenamefont {Dicus},
  \citenamefont {Kolb}, \citenamefont {Teplitz},\ and\ \citenamefont
  {Wagoner}}]{PhysRevD.22.839}%
  \BibitemOpen
  \bibfield  {author} {\bibinfo {author} {\bibfnamefont {D.~A.}\ \bibnamefont
  {Dicus}}, \bibinfo {author} {\bibfnamefont {E.~W.}\ \bibnamefont {Kolb}},
  \bibinfo {author} {\bibfnamefont {V.~L.}\ \bibnamefont {Teplitz}},\ and\
  \bibinfo {author} {\bibfnamefont {R.~V.}\ \bibnamefont {Wagoner}},\ }\href
  {https://doi.org/10.1103/PhysRevD.22.839} {\bibfield  {journal} {\bibinfo
  {journal} {Phys. Rev. D}\ }\textbf {\bibinfo {volume} {22}},\ \bibinfo
  {pages} {839} (\bibinfo {year} {1980})}\BibitemShut {NoStop}%
\end{thebibliography}%
\end{document}